\documentclass[10pt,royal1,ChapterTOCs]{chapmono}
\usepackage{fixltx2e,fix-cm}
\usepackage{graphicx}
\usepackage{makeidx}
\usepackage{multicol}
\usepackage{color}
\usepackage{mathptmx} %uses times font
\usepackage{relsize}

\makeindex

\title{Chapmono Template}

\author{Author}

\begin{document}

\setcounter{page}{7} %to be added title/series pages will occupy earlier pages

\mainmatter
%%Used for multi author book
%%\chapterauthor{Author Name1}\chapteraffiliation{Affiliation text here1}
%%\chapterauthor{Author Name2}\chapteraffiliation{Affiliation text here2}
%%\chapterauthor{Author Name3}\chapteraffiliation{Affiliation text here3}
%%\chapterauthor{Author Name4}\chapteraffiliation{Affiliation text here4}
%%\chapterauthor{Author Name5}\chapteraffiliation{Affiliation text here5}

%%%% General %%%%%%%%%%%%%%
\newcommand{\D}{\mathrm{d}}
\newcommand{\e}{\mathrm{e}}
\newcommand{\half}{\frac{1}{2}}
\newcommand{\vecr}{{\bf r}}
\newcommand{\be}{\begin{equation}}
\newcommand{\ee}{\end{equation}}
\newcommand{\bea}{\begin{eqnarray}}
\newcommand{\eea}{\end{eqnarray}}
\newcommand{\eps}{\varepsilon}

%%%%%%%% Physical Constants%%%%%%%%%
\newcommand{\kb}{k_{\mathrm{B}}}
\newcommand{\kbt}{k_{\mathrm{B}}T}

%%%%%%% Typical lengths of Poisson-Boltzmann Theory %%%%%%%
%\newcommand{\lb}{l_\mathrm{B}}
\newcommand{\lb}{\ell_\mathrm{B}}
\newcommand{\ld}{\lambda_\mathrm{D}}
\newcommand{\kd}{\kappa_\mathrm{D}}
\newcommand{\lgc}{\ell_\mathrm{GC}}
\newcommand{\lgcm}{\lambda^\mathrm{m}_\mathrm{GC}}
\newcommand{\ls}{\lambda_\mathrm{s}}

%%%%%%%% Special notations for the binary mixture problem %%%%%%%%%%
\newcommand{\phia}{\phi_{\mathrm{A}}}
\newcommand{\phib}{\phi_{\mathrm{B}}}
\newcommand{\epsa}{\varepsilon_{\mathrm{A}}}
\newcommand{\epsb}{\varepsilon_{\mathrm{B}}}
\newcommand{\epsz}{\varepsilon_{0}}
\newcommand{\epsr}{\varepsilon_{\mathrm{r}}}
\newcommand{\epsavg}{\varepsilon_{\mathrm{av}}}
\newcommand{\epsn}{\varepsilon(n)}
\newcommand{\epss}{\varepsilon_s}
\newcommand{\alp}{\alpha_{i}}
\newcommand{\alppm}{\alpha_{\pm}}
\newcommand{\elfs}{\psi\,'_{\mathrm{s}}}
\newcommand{\phis}{\phi_{\mathrm{s}}}

\newcommand{\RP}[1]{\textcolor{red} { #1 }}
\newcommand{\DB}[1]{\textcolor{green} { #1 }}

%%%%%%%%%%%%%%%%%%%%%%%%%%%%%%%%%%%%%%%%%%%%%%%%%%%%%%%%%%%%%%%%%%%%%%%%%%%%%%%%%%%%%
\chapterauthor{Tomer Markovich$^\star$, David Andelman$^\star$ and
Rudi Podgornik$^\dagger$}
\chapteraffiliation{{\small $^\star$School of Physics and Astronomy,
Tel Aviv University, Ramat Aviv 69978, Tel Aviv, Israel\\
$^\dagger$Department of Theoretical Physics, J. Stefan Institute
and Department of Physics, Faculty of Mathematics
and Physics, University of Ljubljana, SI-1000 Ljubljana, Slovenia \\
{\it From the forthcoming Handbook of Lipid Membranes: Molecular, Functional, and Materials Aspects.
Edited by Cyrus Safinya and Joachim R\"adler, Taylor \& Francis/CRC Press, 2016}}}

\chapter{Charged Membranes: Poisson-Boltzmann theory, DLVO paradigm and beyond}[]%[Markovich Tomer, Andelman David and Podgornik Rudolf]
\label{ch9}
%\date{}
%%%%%%%%%%%%%%%%%%%%%%%%%%%%%%%%%%%%%%%%%%%%%%%%%%%%%%%%%%%%%%%%%%%%%%%%%%%%%%%%%%%%%

\begin{abstract}
%June 25, 2015, Version submitted

In this chapter we review the electrostatic properties of charged membranes
in aqueous solutions, with or without added salt, employing simple physical models.
The equilibrium ionic profiles close to the membrane are governed by the well-known
{\it Poisson-Boltzmann} (PB) equation. We analyze the effect of different boundary conditions, imposed by the membrane,
on the ionic profiles and  the corresponding osmotic pressure.
The discussion is separated into the single membrane case and that of two interacting membranes.
For the one membrane setup, we show the different solutions of the PB equation and discuss the interplay between
constant-charge and constant-potential boundary conditions.
A modification of the Poisson-Boltzmann theory is presented to treat
the extremely high counter-ion concentration in the vicinity of a charge membrane.
The two membranes setup is reviewed extensively.
For two equally-charged membranes, we analyze the different pressure regimes for the constant-charge boundary condition,
and discuss the difference in the osmotic pressure for various boundary conditions.
The non-equal charged membranes is reviewed as well, and the crossover from repulsion to attraction is calculated analytically
for two limiting salinity regimes  (Debye-H\"uckle and counter-ions only), as well as for general salinity.
We then examine the charge-regulation boundary condition and discuss its effects on the ionic profiles
and the osmotic pressure for two equally-charged membranes.
In the last section, we  briefly review the van der Waals interactions and their effect on the
free energy between two planar membranes.
We explain the simple {\it Hamaker pair-wise summation} procedure,
and introduce the more rigorous Lifshitz theory. The latter is a key ingredient in
the DLVO theory, which combines repulsive electrostatic with attractive van der Waals interactions, and offers a
simple explanation for colloidal or membrane stability. Finally, the chapter ends by a short account of
the limitations of the approximations inherent in the PB theory.

\end{abstract}

\newpage
%% {\bf Notation:} $x$ scale, ${\bf x}$ vector (column vector), $X$ matrix.

%\baselineskip 20pt

%%%%%%%%%%%%%%%%%%%%%%%%%%%%%%%%%%%%%%%%%%%%%%%%%%%%%%%%%%%%%%%%%%%%%%%%%%%%%%%%%%%%%
\section{Introduction}  \label{sec9.1}
%%%%%%%%%%%%%%%%%%%%%%%%%%%%%%%%%%%%%%%%%%%%%%%%%%%%%%%%%%%%%%%%%%%%%%%%%%%%%%%%%%%%%

It is of great importance to understand electrostatic interactions
and their key role in soft and biological matter.
These systems typically consist of aqueous environment in which
charges tend to dissociate and affect a wide variety of functional, structural
and dynamical properties.
Among the numerous effects of electrostatic interactions, it is instructive to
mention their effect on elasticity of flexible charged polymers (polyelectrolytes) and cell membranes,
formation of self-assembled charged micelles, and stabilization
of charged colloidal suspensions that results from the competition
between repulsive electrostatic interactions and attractive van der Waals interactions
(Verwey and Overbeek 1948, Andelman 1995, 2005, Holm, Kekicheff and Podgornik 2000, Dean el al., 2014,
Churaev, Derjaguin and Muller 2014).

In this chapter, we focus on charged membranes.
Biological membranes are complex heterogeneous two-dimensional interfaces separating the living cell
from its extra-cellular surrounding.
Other membranes surround inter-cellular organelles such as the cell nucleus,
golgi apparatus, mitochondria, endoplasmic reticulum and ribosomes.
Electrostatic interactions control many of the membrane structural properties and functions,
%(Verwey and Overbeek 1948, Israelachvili 2011).
{\it e.g.}, rigidity, structural
stability, lateral phase transitions, and dynamics.
Moreover, electric charges are a key player in processes involving more than
one membrane such as membrane adhesion and cell-cell interaction, as well as the
overall interactions of membranes with
other intra- and extra-cellular proteins, bio-polymers and DNA.

How do membranes interact with their surrounding ionic solution?
Charged membranes attract a cloud of oppositely charged mobile ions that forms
a {\it diffusive electric double layer}
(Gouy 1910, 1917, Chapman 1913, Debye and H\"uckel 1923, Verwey and Overbeek 1948, Israelachvili 2011).
The system favors {\it local electro-neutrality},
but while achieving it, entropy is lost.
The competition between electrostatic interactions and entropy of
ions in solution determines the exact distribution of mobile ions
close to charged membranes.
This last point shows the significance of temperature in determining the equilibrium properties,
because temperature controls the strength of entropic effects as compared to electrostatic interactions.
For soft materials, the thermal energy $\kbt$ is also comparable to other characteristic energy
scales associated with elastic deformations and structural degrees of freedom.

It is convenient to introduce a length scale for which the thermal energy is equal to the Coulombic
energy between two unit charges. This is called the {\it Bjerrum length}, defined as:
\begin{eqnarray}
\label{a1}
\lb = \frac{e^2}{4\pi\varepsilon_0\varepsilon_w\kbt} \, ,
\end{eqnarray}
where $e$ is the elementary charge, $\varepsilon_0=8.85\cdot 10^{-12} {\rm [F/m]}$
is the vacuum permittivity\footnote{Throughout this chapter we use the SI unit system.}
and the dimensionless dielectric constant of water is $\varepsilon_w = 80$.
The Bjerrum length is equal to about $0.7 \, {\rm nm}$ at room temperatures, $T = 300 \, {\rm K}$.

A related length is the {\it Gouy-Chapman length} defined as
\begin{eqnarray}
\label{a2}
\lgc = \frac{2\varepsilon_0\varepsilon_w\kbt}{e |\sigma|} = \frac{e}{2\pi\lb |\sigma|} \sim \sigma^{-1} \, .
\end{eqnarray}
At this length scale, the thermal energy is equal to the Coulombic
energy between a unit charge and a planar surface with a constant surface-charge density, $\sigma$.
The Gouy-Chapman length, $\lgc$, is inversely proportional to $\sigma$.
For strongly charged membranes, $\lgc$ is rather small, on the order
of a tenth of nanometer.

In their pioneering work of almost a century ago,
Debye and H\"{u}ckel introduced the important concept
of screening of the electrostatic interactions between two charges in presence of all other
cations and anions of the solution (Debye and H\"{u}ckel 1923).
This effectively limits the range of electrostatic interactions as will be further discussed below.
The characteristic length for which the electrostatic interactions are screened is called the
{\it Debye length}, $\ld$, defined for monovalent 1:1 electrolyte, as
\begin{eqnarray}
\label{a3}
\ld = \kd^{-1} = \left( 8\pi\lb n_b \right)^{-1/2} \simeq \frac{0.3 {\rm [nm]}}{\sqrt{n_b [{\rm M}]} }\, ,
\end{eqnarray}
with $n_b$ being the salt concentration (in molar),
and $\kd$ is the inverse Debye length.
The Debye screening length for 1:1 monovalent salts varies from about $0.3 \, {\rm nm}$ in
strong ionic solutions of $1$M to about $1 {\rm \mu m}$ in pure water,
where the concentration of the dissociated OH$^-$ and H$^+$ ions is $10^{-7}$M.

The aim of this chapter is to review some of the basic considerations underlying
the behavior of charged membranes in aqueous solutions using the three
important length-scales introduced above.
We will not account for the detailed structure of real biological membranes,
which can add considerable complexity,
but restrict ourselves to simple model systems,
relying on several assumptions and simplifications.
The membrane is treated as a flat interface with a continuum surface charge distribution
or constant surface potential.
The mobile charge distributions are continuous and we disregard
the discreteness of surface charges that can lead to multipolar charge distributions.

This chapter is focused only on static properties in thermodynamic equilibrium,
excluding the interesting phenomena of dynamical fluctuations and dynamical responses
to external fields (such as in electrochemistry systems).
We mainly treat the mean-field approximation
of the electric double-layer problem and the solutions of the classical
Poisson Boltzmann (PB) equation.
Nevertheless, some effects of fluctuations and correlations will be briefly discussed in section~\ref{sec9.9}.
We will also discuss the ion finite-size in section~\ref{sec9.4},
where the `Modified PB equation' is introduced.

%%%%%%%  some references %%%%%%%
The classical reference for the electric double layer is the book of Verwey and Overbeek (1948),
which explains the DLVO (Derjaguin-Landau-Verwey-Overbeek) theory for stabilization of charged colloidal systems.
More recent treatments can be found in many textbooks and monographs on colloidal science and interfacial
phenomena, such as Evans and Wennerstr\"om (1999), Israelachvili (2011),
and in two reviews by one of the present authors, Andelman (1995, 2005).
%%%%%%%%%%%%%%%%%%%%

%%%%%%%%%%%%%%%%%%%%%%%%%%%%%%%%%%%%%%%%%%%%%%%%%%%%%%%%%%%%%%%%%%%%%%%%%%%%%%%%%%%%%
\section{Poisson-Boltzmann Theory}  \label{sec9.2}
%%%%%%%%%%%%%%%%%%%%%%%%%%%%%%%%%%%%%%%%%%%%%%%%%%%%%%%%%%%%%%%%%%%%%%%%%%%%%%%%%%%%%

\begin{figure}%1
\centering
\includegraphics[scale=0.5]{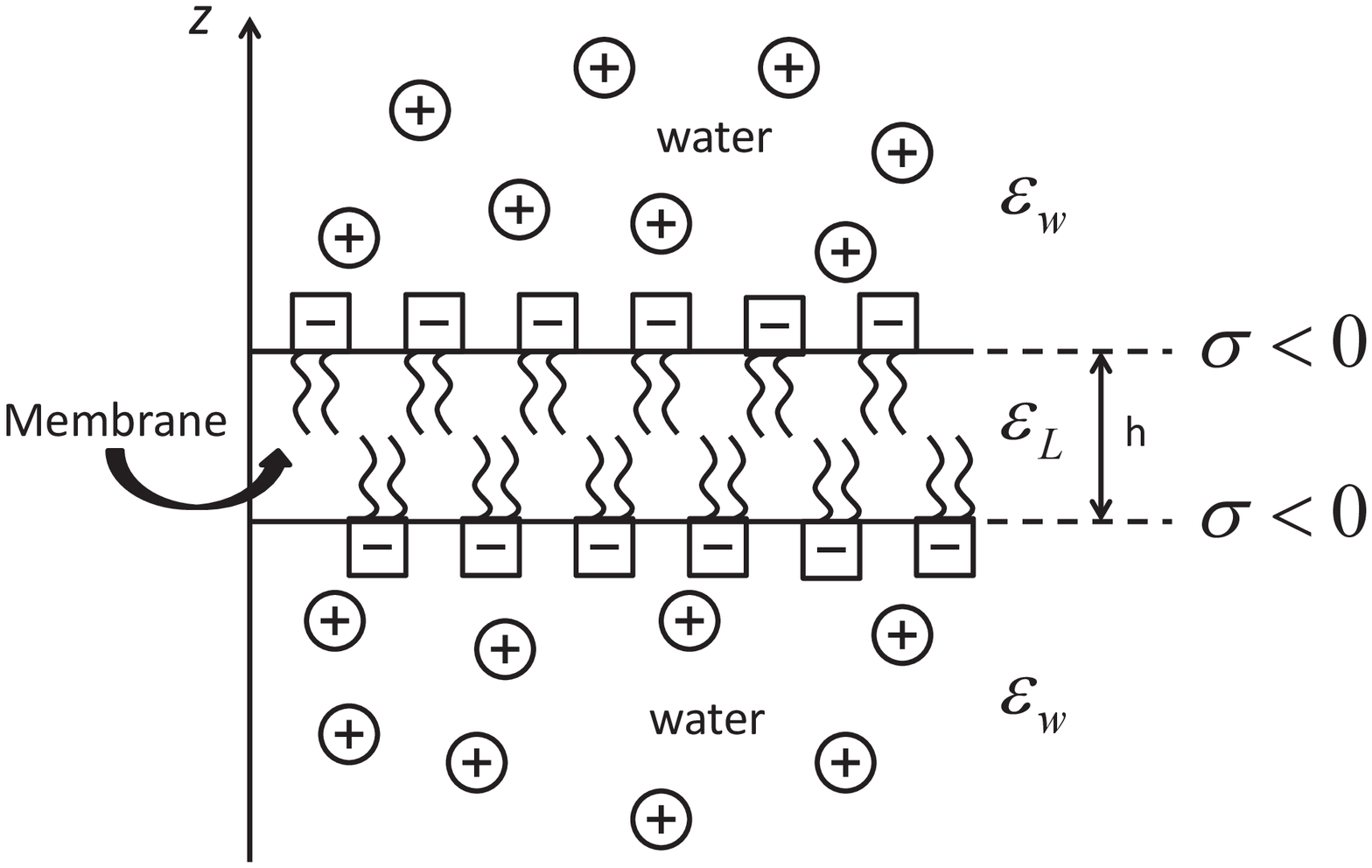}
\caption{\textsf{ A bilayer membrane of thickness $h$
composed of two monolayers (leaflets),
each having a negative charge density, $\sigma<0$.
The core membrane region (hydrocarbon tails)
is modeled as a continuum medium with a dielectric constant $\varepsilon_{\rm L}$,
while the embedding medium (top and bottom) is water and has a dielectric constant, $\varepsilon_{\rm w}$.  }}
\label{figure1}
\end{figure}

In Fig.~\ref{figure1}, a schematic view of a charged amphiphilic (phospholipid) membrane is presented.
A membrane of thickness $h\simeq 4 \, {\rm nm}$ is composed of two monomolecular leaflets packed in a back-to-back configuration.
The constituting molecules are amphiphiles having a charge `head' and a hydrocarbon hydrophobic `tail'.
For phospholipids, the amphiphiles have a double tail.
We model the membrane as a medium of thickness $h$ having a dielectric constant, $\varepsilon_{\rm L}$,
coming essentially from the closely packed hydrocarbon (`oily') tails.
The molecular heads contribute to the surface charges and the entire
membrane is immersed in an aqueous solution characterized by another dielectric constant, $\varepsilon_{\rm w}$,
assumed to be the water dielectric constant throughout the fluid.
The membrane charge can have two origins: either a charge group ({\it e.g.}, ${\rm H}^+$)
dissociates from the polar head-group into the aqueous solution, leaving behind an oppositely
charged group in the membrane; or, an ion from the solution ({\it e.g.}, ${\rm Na}^+$)
binds to a neutral site on the membrane and charges it (Borkovec, J\"onsson and Koper 2001).
These association/dissociation processes are highly sensitive to the ionic strength and pH of the aqueous solution.

\begin{figure}%2
\centering
\includegraphics[scale=0.5]{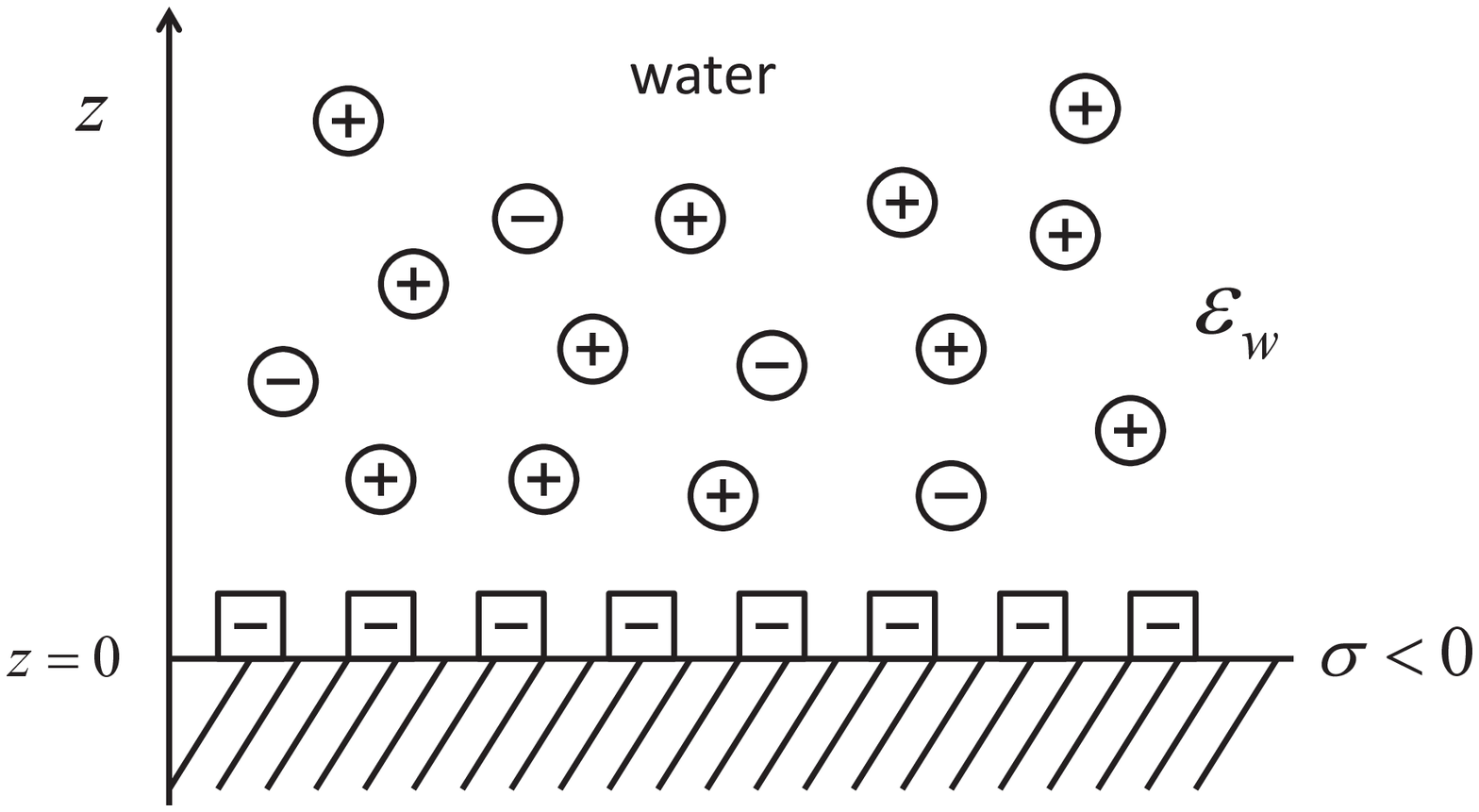}
\caption{\textsf{Schematic illustration of a charged membrane, located at
$z = 0$, with charge density $\sigma$.
Without lost of generality, we take $\sigma < 0$.
For the counter-ion only case, the surface charge is neutralized
by the positive counter-ions.
When monovalent (1:1) electrolyte is added to the reservoir,
its bulk ionic density is $n_{\pm}^{(b)} = n_b$.}
}
\label{figure2}
\end{figure}

When the ionic association/dissociation is slow as compared to the system experimental times,
the charges on the membrane can be considered as fixed and time independent,
while for rapid association/dissociation, the surface charge can vary and
is determined self-consistently from the thermodynamical equilibrium equations.
We will further discuss the two processes of association/dissociation in section~\ref{sec9.7}.
In many situations, the finite thickness of the membrane can be safely taken
to be zero,
% (Kiometzis and Kleinert 1989, Winterhalter and Helfrich 1992),
with the membrane modeled as a planar surface displayed in Fig.~\ref{figure2}.
We will see later under what conditions this simplifying limit is valid.

Let us consider such an ideal membrane represented by a sharp boundary (located at $z = 0$)
that limits the ionic solution to the positive half space.
The ionic solution contains, in general, the two species of mobile ions (anions and cations),
and is modeled as a continuum dielectric medium as explained above.
Thus, the boundary at $z = 0$ marks the discontinuous jump of the
dielectric constant between the ionic solution ($\varepsilon_w$) and the membrane
($\varepsilon_L$), which the ions cannot penetrate.

The PB equation can be obtained using two different approaches.
The first is the one we present below combining the Poisson equation with the Boltzmann distribution,
while the second one (presented later) is done through a minimization of the system free-energy functional.
The PB equation is a mean-field (MF) equation, which can be derived from a field theoretical
approach as the zeroth-order in a systematic expansion of the grand-partition function
(Podgornik and \v{Z}ek\v{s} 1988, Borukhov, Andelman and Orland 1998, 2000,
Netz and Orland 2000, Markovich, Andelman and Podgornik 2014, 2015).

Consider $M$ ionic species, each of them with charge $q_i$,
where $q_i = e z_i$ and $z_i$ is the valency of the $i^{th}$ ionic species.
It is negative ($z_i < 0$) for anions and positive ($z_i > 0$) for cations.
The mobile charge density (per unit volume) is defined as $\rho({\bf r}) = \sum_{i=1}^M q_i n_i({\bf r})$
with $n_i({\bf r})$ being the number density (per unit volume),
and both $\rho$ and $n_i$ are continuous functions of ${\bf r}$.

In MF approximation, each of the ions sees a local environment
constituting of all other ions, which dictates a local electrostatic potential $\psi({\bf r})$.
The potential $\psi({\bf r})$ is a continuous  function that depends
on the total charge density through the Poisson equation:
\begin{eqnarray}
\label{b1}
\nabla^2 \psi({\bf r}) = -\frac{\rho_{\rm tot}({\bf r})}{\varepsilon_0\varepsilon_w} = -\frac{1}{\varepsilon_0\varepsilon_w} \left[ \sum_{i=1}^M q_i n_i({\bf r}) +\rho_f({\bf r})\right]\, ,
\end{eqnarray}
where $\rho_{\rm tot} = \rho + \rho_f$ is the total charge density
and $\rho_f({\bf r})$ is a fixed external charge contribution.
As stated above, the aqueous solution (water) is modeled as a continuum featureless medium.
This by itself represents an approximation because the ions themselves can change the local dielectric
response of the medium (Ben-Yaakov, Andelman and Podgornik 2011, Levy, Andelman and Orland 2012)
by inducing strong localized electric field.
However, we will not include such refined local effects in this review.

The ions dispersed in solution are mobile and are allowed to adjust their positions.
As each ionic species is in thermodynamic equilibrium, its density
obeys the Boltzmann distribution:
\begin{eqnarray}
\label{b2}
n_i({\bf r}) = n_i^{(b)}\e^{-\beta q_i\psi({\bf r})} \, ,
\end{eqnarray}
where $\beta = 1/\kbt$, and $n_i^{(b)}$ is the bulk density of $i^{th}$
species taken at zero reference potential, $\psi = 0$.

%%%%%%%%%%%%%%%%%%%%%%%%%%%%%%%
\begin{shadebox}
\Boxhead{Boltzmann distribution via electrochemical potential}
%{\bf Boltzmann distribution via electrochemical potential.}
A simple derivation of the Boltzmann distribution is obtained through the requirement that the
{\it electrochemical potential} (total chemical potential) $\mu_i^{\rm tot}$,
for each ionic species is constant throughout the system
\begin{eqnarray}
\label{b3}
\mu_i^{\rm tot} = \mu_i({\bf r}) + q_i\psi({\bf r}) = {\rm const} \, ,
\end{eqnarray}
where $\mu_i({\bf r})$ is the intrinsic chemical potential.
For dilute ionic solutions, the $i^{th}$ ionic species entropy is taken as an ideal gas one,
$\mu_i({\bf r}) = \kbt\ln\left[ n_i({\bf r})a^3\right]$.
By substituting $a^3 n_i^{(b)} = \exp(\beta\mu_i^{\rm tot})$ into Eq.~(\ref{b3}),
the Boltzmann distribution of Eq.~(\ref{b2}) follows.
This relation between the bulk ionic density and chemical potential
is obtained by setting $\psi = 0$ in the bulk,
and shows that one can consider the chemical potential, $\mu_i^{\rm tot}$, as a Lagrange multiplier
setting the bulk densities to be $n_i^{(b)}$.
Note that we have introduced a microscopic length scale, $a$,
defining a reference close-packing density, $1/a^3$.
Equation~(\ref{b3}) assumes that the ions are point-like and have
no other interactions in addition to their electrostatic one.

\end{shadebox}
%%%%%%%%%%%%%%%%%%%%%%%%%%%%%%%%%%

We now substitute Eq.~(\ref{b2}) into Eq.~(\ref{b1}) to obtain the {\it Poisson-Boltzmann Equation},
\begin{eqnarray}
\label{b5}
\nabla^2 \psi({\bf r}) = -\frac{1}{\varepsilon_0\varepsilon_w}
\left[ \sum_{i=1}^M q_i n_i^{(b)}\e^{-\beta q_i\psi({\bf r})} + \rho_f({\bf r})\right]\, .
\end{eqnarray}
For binary monovalent electrolytes (denoted as 1:1 electrolyte),
$z_i = \pm 1$, the PB equation reads,
\begin{eqnarray}
\label{b5a}
\nabla^2 \psi({\bf r}) =  \frac{1}{\varepsilon_0\varepsilon_w} \Big[ 2en_b\sinh\left[\beta e \psi({\bf r}) \right] -\rho_f(\vecr) \Big] \, .
\end{eqnarray}
%
%where the only fixed charges, $\rho_f$, are taken to be on the boundaries.

Generally speaking, the PB theory is a very useful analytical approximation
with many applications. It is a good approximation at
physiological conditions (electrolyte strength of about $0.1\,{\rm M}$),
and for other dilute monovalent electrolytes and moderate surface potentials and surface charge.
Although the PB theory produces good results in these situations,
it misses some important features associated with
charge correlations and fluctuations of multivalent counter-ions.
Moreover, close to a charged membrane, the finite size of the surface ionic groups and that
of the counter-ions lead to deviations from the PB results
(see sections~\ref{sec9.4} and \ref{sec9.7} for further details).

As the PB equation is a non-linear equation, it can be solved analytically only for a limited
number of simple boundary conditions. On the other hand, by solving it
numerically or within further approximations or limits, one can obtain ionic
profiles and free energies of complex structures.
For example, the free energy change for a charged globular protein that binds onto an oppositely charged lipid membrane.

%%%%%%%%%%%%%% free energy derivation %%%%%%%%%%%%%%%%%%%%%
In an alternative approach, the PB equation can also be obtained by a minimization of
the system free-energy functional.
One can assume that the internal energy, $U_{\rm el}$, is purely electrostatic,
and that the Helmholtz free-energy, $F = U_{\rm el} - TS$, is composed of an internal energy
and an ideal mixing entropy, $S$, of a dilute solution of mobile ions.

The electrostatic energy, $U_{\rm el}$, is expressed in terms
of the potential $\psi({\bf r})$:
\begin{eqnarray}
\label{b6}
U_{\rm el} = \frac{\varepsilon_0\varepsilon_w}{2}\int_V\D^3r\left| \nabla\psi({\bf r}) \right|^2
=  \half \int_V\D^3r \left[  \sum_{i=1}^M \, q_i n_i({\bf r})\psi(\vecr) + \rho_f({\bf r}) \psi({\bf r}) \right]\, ,
\end{eqnarray}
while the mixing entropy of ions is written in the dilute solution limit as,
\begin{eqnarray}
\label{b7}
S = -\kb \sum_{i=1}^M \int\D^3r \Bigg( n_i({\bf r})\ln\left[ n_i({\bf r}) a^3 \right] - n_i({\bf r})  \Bigg) \, .
\end{eqnarray}
Using Eqs.~(\ref{b6}) and (\ref{b7}), the Helmholtz free-energy can be written as
\begin{eqnarray}
\label{b13}
F &=& \int_V\D^3r \left[ -\frac{\varepsilon_0\varepsilon_w}{2}
\left| \nabla\psi({\bf r}) \right|^2 + \left( \sum_{i=1}^M  q_i n_i({\bf r}) + \rho_f({\bf r}) \right) \psi(\vecr)  \right. \\
\nonumber&+& \left. \kbt \sum_{i=1}^M  \Bigg( n_i({\bf r})\ln\left[ n_i({\bf r}) a^3 \right] - n_i({\bf r})  \Bigg)  \right] \, ,
\end{eqnarray}
where the sum of the first two terms is equal to $U_{\rm el}$ and the third one is $-TS$.
The variation of this free energy with respect to $\psi(\vecr)$, $\delta F/\delta\psi = 0$, gives the Poisson equation, Eq.~(\ref{b1}),
while from the variation with respect to $n_i(\vecr)$, $\delta F/\delta n_i = \mu_i^{\rm tot}$,
we obtain the electrochemical potential of Eq.~(\ref{b3}).
As before, substituting the Boltzmann distribution obtained from Eq.~(\ref{b3}), into the Poisson equation, Eq.~(\ref{b1}),
gives the PB equation, Eq.~(\ref{b5}).

%%%%%%%%%%%%%%%%%%%%%%%%%%%%%%%%%%%%%%%%%%%%%%%%%%%%%%%%%%%%%%%%%%%%%%%%%%%%%%%%%%%%%
\subsection{Debye-H\"uckel Approximation} \label{sec9.2.1}
%%%%%%%%%%%%%%%%%%%%%%%%%%%%%%%%%%%%%%%%%%%%%%%%%%%%%%%%%%%%%%%%%%%%%%%%%%%%%%%%%%%%%

A useful and quite tractable approximation to the non-linear PB equation is its linearized version.
For electrostatic potentials smaller than $25 {\rm mV}$ at room temperature
(or equivalently $e|\psi| < \kbt \, ,\, T\simeq 300 {\rm K}$),
this approximation can be justified and the well-known Debye-H\"uckel (DH) theory is recovered.
Linearization of Eq.~(\ref{b5}) is obtained by expanding
its right-hand side to first order in $\psi$,
\begin{eqnarray}
\label{c1}
\nabla^2 \psi({\bf r}) = -\frac{1}{\varepsilon_0\varepsilon_w} \sum_{i=1}^M q_i n_i^{(b)} +
 8\pi\lb I \psi({\bf r})  - \frac{1}{\varepsilon_0\varepsilon_w}\rho_f({\bf r})  \, ,
\end{eqnarray}
where $I  = \half \sum_{i=1}^M z_i^2 n_i^{(b)}$ is the {\it ionic strength} of the solution.
The first term on the right-hand side of Eq.~(\ref{c1}) vanishes because of electro-neutrality of the bulk reservoir,
\begin{eqnarray}
\label{b6a}
\sum_{i=1}^M q_i n_i^{(b)} = 0 \, ,
\end{eqnarray}
recovering the Debye-H\"uckel equation:
\begin{eqnarray}
\label{c2}
\nabla^2 \psi({\bf r}) =  \kd^2  \psi({\bf r})  - \frac{1}{\varepsilon_0\varepsilon_w} \rho_f({\bf r})  \, ,
\end{eqnarray}
with the inverse Debye length, $\kd$, defined as,
\begin{eqnarray}
\label{c3}
\kd^2 = \ld^{\!-2} = 8\pi\lb I = 4\pi\lb\sum_{i=1}^M z_i^2 n_i^{(b)}\, .
\end{eqnarray}
For monovalent electrolytes, $z_i = \pm 1$, $\kd^2 = 8\pi\lb n_b$ with $n_i^{(b)} = n_b$,
and Eq.~(\ref{a3}) is recovered.
Note that the Debye length, $\ld = \kd^{-1} \sim n_b^{^{-1/2}}$,
is a decreasing function of the salt concentration.

The DH treatment gives a simple tractable description of the pair interactions
between ions.
It is related to the Green function associated with
the electrostatic potential around a point-like ion, and can be calculated
by using Eq.~(\ref{c2}) for a point-like charge, $q$,  placed at the origin, $\vecr = 0$,
$\rho_f({\bf r}) = q\delta(r)$,
\begin{eqnarray}
\label{c4}
\left( \nabla^2 - \kd^2 \right) \psi({\bf r}) = -\frac{q}{\varepsilon_0\varepsilon_w} \delta(\vecr)  \, ,
\end{eqnarray}
where $\delta(\vecr)$ is the Dirac $\delta$-function.
The solution to the above equation can be written in spherical coordinates as,
\begin{eqnarray}
\label{c5}
\psi(r) = \frac{q}{4\pi\varepsilon_0\varepsilon_w r} \, \e^{-\kd r}  \, .
\end{eqnarray}
It manifests the exponential decay of the electrostatic potential with a characteristic length scale, $\ld = 1/\kd$.
In a crude approximation, this exponential decay is replaced by a
Coulombic interaction, which is only slightly screened for $r \leq \ld$ and, thus,
varies as $\sim r^{-1}$, while for $r > \ld$, $\psi(r)$ is strongly screened and can sometimes be completely neglected.

%%%%%%%%%%%%%%%%%%%%%%%%%%%%%%%%%%%%%%%%%%%%%%%
\section{One Planar Membrane} \label{sec9.3}
%%%%%%%%%%%%%%%%%%%%%%%%%%%%%%%%%%%%%%%%%%%%%%%

We consider the PB equation for a single membrane assumed to be planar and charged,
and discuss separately two cases:
(i) a charged membrane in contact with a solution containing
only counter-ions,
and (ii) a membrane in contact with a monovalent electrolyte reservoir.

As the membrane is taken to have an infinite extent in the lateral $(x,y)$ directions,
the PB equation is reduced to an effective one-dimensional equation,
where all local quantities, such as the electrostatic potential, $\psi(\vecr) = \psi(z)$,
and ionic densities, $n(\vecr) = n(z)$, depend only on the $z$-coordinate perpendicular to the planar membrane.

For a binary monovalent electrolyte (1:1 electrolyte, $z_i = \pm 1$),
the PB equation from Eq.~(\ref{b5}), reduces in its effective one-dimensional form
to an ordinary differential equation depending only on the $z$-coordinate:
\begin{eqnarray}
\label{e1}
\Psi^{\, \prime\prime}(z) =  \kd^2 \sinh\Psi(z) \, ,
\end{eqnarray}
where $\Psi \equiv \beta e \psi$ is the rescaled dimensionless potential and we have assumed that
the external charge, $\rho_f$, is restricted to the system boundaries and will only affect the boundary conditions.

We will consider two boundary conditions in this section.
A fixed surface potential (Dirichlet boundary condition), $\Psi_s \equiv \Psi(z=0) = {\rm const}$,
and constant surface charge (Neumann boundary condition),
$\sigma \propto \Psi^{\, \prime}_s  = {\rm const}$.
A third and more specialized boundary condition of charge regulation will be treated
in detail in section~\ref{sec9.7}.
In the constant charge case, the membrane charge is modeled via a fixed surface charge density,
$\rho_f = \sigma\delta(z)$ in Eq.~(\ref{b5a}).
A variation of the Helmholtz free energy, $F$, of Eq.~(\ref{b13})
with respect to the surface potential, $\Psi_s$, $\delta F/\delta \Psi_s=0$,
is equivalent to constant surface charge boundary:
\begin{eqnarray}
\label{e2}
\frac{\D\Psi}{\D z}\Bigg|_{z=0} = - 4\pi\lb \sigma/e \, .
\end{eqnarray}
%
%to be constant surface charge (Neumann boundary conditions).
%
Although we focus in the rest of the chapter on monovalent electrolytes,
the extension to multivalent electrolytes is straightforward.

The boundary condition of Eq.~(\ref{e2}) is valid if
the electric field does not penetrate the `oily' part of the membrane.
%For typical values of membrane thickness and $\varepsilon_w/\varepsilon_L \simeq 40$,
This assumption can be justified (Kiometzis and Kleinert 1989, Winterhalter and Helfrich 1992),
as long as
$\varepsilon_L/\varepsilon_w \simeq 1/40 \ll h/\ld$,
where $h$ is the membrane thickness (see Fig.~\ref{figure1}).
All our results for one or two flat membranes, sections~\ref{sec9.3}-\ref{sec9.4} and \ref{sec9.5}-\ref{sec9.6}, respectively,
rely on this decoupled limit where the two sides (monolayers) of the membrane are
completely decoupled and the electric field inside the membrane is negligible.

%%%%%%%%%%%%%%%%%%%%%%%%%%%%%%%%%%%%%%%%%%%%%%%
\subsection{Counter-ions Only} \label{sec9.3.1}
%%%%%%%%%%%%%%%%%%%%%%%%%%%%%%%%%%%%%%%%%%%%%%%

A single charged membrane in contact with a cloud of counter-ions in solution
is one of the simplest problems that has an analytical solution.
It has been formulated and solved in the beginning of the 20th century
by Gouy (1910, 1917) and Chapman (1913).
The aim is to find the profile of a counter-ion cloud
forming a diffusive electric double-layer close
to a planar membrane (placed at $z = 0$) with a fixed
surface charge density (per unit area), $\sigma$, as in Fig.~\ref{figure2}.

Without loss of generality, the single-membrane problem is treated here for
negative (anionic) surface charges ($\sigma < 0$) and positive monovalent counter-ions
(cations) in the solution, $q_+ = e$ and $n(z) = n_+(z)$,
such that the charge neutrality condition,
\begin{eqnarray}
\label{f0}
\sigma = -e\int_{0}^{\infty}n(z)\D z \, ,
\end{eqnarray}
is fulfilled.

The PB equation for monovalent counter-ions is written as
\begin{eqnarray}
\label{f1}
\Psi^{\, \prime\prime}(z) =  - 4\pi\lb n_0 \e^{- \Psi(z)} \, ,
\end{eqnarray}
where $n_0$ is the reference density, taken at zero potential
in the absence of a salt reservoir.
The PB equation, Eq.~(\ref{f1}), with the boundary condition for one
charged membrane, Eq.~(\ref{e2}), and vanishing electric field at infinity,
can be integrated analytically twice, yielding
\begin{eqnarray}
\label{f4a}
\Psi(z) = 2 \ln\left(z + \lgc\right) + \Psi_0 \, ,
\end{eqnarray}
so that the density is
\begin{eqnarray}
\label{f4}
n(z) = \frac{1}{2\pi\lb}\frac{1}{(z + \lgc)^2} \, ,
\end{eqnarray}
where $\Psi_0$ is a reference potential and $\lgc$ is the Gouy-Chapman length defined in Eq.~(\ref{a2}).
For example, for a choice of $\Psi_0 = -2\ln(\lgc)$, the potential at $z=0$ vanishes and Eq.~(\ref{f4a}) reads
\begin{eqnarray}
\label{f5a}
\Psi(z) = 2 \ln\left(1 + z/\lgc\right) \, .
\end{eqnarray}
Although the entire counter-ion profile is diffusive as it decays algebraically,
half of the counter-ions ($\half |\sigma|$ per unit area) accumulates in a layer
of thickness $\lgc$ close to the membrane,
\begin{eqnarray}
\label{f5b}
e\int_0^{\lgc} n(z)\D z = \half|\sigma| \, .
\end{eqnarray}

\begin{figure}%3
\centering
\includegraphics[scale=0.7]{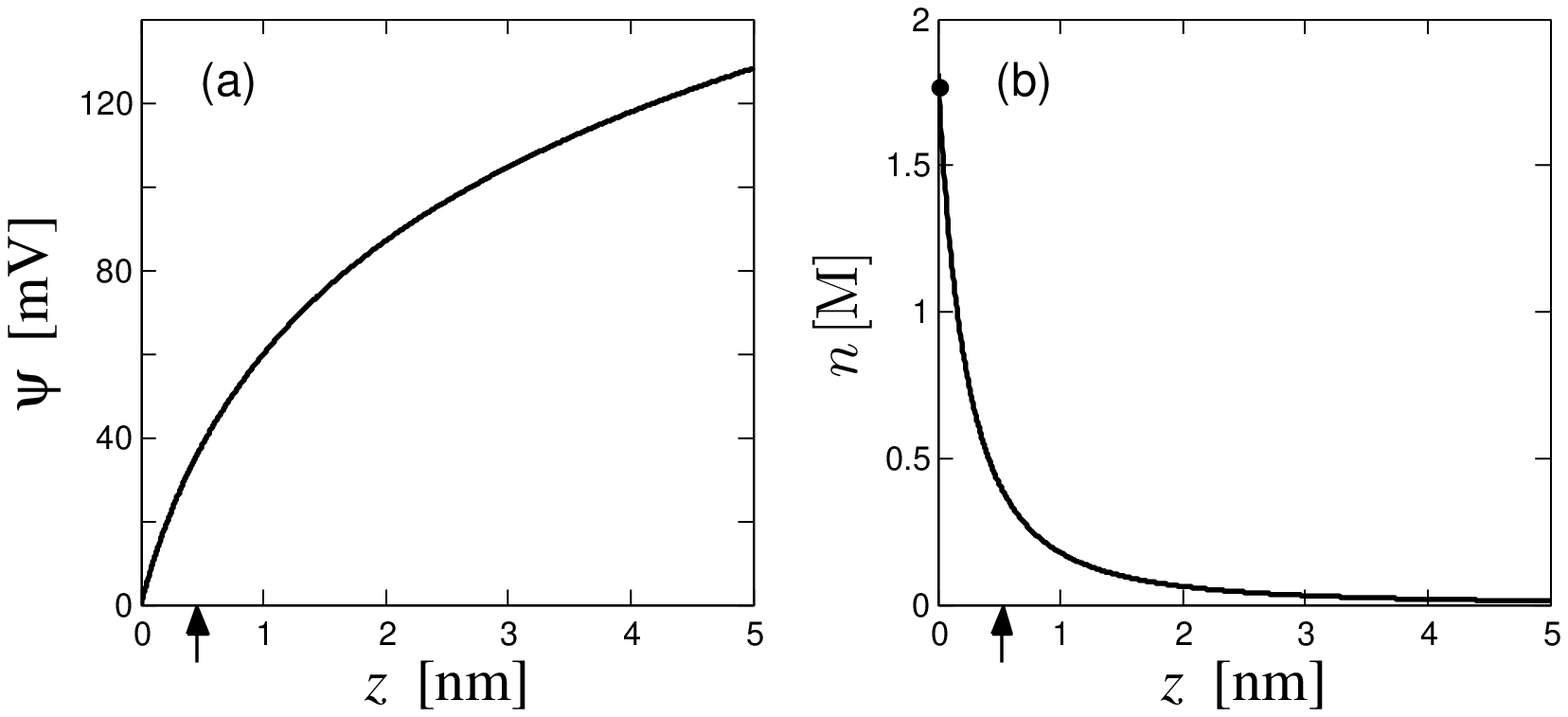}
\caption{ \textsf{The electric double layer for a single charged membrane in contact
with an aqueous solution of neutralizing monovalent counter-ions.
In (a) the electrostatic potential $\psi(z)$ (in {\rm mV})
is plotted as function of the distance from the membrane, $z$, Eq.~(\ref{f5a}).
The charged membrane is placed at $z = 0$ with $\sigma = -e/2 \, {\rm nm}^2 < 0$.
The zero of the potential is chosen to be at the membrane, $\psi(z=0) = 0$.
In (b) the density profile of the counter-ions, $n$ (in {\rm M}),
is plotted as function of the distance $z$.
Its value at the membrane is $n(z=0) = n_s \simeq 1.82 \, {\rm M}$
and the Gouy-Chapman length, $\lgc \simeq 0.46 \, {\rm nm}$, is marked by an arrow.  }}
\label{figure3}
\end{figure}

As an example, we present in Fig.~\ref{figure3} the potential $\psi$ (in mV)
and ionic profile $n$ (in M) for a surface density of $\sigma = -e/2 \, {\rm nm}^2$,
leading to a Gouy-Chapman length, $\lgc \simeq 0.46 \, {\rm nm}$.
The figure clearly shows the build-up of the diffusive layer of
counter-ions attracted by the negatively charged membrane,
reaching a limiting value of $n_s = n(0) \simeq 1.82\,{\rm M}$.
Note that the potential has a weak logarithmic divergence as $z\to\infty$.
This divergency is a consequence of the vanishing ionic reservoir (counter-ions only)
with counter-ion density obeying the Boltzmann distribution.
However, the physically measured electric field, $E = -\D\psi/\D z$,
properly decays to zero as $\sim 1/z$, at $z\to\infty$.

%%%%%%%%%%%%%%%%%%%%%%%%%%%%%%%%%%%%%%%%%%%%%%%
\subsection{Added Electrolyte} \label{sec9.3.2}
%%%%%%%%%%%%%%%%%%%%%%%%%%%%%%%%%%%%%%%%%%%%%%%

Another case of experimental interest is that of a single charged
membrane at $z=0$ in contact with an electrolyte reservoir.
For a symmetric electrolyte, $n_+^{(b)} = n_+^{(b)}  \equiv n_b$,
and the same boundary condition of constant surface charge $\sigma$, Eq.~(\ref{e2}),
holds at the $z=0$ surface.
The negatively charged membrane attracts the counter-ions and repels the co-ions.
As will be shown below, the potential decays to zero from below at large $z$; hence, it is always negative.
Since the potential is a monotonic function, this also implies that $\Psi^{\, \prime}(z)$ is always positive.
At large $z$, where the potential decays to zero,
the ionic profiles tend to their bulk (reservoir) densities, $n^{\pm}\big|_{\infty} = n_b$.

The PB equation for monovalent electrolyte, Eq.~(\ref{e1}),
with the boundary conditions as explained above can be solved analytically.
The first integration of the PB equation for 1:1 electrolyte yields
\begin{eqnarray}
\label{h1}
\frac{\D\Psi}{\D z} = -2\kd\sinh(\Psi/2) \, ,
\end{eqnarray}
where we have used $\D\Psi/\D z(z\to\infty)=0$ that is implied by the Gauss law and electro-neutrality,
and chose the bulk potential, $\Psi(z\to\infty)=0$, as the reference potential.
A further integration yields
\begin{eqnarray}
\label{g1a}
\Psi = - 4\tanh^{-1}\left(\gamma \e^{-\kd z}  \right) = - 2\ln\left( \frac{1+\gamma\e^{-\kd z}}{1-\gamma\e^{-\kd z}}  \right) \, ,
\end{eqnarray}
where $\gamma$ is an integration constant, $0 < \gamma < 1$. Its value is determined by the boundary condition at $z=0$.

%
%
%%
%\begin{eqnarray}
%\label{g1a}
%\Psi(z) = - 2\ln\left( \frac{1+\gamma\e^{-\kd z}}{1-\gamma\e^{-\kd z}}  \right)\, ,
%\end{eqnarray}
%%
%and the ionic densities,
%
The two ionic profiles, $n_{\pm}(z)$, are calculated from the Boltzmann distribution, Eq.~(\ref{b2}),
and from Eq.~(\ref{g1a}), yielding:
\begin{eqnarray}
\label{g1b}
n_{\pm}(z) = n_b\left(  \frac{1\pm\gamma\e^{-\kd z}}{1\mp\gamma\e^{-\kd z}} \right)^{\!\!\!2} \, .
\end{eqnarray}
For constant surface charge, the parameter $\gamma$ is obtained
by substituting the potential from Eq.~(\ref{g1a})
into the boundary condition at $z=0$, Eq.~(\ref{e2}).
This yields a quadratic equation, $\gamma^2 + 2\kd\lgc\gamma - 1 = 0$,
with $\gamma$ as its positive root:
\begin{eqnarray}
\label{g2}
\gamma = -\kd\lgc + \sqrt{(\kd\lgc)^2 + 1} \, .
\end{eqnarray}

\begin{figure}%4
\centering
\includegraphics[scale=0.7]{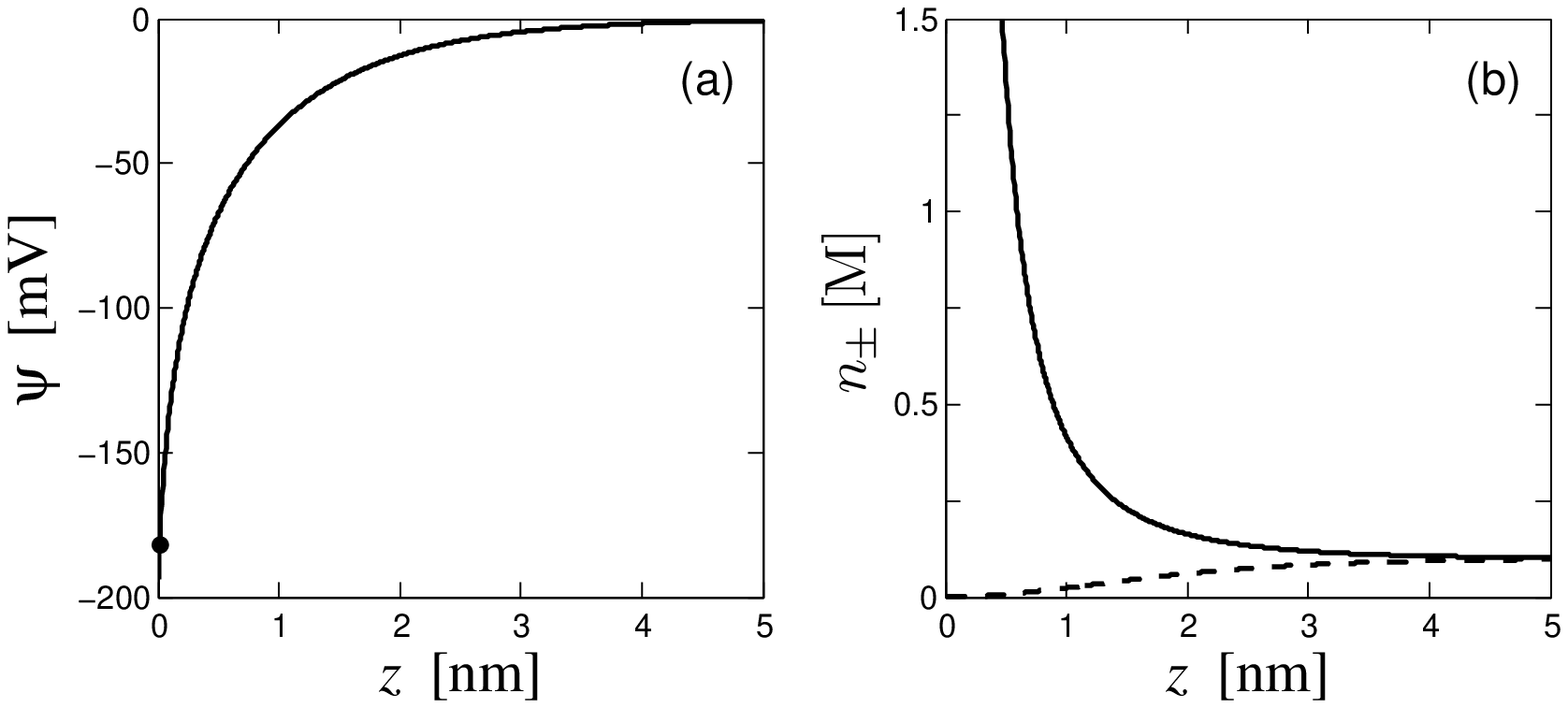}
\caption{\textsf {The electric double layer for a single charged membrane in contact
with a 1:1 monovalent electrolyte reservoir of concentration $n_b = 0.1 \, {\rm M}$,
corresponding to $\ld \simeq 0.97 \, {\rm nm}$.
The membrane located at $z = 0$ is negatively charged with $\sigma = -5e/{\rm nm}^2$,
yielding $\lgc \simeq 0.046 \, {\rm nm}$.
Note that the value of $\sigma$ is ten times larger than the value used in Fig.~\ref{figure3}.
In (a) we plot the electrostatic potential, $\psi(z)$
as function of $z$, the distance from the membrane.
The value of the surface potential is $\psi_s \simeq -194 \, {\rm mV}$.
In (b) the density profiles of counter-ions (solid line) and co-ions (dashed line), $n_{\pm}$ (in ${\rm M}$),
are plotted as function of the distance from the membrane, $z$.
The positive counter-ion density at the membrane is $n_+(z=0) \simeq 182 \, {\rm M}$
(not shown in the figure).
%(for presentation purposes, of the co-ions profile we show the counter-ion profile only up to $1.5 \, {\rm M}$).
 }}
\label{figure4}
\end{figure}

For constant surface potential, the parameter $\gamma$ can be obtained by setting $z=0$ in Eq.~(\ref{g1a}),
\begin{eqnarray}
\label{g3}
\Psi_s = e\psi_s/\kbt = -4 \tanh^{-1}\gamma \, .
\end{eqnarray}
We use the fact that the surface potential $\Psi_s$
is uniquely determined by the two lengths, $\lgc \sim \sigma^{-1}$ and $\ld$,
and write the electrostatic potential as
\begin{eqnarray}
\label{g4a}
\Psi(z) =  -2 \ln\left[ \frac{1 - \tanh(\Psi_s/4)\e^{-\kd z}}{ 1 + \tanh(\Psi_s/4)\e^{-\kd z}} \right]   \, ,
\end{eqnarray}
where $\Psi_s < 0$, in accord with our choice of $\sigma < 0$.
In Fig~\ref{figure4} we show typical profiles for the electrostatic potential
and ionic densities, for $\sigma = -5e/{\rm nm}^2$ ($\lgc \simeq 0.046 \, {\rm nm}$).
Note that this surface charge density is ten times larger than $\sigma$ of Fig~\ref{figure3}.
For electrolyte bulk density of $n_b = 0.1$M,
the Debye screening length is $\ld \simeq 0.97 \, {\rm nm}$.

The DH (linearized) limit of the PB equation, Eq.~(\ref{c2}),
is obtained for small surface charge and/or high electrolyte strength, $\kd\lgc \gg 1$.
This limit yields $\gamma \simeq (2\kd\lgc)^{-1}$ and the potential can be approximated as
\begin{eqnarray}
\label{g4}
\Psi \simeq \Psi_s\e^{-\kd z} \simeq - \frac{2}{\kd\lgc}\e^{-\kd z} \, .
\end{eqnarray}
As expected for the DH limit, the solution is exponentially screened
and falls off to zero for $z \gg \kd^{-1} = \ld$.

The opposite
%The above Eqs.~(\ref{g1b})-(\ref{g2}) reduce to the
counter-ion only case,
considered earlier in section~\ref{sec9.3.1} is obtained
by formally taking the $n_b \to 0$ limit in Eqs.~(\ref{g1a})-(\ref{g2})
or, equivalently, $\kd\lgc \ll 1$.
This means that $\gamma \simeq 1-\kd\lgc$ and from Eq.~(\ref{g1a})
we recover Eq.~(\ref{f4}) for the counter-ion density, $n(z) = n_+(z)$,
while the co-ion density, $n_-(z)$, vanishes.

For a system in contact with an electrolyte reservoir,
the potential always has an exponentially screened form
in the distal region (far from the membrane).
This can be seen by taking $z\to \infty$ while keeping $\kd\lgc$ finite in Eq.~(\ref{g1a})
\begin{eqnarray}
\label{g5}
\Psi(z) \simeq - 4\gamma\e^{-\kd z} \, .
\end{eqnarray}
Moreover, it is possible to extract from the distal form an effective surface charge density, $\sigma_{\rm eff}$,
by comparing the coefficient $4\gamma$ of Eq.~(\ref{g5}) with an effective coefficient $2/(\kd\lgc)$
from the DH form, Eq.~(\ref{g4}),
\begin{eqnarray}
\label{g6}
|\sigma_{\rm eff}| = 2\gamma\kd\lgc|\sigma| = \frac{e\kd}{\pi\lb}\gamma \, .
\end{eqnarray}
Note that $\gamma = \gamma(\kd\lgc)$ is calculated for the nominal parameter values in Eq.~(\ref{g2}).
The same concept of an effective $\sigma$ is useful in several
situations other than the simple planar geometry considered here.

%%%%%%%%%%%%%%%%%%%%%%%%%%%%%%%%%%%%%%%%%%%%%%%
\subsection{The Grahame Equation} \label{sec9.3.3}
%%%%%%%%%%%%%%%%%%%%%%%%%%%%%%%%%%%%%%%%%%%%%%%

In the planar geometry, for any amount of salt, the non-linear PB equation can be integrated analytically,
resulting in a useful relation known as the Grahame equation (Grahame 1947).
This equation is a relation between the surface charge density, $\sigma$,
and the limiting value of the ionic density profile at the membrane, $n_{\pm}^{(s)} \equiv n_{\pm}(z=0)$.
The first integration of the PB equation for a 1:1 electrolyte yields, Eq.~(\ref{h1}),
$\D\Psi / \D z = - 2\kd \sinh( \Psi/2)$.
%%
%\begin{eqnarray}
%\label{h1}
%\frac{\D\Psi}{\D z} = - 2\kd \sinh( \Psi/2) \, .
%\end{eqnarray}
%%
Using the boundary condition, Eq.~(\ref{e2}), and simple hyperbolic function identities
gives a relation between $\sigma$ and $\Psi_s$
\begin{eqnarray}
\label{h2}
\pi\lb\left(\frac{\sigma}{e}\right)^2 = n_b\left( \cosh \Psi_s - 1 \right) \, ,
\end{eqnarray}
and via the Boltzmann distribution of $n_{\pm}$, the Grahame equation is obtained
\begin{eqnarray}
\label{h3}
\sigma^2 = \frac{e^2}{2\pi\lb}\left( n_+^{(s)} + n_-^{(s)} - 2n_b  \right) \, .
\end{eqnarray}
This equation implies a balance of stresses on the surface,
with the Maxwell stress of the electric field compensating the van 't Hoff ideal pressure of the ions.

For large and negative surface potential, $|\Psi_s| \gg 1$, the co-ion density, $n_-^{(s)} \sim \exp(-|\Psi_s|)$,
can be neglected and Eq.~(\ref{h3}) becomes
\begin{eqnarray}
\label{h3b}
\sigma^2 = \frac{e^2}{2\pi\lb}\left( n_+^{(s)} - 2n_b  \right) \, .
\end{eqnarray}
For example, for a surface charge density of $\sigma = -5e/{\rm nm}^2$ (as in Fig.~\ref{figure4})
and an ionic strength of $n_b = 0.1 \, {\rm M}$, the limiting value of the counter-ion
density at the membrane is $n_+^{(s)} \simeq 182$~M,
and that of the co-ions is $n_-^{(s)} \simeq 5 \cdot 10^{-5}$~M.
The very high and unphysical value of $n_+^{(s)}$
should be understood as an artifact of the continuum PB theory.
In physical situations, the ions accumulate in the membrane vicinity till their concentration
saturates due to the finite ionic size and other ion-surface interactions.
We will further explore this point in sections~\ref{sec9.4} and \ref{sec9.7}.

The differential capacitance is another useful quantity to calculate and it
gives a physical measurable surface property.
By using Eq.~(\ref{h2}), we obtain
\begin{eqnarray}
\label{h4}
C_{\rm PB} = \frac{\D\sigma}{\D\psi_s} = \frac{e}{\kbt} \frac{\D\sigma}{\D\Psi_s}
= \varepsilon_0\varepsilon_w\kd\cosh\left( \Psi_s /2 \right) \, .
\end{eqnarray}
As shown in Fig.~\ref{figure5b}, the PB differential capacitance,
has a minimum at the potential of zero charge, $\Psi_s = 0$,
and increases exponentially for $|\Psi_s| \gg 1$.

%%%%%%%%%%%%%%%%%%%%%%%%%%%%%%%%%%%%%%%%%%%%%%%
\section{Modified Poisson-Boltzmann (mPB) Theory}  \label{sec9.4}
%%%%%%%%%%%%%%%%%%%%%%%%%%%%%%%%%%%%%%%%%%%%%%%

The density of accumulated counter-ions at the
membrane might reach unphysical high values (see Fig.~\ref{figure4}).
This unphysical situation is avoided by accounting for the solvent entropy.
Including this additional term yields a modified free-energy and PB equation (mPB).
Taking this entropy into account yields a modified free-energy,
written here for monovalent electrolyte:
\begin{eqnarray}
\label{i1}
& &\beta F = \int_V\D^3r \Bigg[ -\frac{1}{8\pi\lb} \left| \nabla\Psi({\bf r}) \right|^2 + \left[ n_+({\bf r}) - n_-({\bf r}) \right] \Psi(\vecr)  \Bigg. \\
\nonumber& &+ \,  \Bigg. n_+\ln\left( n_+ a^3 \right) + n_-\ln\left( n_- a^3 \right)
+ \frac{1}{a^3} \left( 1 - a^3 n_+ - a^3 n_- \right) \ln\left( 1 - a^3 n_+ - a^3 n_- \right)   \Bigg]  \, .
\end{eqnarray}
This is the free energy of a Coulomb lattice-gas
(Borukhov, Andelman and Orland 1997, 2000, Kilic, Bazant and Ajdari 2007).
%
%which is different from the PB free energy where the solvent entropy is neglected as the ionic densities are low.
%When the ionic densities become large, the modified entropy of the Coulomb lattice gas becomes important.
%
Taking the variation of the above free energy with respect to $n_{\pm}$, $\delta F/\delta n_{\pm} = \mu_{\pm}$,
gives the ionic profiles
\begin{eqnarray}
\label{i2}
n_{\pm}(z) = \frac{n_b\e^{\mp \Psi}}{1 - 2\phi_b + 2\phi_b\cosh \Psi} \, ,
\end{eqnarray}
with $\phi_b = n_ba^3$ being the bulk volume fraction of the ions.
For simplicity, $a$ is taken to be the same molecular size of all ionic species and the solvent.

%%%%%%%%%%%%%%%%%%%%%%%%%%%%%
\begin{shadebox}
\Boxhead{Entropy derivation of the mPB}
Let us start with a homogenous system containing an ionic solution inside a volume $v$,
with $N_+$ cations, $N_-$ anions and $N_w$ water molecules, such that $N_+ + N_- + N_w = N$.
The number of different combinations of cations, anions and water molecules is
$N!/(N_+!N_-!N_w!)$.
Therefore, the entropy is
\begin{eqnarray}
\label{ib1}
\nonumber S_v &=& -\kb \log\left( \frac{N!}{N_+!N_-!N_w!} \right) \simeq -\kb \Big[ N_+\log\left(\frac{N_+}{N}\right) +
N_-\log\left(\frac{N_-}{N}\right)  \\
&+& \left(N - N_+ - N_-\right)\log\left(1 - \frac{N_+}{N} - \frac{N_-}{N} \right)  \Big] \, ,
\end{eqnarray}
where we have used Stirling's formula for $N_{\pm},N_w \gg 1$.

We now consider a system of volume $V \gg v$.
The entropy of such system can be written in the continuum limit as
\begin{eqnarray}
\label{ib2}
\nonumber S_V &=& \int\frac{\D^3r}{v} \, S_v =  \kb \int\D^3 r \Bigg[ n_+\ln\left( n_+ a^3 \right) + n_-\ln\left( n_- a^3 \right) \\
&+& \frac{1}{a^3} \left( 1 - a^3 n_+ - a^3 n_- \right) \ln\left( 1 - a^3 n_+ - a^3 n_- \right) \Bigg]  \, ,
\end{eqnarray}
where $n_{\pm}=N_{\pm}/v$ and $n_{w}=N_{w}/v$, are the densities of the cations, anions and water molecules, respectively,
and $N=v/a^3$ is the total number of molecules in the volume $v$.
In this last equation we have used the lattice-gas formulation, in which
the solution is modeled as a cubic lattice with unit cell of size $a\times a\times a$.
Each unit cell contains only one molecule, $a^3(n_+ + n_- + n_w) = 1$.
\end{shadebox}
%%%%%%%%%%%%%%%%%%%%%%%%%%%%%

In the above equation we have also used the equilibrium relation
\begin{eqnarray}
\label{i2a}
\e^{\beta\mu_{\pm}} = n_ba^3/(1-2n_ba^3) = \frac{\phi_b}{(1-2\phi_b)} \, ,
\end{eqnarray}
valid in the bulk where $\Psi = 0$.
Variation with respect to $\Psi$, $\delta F/\delta\Psi = 0$, yields the mPB equation for 1:1 electrolyte:
\begin{eqnarray}
\label{i3}
\nabla^2\Psi(\vecr) = - 4\pi\lb \left[ n_{+}(\vecr) - n_{-}(\vecr) \right] =  \frac{\kd^2\sinh \Psi}{1 - 2\phi_b + 2\phi_b\cosh \Psi} \, .
\end{eqnarray}

For small electrostatic potentials, $|\Psi| \ll 1$, the ionic distribution, Eq.~(\ref{i2}),
reduces to the usual Boltzmann distribution,
but for large electrostatic potentials, $|\Psi| \gg 1$,
this model gives very different results with respect to the PB theory.
In particular, the ionic concentration is unbound in the standard PB theory,
whereas it is bound for the mPB by the close-packing density, $1/a^3$.
This effect is important close to strongly charged membranes immersed in an electrolyte solution,
while the regular PB equation is recovered in the dilute bulk limit, $n_b a^3 \ll 1$,
for which the solvent entropy can be neglected.

\begin{figure}%5
\centering
\includegraphics[scale=0.7]{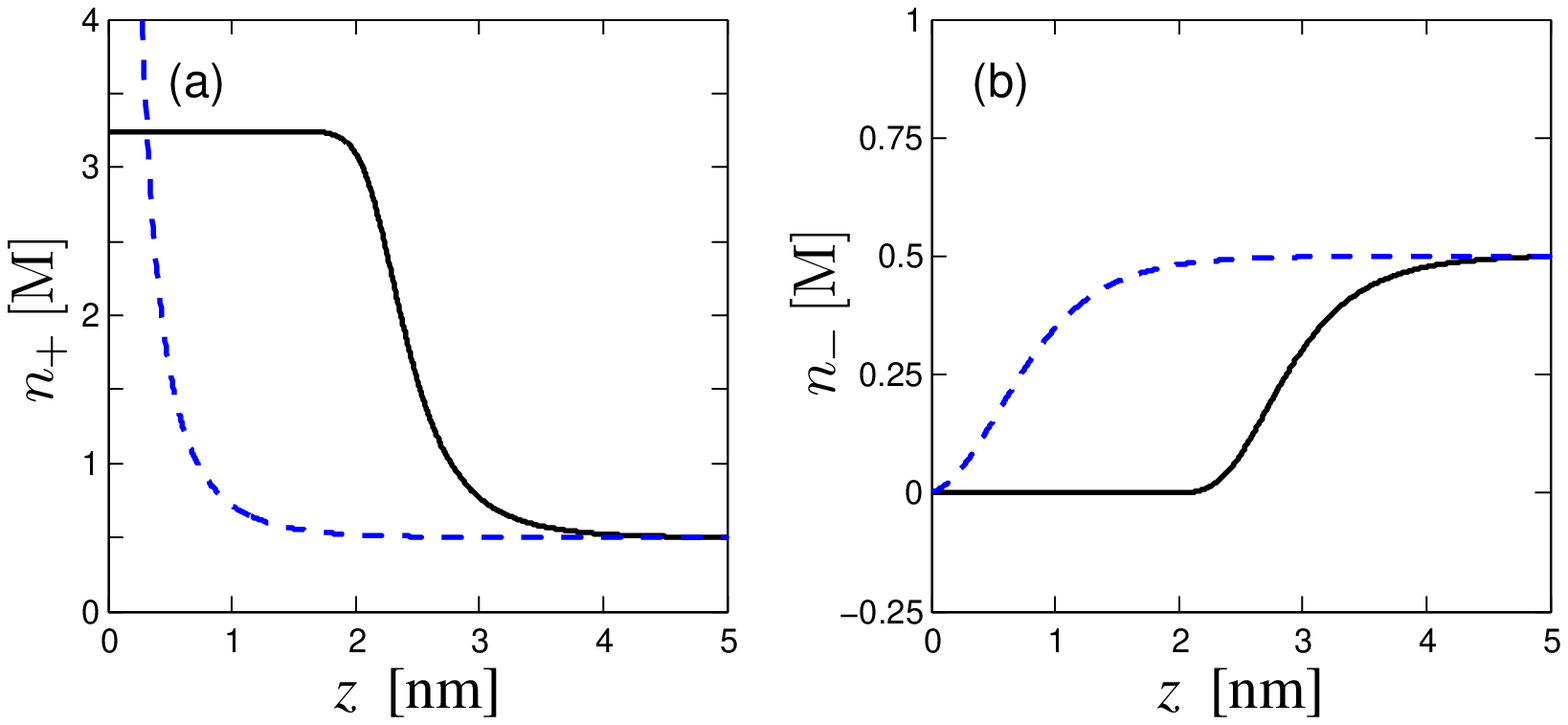}
\caption{\textsf{Comparison of the modified PB (mPB) profiles
(black solid lines) with the regular PB one (dashed blue lines).
In (a) we show the counter-ion profile, and in (b) the co-ion profile.
The parameters used are: ion size $a = 0.8 \, {\rm nm}$,
surface charge density $\sigma = -5e/{\rm nm}^2$
and 1:1 electrolyte ionic strength $n_b = 0.5 \, {\rm M}$.
Note that while the PB value at the membrane is $n_s^+ \simeq 182\, {\rm M}$,
the mPB density saturates at $n_s^+ \simeq 3.2\, {\rm M}$.}
}
\label{figure5}
\end{figure}

For large electrostatic potentials, the
contribution of the co-ions is negligible and the
counter-ion concentration follows a distribution reminiscent
of the Fermi$-$Dirac distribution
\begin{eqnarray}
\label{i4}
n_-(\vecr) \simeq \frac{1}{a^3} \frac{1}{1+\e^{-(\Psi + \beta\mu)}} \, ,
\end{eqnarray}
where electro-neutrality dictates $\mu = \mu_{\pm}$.
In Fig~\ref{figure5} we show for comparison the modified and regular PB profiles for
a 1:1 electrolyte. To emphasize the saturation effect of the mPB theory,
we chose in the figure a large ion size, $a = 0.8 \, {\rm nm}$.

The mPB theory also implies a modified Grahame equation that relates the surface charge
density to the ion surface density, $n_{\pm}^{(s)}$.
First, we find the relation between $\sigma$ and the surface potential, $\Psi_s$,
\begin{eqnarray}
\label{i5}
\left(\frac{\sigma}{e}\right)^2 &=& \frac{1}{2\pi a^3\lb}\ln\Big[ 1 + 2\phi_b\left( \cosh \Psi_s - 1 \right)  \Big] \, .
\end{eqnarray}
This equation represents a balance of stresses on the surface,
where the Maxwell stress of the electric field is equal to the lattice-gas pressure of the ions.
The surface potential can also be calculated
\begin{eqnarray}
\label{i6}
\Psi_s = \cosh^{-1}\left( \frac{\e^{\xi} - 1 + 2\phi_b}{2\phi_b}\right)   \, ,
\end{eqnarray}
with the dimensionless parameter $\xi = a^3/(2\pi\lb\lgc^2)$.

For large surface charge or large surface potential,
the co-ions concentration at the membrane is negligible, $n_-^{(s)} \ll 1$,
and the surface potential, Eq.~(\ref{i6}) is approximated by
\begin{eqnarray}
\label{i7a}
\Psi_s \simeq   \ln\left(\e^{\xi} - 1 + 2\phi_b\right) - \ln\left(\phi_b\right) \, ,
\end{eqnarray}
and from Eq.~(\ref{i5}) we obtain the Grahame equation,
\begin{eqnarray}
\label{i7}
\left(\frac{\sigma}{e}\right)^2 \simeq \frac{1}{2\pi a^3\lb}\ln\left(\frac{1-2\phi_b}{1-a^3n_+^{(s)}}\right) \, .
\end{eqnarray}
Note that in the dilute limit $\phi_b \ll 1$,
the Grahame equation reduces to the regular PB case, Eq.~(\ref{h3}).

\begin{figure}%5b
\centering
\includegraphics[scale=0.7]{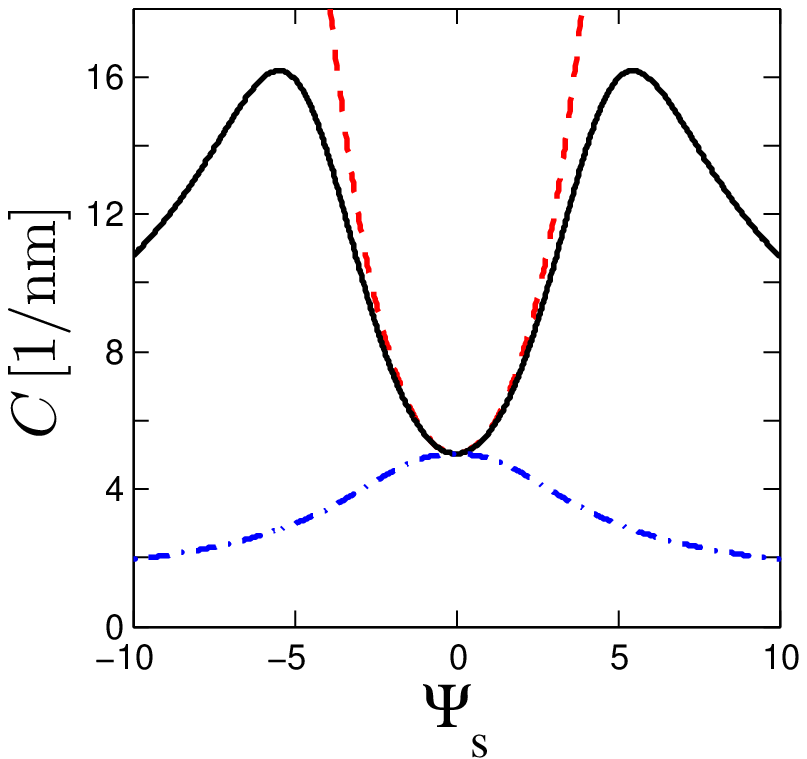}
\caption{\textsf{
Comparison of the differential capacitance, $C$,
calculated from the regular PB theory (dashed red line), Eq.~(\ref{h4}), $n_b \simeq 0.4 \, {\rm mM}$
(chosen so that it corresponds to $\phi_b = 0.01$ and $a = 0.3 \, {\rm nm}$),
and from the mPB theory, Eq.~(\ref{i8}).
The mPB differential capacitance is calculated for $a = 0.3 \, {\rm nm}$.
For low $\phi_b = 0.01$, it shows a camel shape (black solid line),
while for high $\phi_b = 0.2$, it shows a unimodal (dash-dotted blue line).}
}
\label{figure5b}
\end{figure}

It is also straightforward but more cumbersome to calculate the differential capacitance, $C = \D\sigma/\D\psi_s$,
for the mPB theory.
From Eq.~(\ref{i5}) we obtain,
\begin{eqnarray}
\label{i8}
C_{\rm mPB} = \frac{C_{\rm PB}}{1 + 4\phi_b\sinh^2\left( \Psi_s/2\right)}
\sqrt{\frac{4\phi_b\sinh^2\left( \Psi_s/2\right)}{\ln\left[ 1 + 4\phi_b\sinh^2\left(\Psi_s/2\right)\right]  }} \, .
\end{eqnarray}
Although it can be shown that for $\phi_b \to 0$ the mPB differential capacitance
reduces to the standard PB result, the resulting $C_{\rm mPB}$ is quite different for any finite value of $\phi_b$.
The main difference is that instead of an exponential divergence of $C_{\rm PB}$ at large potentials,
$C_{\rm mPB}$ decreases for high-biased $|\Psi_s| \gg 1$.
For rather small bulk densities, $\phi_b < 1/6$, the $C_{\rm mPB}$ shows a behavior called
{\it camel-shape} or {\it double-hump}.
This behavior is also observed in experiments at relatively low salt concentrations.
As shown in Fig.~\ref{figure5b}, the double-hump $C_{\rm mPB}$ has a minimum
at $\Psi_s = 0$ and two maxima.
The peak positions can roughly be estimated
by substituting the closed-packing concentration, $n = 1/a^3$,
into the Boltzmann distribution, Eq.~(\ref{b2}), yielding $\Psi_s^{\rm max} \simeq \mp\ln (\phi_b)$.
Using parameter values as in Fig.~\ref{figure5b}, $\Psi_s^{\rm max}$ is estimated as $\pm4.6$ as
compare to the exact values, $\Psi_s^{\rm max} = \pm5.5$.

Furthermore, it can be shown that for high salt densities, $\phi_b > 1/6$,
$C_{\rm mPB}$ exhibits (see also Fig.~\ref{figure5b}) a unimodal maximum
close to the potential of zero charge,
rather than a minimum as does $C_{\rm PB}$.
Such results that take into account finite ion size for the differential capacitance
are of importance in the theory of confined ionic liquids (Kornyshev 2007, Nakayama and Andelman 2015).

%%%%%%%%%%%%%%%%%%%%%%%%%%%%%%%%%%%%%%%%%%%%%%%
\section{Two Membrane System: Osmotic Pressure} \label{sec9.5}
%%%%%%%%%%%%%%%%%%%%%%%%%%%%%%%%%%%%%%%%%%%%%%%

\begin{figure}%6
\centering
\includegraphics[scale=0.5]{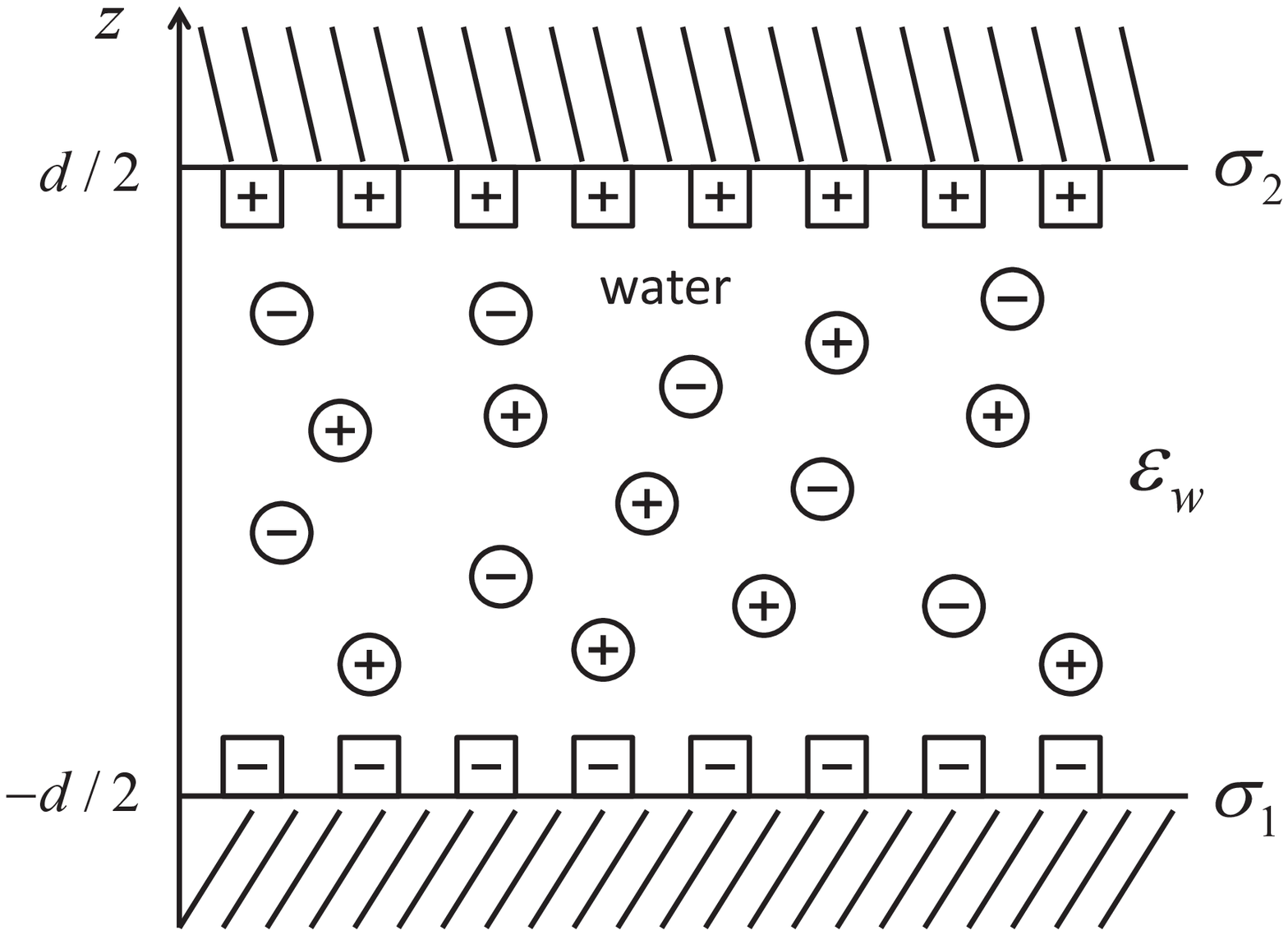}
\caption{ \textsf{Schematic drawing of two asymmetric membranes.
The planar membrane located at $z = -d/2$ carries a charge density $\sigma_1$,
while the membrane at $z = d/2$ has a charge density of $\sigma_2$.
%With no-added salt, the counter-ions in the gap neutralize the two membrane charges, $\sigma_1 + \sigma_2$.
The antisymmetric membrane setup is a special case with $\sigma_1 = -\sigma_2$,
while in the symmetric case, $\sigma_1 = \sigma_2$.}
}
\label{figure6}
\end{figure}

We consider now the PB theory of two charged membranes as shown in Fig.~\ref{figure6}.
The two membranes can, in general, have different surface charge densities:
$\sigma_1$ at $z = -d/2$ and $\sigma_2$ at $z = d/2$.
The boundary conditions of the two-membrane system are written as
$\rho_f = \sigma_1\delta(z + d/2) + \sigma_2\delta(z - d/2)$,
and using the variation of the free energy, $\delta F/\delta\Psi_s = 0$:
\begin{eqnarray}
%\label{j1a}
\nonumber\Psi^{\, \prime}\Big|_{-d/2} &=& -4\pi\lb \frac{\sigma_1}{e} \, , \\ \nonumber \\
\label{j1b}
\Psi^{\, \prime}\Big|_{d/2} &=& 4\pi\lb \frac{\sigma_2}{e}  \, .
\end{eqnarray}

It is of interest to calculate the force (or the osmotic pressure)
between two membranes interacting across the ionic solution.
The osmotic pressure is defined as $\Pi = P_{\rm in} - P_{\rm out}$,
where $P_{\rm in}$ is the inner pressure and $P_{\rm out}$ is the pressure exerted by the reservoir
that is in contact with the two-membrane system.
Sometimes the osmotic pressure is referred to as the disjoining pressure,
introduced first by Derjaguin (Churaev, Derjaguin and Muller 2014).

Let us start by calculating the inner and outer pressures from the Helmholtz free energy.
The pressure ($P_{\rm in}$ or $P_{\rm out}$) is the variation of the
free-energy with the volume:
\begin{eqnarray}
\label{j1}
P=-\frac{\partial F}{\partial V} = -\frac{1}{A}\frac{\partial F}{\partial d}\, ,
\end{eqnarray}
with $V = Ad$, being the system volume, $A$ the lateral membrane area, and $d$ is the inter-membrane distance.
As the interaction between the two membranes can be either attractive ($\Pi < 0$) or repulsive ($\Pi > 0$),
we will analyze the criterion for the crossover ($\Pi = 0$) between these two regimes
as function of the surface charge asymmetry and inter-membrane distance.

%%%%%%%%%%%%%%%%%%%%%%%
\begin{shadebox}
\Boxhead{ General derivation of the pressure}
The Helmholtz free-energy obtained from Eq.~(\ref{b13})
can be written in a general form as, $F = A\int f[\Psi(z),\Psi^{\, \prime}(z)] \D z$,
where we use the Poisson equation to obtain the relation, $n_{\pm} = n_{\pm}(\Psi^{\, \prime})$.
As the integrand $f$ depends only implicitly on the $z$ coordinate through $\Psi(z)$,
one can obtain from the Euler-Lagrange equations the following relation
(Ben-Yaakov et al. 2009).
\begin{eqnarray}
\label{j3}
\nonumber & &f - \frac{\partial f}{\partial\Psi^{\, \prime}}\Psi^{\, \prime} = {\rm const} \\
& &\qquad=  \frac{\kbt}{8\pi\lb}{\Psi^{\, \prime}}^2 + \kbt\sum_{i=1}^M z_i n_i \Psi
+ \kbt\sum_{i=1}^M \left[n_i\ln\left(n_i a^3\right) - n_i \right]   \, ,
\end{eqnarray}
where the sum is over $i=1,...,M$ ionic species.
Let us understand the meaning of the constant on the right-hand side of the above equation.
For uncharged solutions, the Helmholtz free-energy per unit volume
contains only the entropy term, $f = \kbt\sum_i \left[n_i\ln\left(n_i a^3\right) - n_i \right]$,
and from Eq.~(\ref{j3}), we obtain $f = {\rm const}$.
A known thermodynamic relation is $P = \sum_i \mu_i^{\rm tot}n_i -f$,
with the total chemical potential defined as before, $\partial f/\partial n_i = \mu_i^{\rm tot}$,
implying that the right-hand side constant is $\sum_i \mu_i^{\rm tot}n_i - P$.
%
%Using the equilibrium equation for the densities, $\partial f/\partial n_i = \mu_i^{\rm tot}$,
%we obtain for the pressure, $P = -f + \sum_i \mu_i^{\rm tot}n_i $.
%
However, even for charged liquid mixtures, the electrostatic
potential vanishes in the bulk, away from the boundaries,
and reduces to the same value as for uncharged solutions.
Therefore, we conclude that the right-hand side constant is $\sum_i \mu_i^{\rm tot}n_i - P$, yielding
\begin{eqnarray}
\label{j7}
P = -\frac{\kbt}{8\pi\lb}{\Psi^{\, \prime}}^2 + \kbt\sum_{i=1}^M n_i \, .
\end{eqnarray}
If the electric field and ionic densities are calculated right at the surface,
we obtain the {\it contact theorem} that gives the osmotic pressure acting on the surface.
Another and more straightforward way to calculate the pressure, is
to calculate the incremental difference in free energy, $F$,
for an inter-membrane separation $d$, {\it i.e.} $\left[ F(d+\delta d) - F(d) \right]/\delta d$.
The calculation of $F(d+\delta d)$ can be done by
including an additional slab of width $\delta d$ in the space between the two membranes at an arbitrary position.
We remark that the validity  of the contact theorem itself is not limited to the PB theory,
but is an exact theorem of statistical mechanics
(Henderson and Blum 1981, Evans and Wennerstr\"om 1999, Dean and Horgan 2003).
%%%%%%%%%%%%%%%%%%%%%%%
\end{shadebox}

We are interested in the osmotic pressure, $\Pi$.
For an ionic reservoir in the dilute limit, Eq.~(\ref{j7}) gives $P_{\rm out} = \kbt\sum_i n_i^{(b)}$,
where $n_i^{(b)}$ is the $i^{\rm th}$ ionic species bulk density.
Thus, the osmotic pressure can be written as
\begin{eqnarray}
\label{j8}
\Pi = -\frac{\kbt}{8\pi\lb}{\Psi^{\, \prime}}^2(z) + \kbt\sum_{i=1}^M \left( n_i(z) - n_i^{(b)} \right) = {\rm const} \, ,
\end{eqnarray}
and for monovalent 1:1 ions:
\begin{eqnarray}
\label{j8b}
\Pi = -\frac{\kbt}{8\pi\lb}{\Psi^{\, \prime}}^2(z) + 2\kbt n_b \Big( \cosh\Psi(z) - 1  \Big) = {\rm const} \, .
\end{eqnarray}
At any position $z$ between the membranes, the osmotic pressure has two contributions.
The first is a negative Maxwell electrostatic pressure proportional to ${\Psi^{\, \prime}}^2$.
The second is due to the entropy of mobile ions and measures the local entropy change
(at an arbitrary position, $z$) with respect to the ion entropy in the reservoir.

%%%%%%%%%%%%%%%%%%%%%%%%%%%%%%%%%%%%%%%%%%%%%%%
\section{Two Symmetric Membranes, $\sigma_1 = \sigma_2$} \label{sec9.5a}
%%%%%%%%%%%%%%%%%%%%%%%%%%%%%%%%%%%%%%%%%%%%%%%

For two symmetric charged membranes, $\sigma_1 = \sigma_2 \equiv \sigma$ at $z = \pm d/2$,
the electrostatic potential is symmetric about the mid-plane
yielding a zero electric field, $E = 0$ at $z = 0$.
It is then sufficient to consider the interval $[0,d/2]$ with the boundary conditions,
%$\Psi^{\, \prime}(z=d/2) = \Psi_s^{\, \prime} = 4\pi\lb \sigma/e$, and $\Psi^{\, \prime}(z=0) = \Psi_m^{\, \prime} = 0$.
%
\begin{eqnarray}
\label{k1}
\nonumber & &\Psi^{\, \prime}\Big|_{z=d/2} = \Psi_s^{\, \prime} = 4\pi\lb \sigma/e \, ,\\
& &\Psi^{\, \prime}\Big|_{z=0} = \Psi_m^{\, \prime} = 0 \, .
\end{eqnarray}

As $\Pi$ is constant (independent of $z$)  between the membranes,
one can calculate the disjoining pressure, $\Pi$, from Eq.~(\ref{j8}), at any position $z$, between the membranes.
A simple choice will be to evaluate it at $z=0$ (the mid-plane), where the
electric field vanishes for the symmetric $\sigma_1 = \sigma_2$ case,
\begin{eqnarray}
\label{k2}
\Pi = \kbt\sum_{i=1}^M \left( n_i^{(m)} - n_i^{(b)} \right) = \kbt\sum_{i=1}^M n_i^{(b)}\left( \e^{-z_i\Psi_m} - 1 \right) > 0\, ,
\end{eqnarray}
and for monovalent ions, $z_i = \pm 1$, we get
\begin{eqnarray}
\label{k2a}
\Pi = 4\kbt n_b\sinh^2(\Psi_m/2) > 0 \, ,
\end{eqnarray}
where $n_i^{(m)} = n_i(z=0)$ is the mid-plane concentration of the $i^{\rm th}$ species.
It can be shown that the electro-neutrality condition implies that
the osmotic pressure is always {\it repulsive} for any shape of boundaries
(Sader and Chan 1999, Neu 1999) as long as we have two symmetric membranes $(\sigma_1=\sigma_2)$.

Note that the Grahame equation can be derived also for the two-membrane case with added electrolyte.
One way of doing it is by comparing the pressure of Eq.~(\ref{j8}) evaluated at one of the membranes, $z = \pm d/2$,
and at the mid-plane, $z=0$.
The pressure is constant between the two membranes, thus,
by equating these two pressure expressions, the Grahame equation emerges
\begin{eqnarray}
\label{k4}
\left(\frac{\sigma}{e}\right)^2 = \frac{1}{2\pi\lb}\sum_{i=1}^M \left( n_i^{(s)} - n_i^{(m)} \right) \, .
\end{eqnarray}
By taking the limit of infinite separation between the two-membranes and $n_i^{(m)} \to n_i^{(b)}$,
the Grahame equation for a single membrane, Eq.~(\ref{h3}), is recovered.

%%%%%%%%%%%%%%%%%%%%%%%%%%%%%%%%%%%%%%%%%%%%%%%%%%%%%%%%%%%%%%%%%%%%%%%%%%%%%%%%%%%%%
\subsection{Counter-ions Only} \label{sec9.5.1}
%%%%%%%%%%%%%%%%%%%%%%%%%%%%%%%%%%%%%%%%%%%%%%%%%%%%%%%%%%%%%%%%%%%%%%%%%%%%%%%%%%%%%

In the absence of an external salt reservoir,
the only ions in the solution
for a symmetric two-membrane system,
are positive monovalent $(z=+1)$ counter-ions with density $n(z)$
that neutralizes the surface charge,
\begin{eqnarray}
\label{l1}
2\sigma = -e\int_{-d/2}^{d/2}n(z)\D z \, .
\end{eqnarray}

\begin{figure}%8
\centering
\includegraphics[scale=0.7]{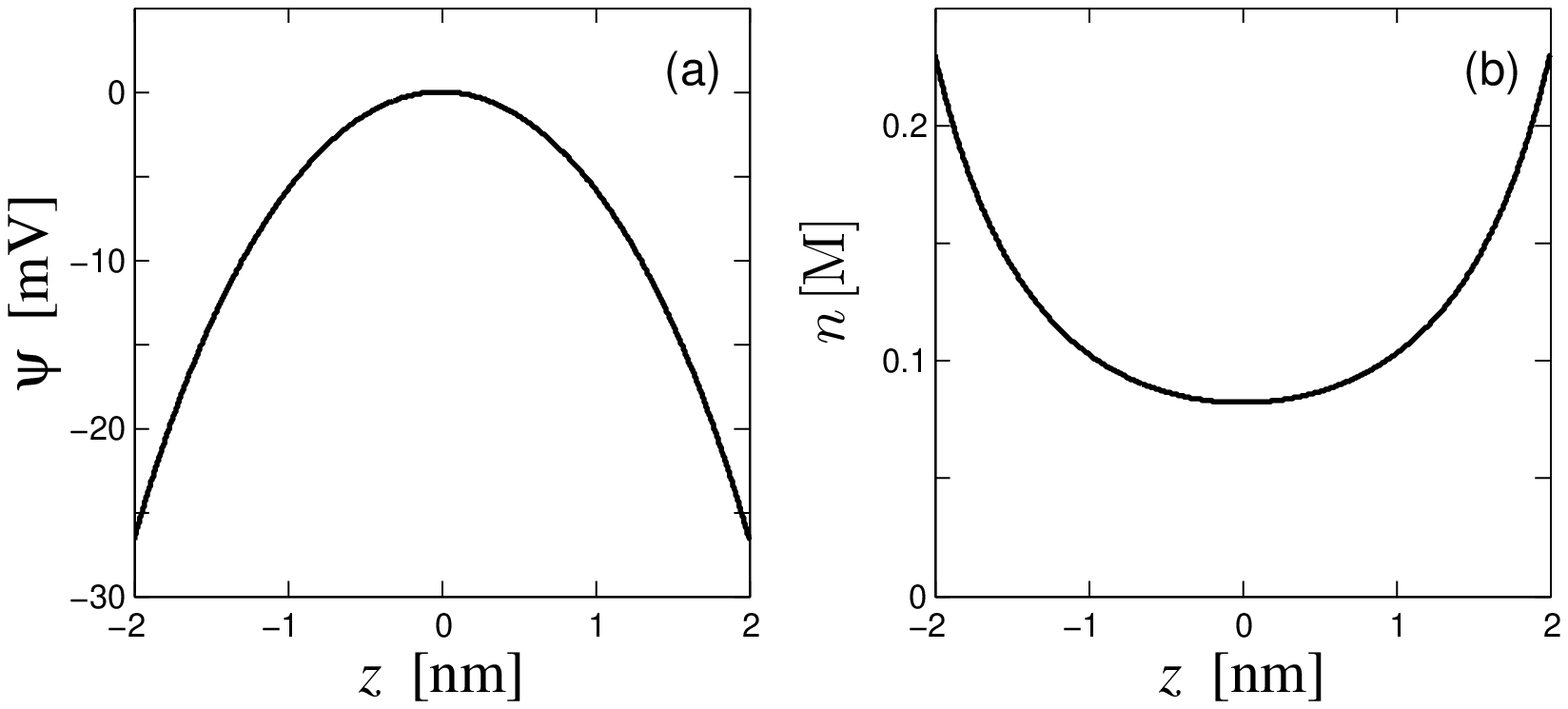}
\caption{ \textsf{The counter-ion only case for two identically charged membranes
located at $z = \pm d/2$ with $d = 4 \, {\rm nm}$ and $\sigma = -e/7 \, {\rm nm^2}$ on each membrane ($\lgc \simeq 1.6 \, {\rm nm}$).
In (a) we plot the electrostatic potential, $\psi$,
and in (b) the counter-ion density profile, $n$.
The plots are obtained from Eqs.~(\ref{l2a})-(\ref{l3}).}
}
\label{figure8}
\end{figure}

The PB equation has an analytical solution for this case.
Integrating twice the PB equation, Eq.~(\ref{e1}),
with the appropriate boundary conditions, Eq.~(\ref{k1}),
yields an analytical expression for the electrostatic potential:
\begin{eqnarray}
\label{l2a}
\Psi(z) = \ln\left( \cos^2 K z  \right) \, ,
\end{eqnarray}
and consequently the counter-ion density is
\begin{eqnarray}
\label{l2}
n(z) = n_m\e^{-\Psi(z)} = \frac{n_m}{\cos^2 \left( K z \right)} \, .
\end{eqnarray}
In the above we have defined $n_m = n(z=0)$ and chose arbitrarily $\Psi_m = 0$.
We also introduced a new length scale, $K^{-1}$, related to $n_m$ by
\begin{eqnarray}
\label{l2b}
K^2 = 2\pi\lb n_m \, .
\end{eqnarray}
Notice that $K$ plays a role similar to the inverse Debye length $\kd = \sqrt{8\pi\lb n_b}$,
with the mid-plane density replacing the bulk density, $n_b \to n_m$.
Using the boundary condition at $z = d/2$, we get a transcendental relation for $K$
\begin{eqnarray}
\label{l3}
K d\tan\left( Kd/2 \right) = \frac{d}{\lgc} \, .
\end{eqnarray}
In Fig.~\ref{figure8} we show a typical counter-ion profile with its
corresponding electrostatic potential for $\sigma = -e/7 \, {\rm nm^2}$ and $d = 4 \, {\rm nm}$.

The osmotic pressure, Eq.~(\ref{j8}), calculated for the counter-ion only case, is
\begin{eqnarray}
\label{l4}
\Pi = \frac{\kbt }{2\pi\lb} \, K^2 \, .
\end{eqnarray}
For weak surface charge, $d/\lgc \ll 1$,
one can approximate $(K d)^2 \simeq 2d/\lgc \ll 1$, and the pressure is given by
\begin{eqnarray}
\label{l5}
\Pi \simeq -\frac{2\kbt\sigma}{e}\frac{1}{d} = \frac{\kbt}{\pi\lb\lgc } \frac{1}{d} \, \sim \, \frac{1}{d} \, .
\end{eqnarray}
The $\Pi \sim 1/d$ behavior is similar to an ideal-gas equation of state, $P = N\kbt/V$ with $V = Ad$
and $N$ the total number of counter-ions.
The density (per unit volume) of the counter-ions is almost
constant between the two membranes and is equal to $2|\sigma|/(ed)$.
This density neutralizes the surface charge density, $\sigma$, on the two membranes.
The main contribution to the pressure comes from an ideal-gas like pressure of the
counter-ion cloud. This regime can be reached experimentally for small inter-membrane separation, $d < \lgc$.
For example, for $e/\sigma$ in the range of $1-100 \, {\rm nm}^2$,
$\lgc \sim 1/\sigma$ varies between $0.2 \, {\rm nm}$ \ and $20 \, {\rm nm}$.

For the opposite case of strong surface charge, $d/\lgc \gg 1$,
one gets $K d \simeq \pi$ from Eq.~(\ref{l3}).
This is the Gouy-Chapman regime.
It is very different from the weak surface-charge,
as the density profile between the two membranes varies
substantially leading to $n_s \gg n_m$, and to a pressure
\begin{eqnarray}
\label{l6}
\Pi \simeq \frac{\pi\kbt}{2\lb d^2} \, \sim \, \frac{1}{d^2} \, .
\end{eqnarray}

It is interesting to note that the above pressure expression does not depend
explicitly on the surface charge density.
This can be rationalized as follows.
Counter-ions are accumulated close to the surface,
at an average separation $\lgc \sim 1/|\sigma|$.
Therefore, creating a surface dipole density of $|\sigma|\lgc$.
The interaction energy per unit area is proportional to the
electrostatic energy between two such planar dipolar layers,
%which varies as $z^{-3}$.
%This is a dipolar Green function integrated over an infinite surface.
which scales as $1/d$ for the free energy density and $d^{-2}$ for the pressure.
The surface charge density dependence itself vanishes because
the effective dipolar-moment surface density, $|\sigma|  \lgc$, is charge-independent.
In the Gouy-Chapman regime, the electrostatic interactions
are most dominated as they are long-ranged and unscreened.
Of course, even in pure water the effective Debye screening length
is about $1 {\rm \mu m}$, and the electrostatic interactions
will be screened for larger distances.

%%%%%%%%%%%%%%%%%%%%%%%%%%%%%%%%%%%%%%%%%%%%%%%%%%%%%%%%%%%%%%%%%%%%%%%%%%%%%%%%%%%%%
\subsection{Added Electrolyte} \label{sec9.5.2}
%%%%%%%%%%%%%%%%%%%%%%%%%%%%%%%%%%%%%%%%%%%%%%%%%%%%%%%%%%%%%%%%%%%%%%%%%%%%%%%%%%%%%

When two charged membranes are placed in contact with an electrolyte reservoir,
the co-ions and counter-ions between the membranes have a non-homogenous density profile.
The PB equation does not have a closed-form analytical solution for
two (or more) ionic species, even when we restrict ourselves to a 1:1 symmetric
and monovalent electrolyte. Instead, the solution can be expressed in
terms of elliptic functions.

The PB equation for a monovalent 1:1 electrolyte, Eq.~(\ref{e1}),
is $\Psi^{\, \prime\prime}(z) = \kd^2\sinh\Psi$, while the same boundary conditions
as in Eq.~(\ref{k1}) is satisfied.
The first integration from the mid-plane $(z=0)$
to an arbitrary point between the membranes, $z \in [-d/2,d/2]$ , gives
\begin{eqnarray}
\label{m1}
\frac{\D\Psi}{\D z} = -\kd\sqrt{ 2\cosh \Psi(z) - 2\cosh \Psi_m } \, .
\end{eqnarray}
As explained in the beginning of section~\ref{sec9.5a}, $\Psi_m^{\, \prime} = 0$ for two symmetric membranes
and the second integration leads to an elliptic integral (see box below)
\begin{eqnarray}
\label{m2}
z = -\ld\int_{\Psi_m}^{\Psi}\frac{\D\eta}{\sqrt{ 2\cosh \eta - 2\cosh \Psi_m }} \, .
\end{eqnarray}
Inverting the relation $z=z(\Psi)$ leads to the expression for the profile, $\Psi(z)$.

%%%%%%%%%%%%%%%%%%%%%%%%%%%%%%%
\begin{shadebox}\label{elliptic_box}
\Boxhead{The electrostatic potential via Jacobi elliptic functions}
It is possible to write Eq.~(\ref{m2}) in terms of an
incomplete elliptic integral of the first kind
\begin{eqnarray}
\label{m2a}
F\left(\theta|a^2\right) \equiv \int_0^{\theta} \frac{\D\eta}{\sqrt{1-a^2\sin^2{\eta}}}  \, .
\end{eqnarray}
After change of variables and some algebra we write Eq.~(\ref{m2}) with the help of Eq.~(\ref{m2a}) as:
\begin{eqnarray}
\label{m2b}
z = 2\ld\sqrt{m}\left[ F\left(\frac{\pi}{2}\Big|m^2\right) - F\left(\varphi|m^2\right) \right]\, ,
\end{eqnarray}
with $m = \exp\left(\Psi_m\right)$ and
$\varphi = \sin^{-1}\left[ \exp\left( \left[\Psi - \Psi_m\right]/2 \right) \right]$.

The electrostatic potential, which is the inverse relation of Eq.~(\ref{m2b}),
can then be written in terms of the Jacobi elliptic function, ${\rm cd}(u|a^2)$,
\begin{eqnarray}
\label{m2c}
\Psi = \Psi_m + 2\ln\left[ {\rm cd} \left( \frac{z}{2\ld\sqrt{m}} \Big| m^2 \right) \right] \, .
\end{eqnarray}
In writing this equation we have used the definition of the Jacobi elliptic functions:
\begin{eqnarray}
\label{m2d}
& &{\rm sn}(u|a^2) = \sin\alpha \, , \\
& &{\rm cn}(u|a^2) = \cos\alpha = \sqrt{1-{\rm sn}^2(u|a^2)} \, , \\
& &{\rm dn}(u|a^2) = \sqrt{1-a^2 \ {\rm sn}^2(u|a^2)} \, , \\
& &{\rm cd}(u|a^2) = \frac{{\rm cn}(u|a^2)}{{\rm dn}(u|a^2)} \, ,
\end{eqnarray}
with $u \equiv F\left(\theta|a^2\right)$.

\end{shadebox}
%%%%%%%%%%%%%%%%%%%%%%%%%%%%%%%%%%%%%%%%%%

Using one of the boundary conditions, Eq.~(\ref{k1}),
with the first integration, Eq.~(\ref{m1}), yields
\begin{eqnarray}
\label{m3}
\cosh \Psi_s = \cosh \Psi_m + 2 \left(\frac{\ld}{\lgc} \right)^2 \, .
\end{eqnarray}
The above equation also gives a relation between $\sigma$
and the mid-plane potential, $\Psi_m$, in terms of Jacobi elliptic functions
(see box above),
\begin{eqnarray}
\label{m3a}
\frac{\sigma}{e} = \frac{\kd}{4\pi\lb}\frac{m^2-1}{\sqrt{m}}\frac{{\rm sn}(u_s|m^2)}{{\rm cn}(u_s|m^2){\rm dn}(u_s|m^2)} \, ,
\end{eqnarray}
with $u_s \equiv d/(4\ld\sqrt{m})$
%with $u_s \equiv F(\varphi_s|m^2)$ and
%$\varphi_s = \sin^{-1}\left[ \exp\left( \left[\Psi_s - \Psi_m\right]/2 \right) \right]$
and $m=\exp(\Psi_m)$ as defined after Eq.~(\ref{m2b}).
For fixed surface charge, this relation gives the mid-plane potential, $\Psi_m$,
and the osmotic pressure can then be calculated from Eq.~(\ref{k2}).
The other boundary condition can also be expressed as an elliptic integral
\begin{eqnarray}
\label{m4}
\frac{d}{2\ld} = -\int_{\Psi_m}^{\Psi_s}\frac{\D\eta}{\sqrt{ 2\cosh \eta - 2\cosh \Psi_m }}
= 2\sqrt{m}\left[ F\left(\frac{\pi}{2}\Big|m^2\right) - F\left(\varphi_s|m^2\right) \right]  \, ,
\end{eqnarray}
where $\varphi_s = \sin^{-1}\left[ \exp\left( \left[\Psi_s - \Psi_m\right]/2 \right) \right]$.

\begin{figure}%9
\centering
\includegraphics[scale=0.7]{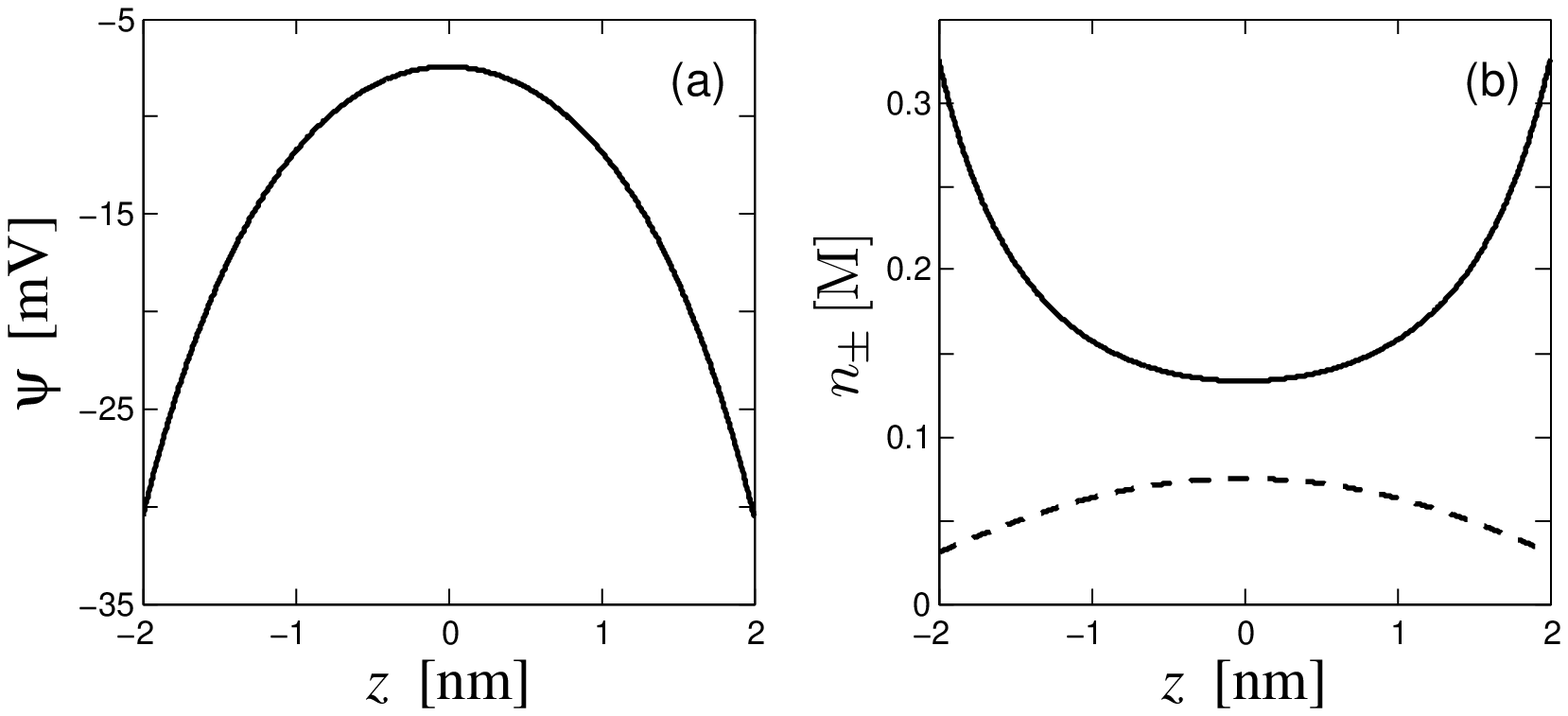}
\caption{ \textsf{Monovalent 1:1 electrolyte with $n_b = 0.1 \, {\rm M}$ ($\ld \simeq 0.97 \, {\rm nm}$),
between two identically charged membranes with $\sigma = -e/7 \, {\rm nm}^2$ each
($\lgc \simeq 1.6 \, {\rm nm}$), located at $z = \pm d/2$ with $d = 4 \, {\rm nm}$.
In (a) we plot the electrostatic potential, $\psi$,
and in (b) we show the co-ion (dashed line) and counter-ion (solid line) density profile, $n_{\pm}$.
The plots are obtained from Eqs.~(\ref{m2}), (\ref{m3}) and (\ref{m4}).}
}
\label{figure9}
\end{figure}

The three equations, Eqs.~(\ref{m2}), (\ref{m3}) and (\ref{m4}), completely determine the potential $\Psi(z)$,
the two species density profiles, $n_{\pm}(z) = n_b \exp(\mp \Psi)$
and their mid-plane values $n_{\pm}^{(m)} = n_b \exp(\mp \Psi_m)$,
as function of the three parameters: the
inter-membrane spacing $d$, the surface charge density $\sigma$ (or equivalently $\lgc$), and
the electrolyte bulk ionic strength $n_b$ (or equivalently $\ld$).
The exact form of the profiles and pressure can be obtained either from the numerical
solution of Eqs.~(\ref{m2}), (\ref{m3}) and (\ref{m4}) or by the usage of the elliptic functions.
For example, we calculate numerically the counter-ion, co-ion and potential profiles
as shown in Fig.~\ref{figure9},
where the three relevant lengths are $d = 4 \, {\rm nm}$, $\lgc \simeq 0.4d$ and $\ld \simeq 0.25d$.
%$d \simeq 2.5\lgc \simeq 4\ld$.

%%%%%%%%%%%%%%%%%%%%%%%%%%%%%%%%%%%%%%%%%%%%%%%%%%%%%%%%%%%%%%%%%%%%%%%%%%%%%%%%%%%%%
\subsection{Debye-H\"{u}ckel Regime} \label{sec9.5.3}
%%%%%%%%%%%%%%%%%%%%%%%%%%%%%%%%%%%%%%%%%%%%%%%%%%%%%%%%%%%%%%%%%%%%%%%%%%%%%%%%%%%%%

Broadly speaking (see section~\ref{sec9.2.1}), the PB equation
can be linearized when the surface potential is small, $|\Psi_s| \ll 1$.
In this case, the potential is small everywhere because it is a monotonous function that vanishes in the bulk.
The DH solution has the general form
\begin{eqnarray}
\label{n1a}
\Psi(z) = A\cosh{\kd z} + B\sinh{\kd z} \, ,
\end{eqnarray}
and the boundary conditions of Eq.~(\ref{k1}) dictate the specific solution
\begin{eqnarray}
\label{n1}
\Psi(z) = -\frac{2}{\kd\lgc}\frac{\cosh{\kd z}}{\sinh{(\kd d/2)}} \, .
\end{eqnarray}
In the DH regime, the potential is small, and the disjoining pressure, Eq.~(\ref{k2}),
can be expanded to second order in $\Psi_m$.
As the first order vanishes from electro-neutrality, we obtain,
\begin{eqnarray}
\label{n3}
\Pi \simeq \kbt n_b\Psi_m^2 = \frac{\kbt}{2\pi\lb\lgc^2}\frac{1}{\sinh^2\left(\kd d/2\right)}  \, .
\end{eqnarray}

The DH regime can be further divided into two sub-cases: ${\rm DH}_1$ and ${\rm DH}_2$.
For large separations, $d \gg \ld$, the above expression reduces to
\begin{eqnarray}
\label{n4}
\Pi \simeq \frac{2\kbt}{\pi\lb\lgc^2}\e^{-\kd d} \, .
\end{eqnarray}
This ${\rm DH}_1$ sub-regime is valid for $d \gg \ld$ and $\lgc \gg \ld$. \\
In the other limit of small separations, the pressure is approximated by
\begin{eqnarray}
\label{n5}
\Pi \simeq \frac{2\kbt}{\pi\lb\left(\kd\lgc\right)^2}\frac{1}{d^2} \, .
\end{eqnarray}
The limits of validity for this ${\rm DH}_2$  sub-regime are: $d \ll \ld$ and $\lgc \gg \ld^2/d$ (see Table~\ref{table1}).

%%%%%%%%%%%%%%%%%%%%%%%%%%%%%%%%%%%%%%%%%%%%%%%%%%%%%%%%%%%%%%%%%%%%%%%%%%%%%%%%%%%%%
\subsection{Intermediate Regime} \label{sec9.5.3.1}
%%%%%%%%%%%%%%%%%%%%%%%%    Intermediate regime  %%%%%%%%%%%%

When $d$ is the largest length-scale in the system,  $d\gg \ld$ and $d\gg \lgc$,
%For large distances, $d \gg \ld$ and $d \gg \lgc$,
the interaction  between the membranes is weak,
and one can use the superposition principle. This defines the {\it distal} region, where
the mid-plane potential is obtained by adding the contributions from two identical charged single surfaces,
located at $z=\pm d/2$.

%%%%%%%%%%%%%%%%%%%%%%%%%%%%%%%%%%%%%%%%%%%%%%%%%%%
\begin{table} [!h]
\center
\begin{tabular}{ c  c  c}
  \hline\noalign{\medskip}
  Pressure regime \,\,  &  \,\,  $\Pi$ \,\, & \,\, Range of validity \,\,
   \\ \noalign{\medskip}  \hline   \\
  Ideal-Gas    (IG)  &   $\mathlarger{\frac{\kbt}{\pi\lb\lgc}\frac{1}{d}}$         &  $\ld/d \gg \lgc/\ld \gg d/\ld$
  \\ \\ \hline \\
  Gouy-Chapman (GC)  &   $\mathlarger{\frac{\pi\kbt}{2\lb}\frac{1}{d^2}}$         &  $1 \gg d/\ld \gg \lgc/\ld$
  \\ \\ \hline \\
  Intermediate       &   $\mathlarger{\frac{8\kbt}{\pi\lb\ld^2}} \, \e^{-d/\ld}$       &  $ d/\ld \gg 1 \gg \lgc/\ld$
  \\ \\ \hline \\
  Debye-H\"uckel (${\rm DH_1}$) &   $\mathlarger{\frac{2\kbt}{\pi\lb\lgc^2}} \, \e^{-d/\ld}$
  &  $ d/\ld \gg 1 \,\, ; \,\,  \lgc/\ld \gg 1$
  \\ \\ \hline \\
  Debye-H\"uckel (${\rm DH_2}$) &   $\mathlarger{\frac{2\kbt\ld^2}{\pi\lb\lgc^2}\frac{1}{ d^2}}$
  &  $1 \gg d/\ld \gg \ld/\lgc$
  \\ \\ \hline
\end{tabular}
\caption{\textsf{ The five pressure regimes of the symmetric two-membrane system. }}
\label{table1}
\end{table}
%%%%%%%%%%%%%%%%%%%%%%%%%%%%%%%%%%%%%%%%%%%%%%%%%%%%%

In the distal region, the midplane potential is obtained from Eq.~(\ref{g5})  by the above-mentioned superposition,
\begin{eqnarray}
\label{n5a}
\Psi_m = -8\gamma\e^{-\kd d/2} \, .
\end{eqnarray}
Since $\Psi_m$ is small, the pressure expression,  Eq.~(\ref{k2}), can be expanded
to second order in $\Psi_m$, as was done in Eq.~(\ref{n3}), giving
\begin{eqnarray}
\label{n5b}
\Pi \simeq \kbt n_b \Psi_m^2 = 64\kbt\gamma^2n_b\e^{-\kd d} \, .
\end{eqnarray}
This osmotic pressure expression is valid for large distances, $d \gg \ld$ and $d \gg \lgc$,
and partially holds for the DH$_1$ regime.

The {\it intermediate} regime is obtained by further assuming strongly charged surfaces, $\ld \gg \lgc$.
In this limit, $\gamma =\tanh(-\Psi_s/4) \simeq 1$,
and the osmotic pressure is written as
\begin{eqnarray}
\label{n5b}
\Pi \simeq \frac{8\kbt\kd^2}{\pi\lb} \e^{-\kd d} \, .
\end{eqnarray}
The intermediate regime is valid for $d \gg \ld \gg \lgc$ (see Table~\ref{table1}).

%%%%%%%%%%%%%%%%%%%%%%%%%%%%%%%%%%%%%%%%%%%%%%%%%%%%%%%%%%%%%%%%%%%%%%%%%%%%%%%%%%%%%
\subsection{Other Pressure Regimes} \label{sec9.5.3.2}
%%%%%%%%%%%%%%  Other pressure regimes   %%%%%%%%%%%%%%%%

The pressure expression can be derived analytically in two other limits, which
represent the two regimes obtained for the counter-ions only case:
the Ideal-Gas regime (IG), Eq.~(\ref{l5}),
\begin{eqnarray}
\label{op1}
\Pi \simeq \frac{\kbt}{\pi\lb\lgc}\frac{1}{d} \, ,
\end{eqnarray}
valid for $\ld^2/d \gg \lgc \gg d$,
and the Gouy-Chapman regime (GC), Eq.~(\ref{l6}),
\begin{eqnarray}
\label{op2}
\Pi \simeq \frac{\pi\kbt}{2\lb}\frac{1}{d^2} \, ,
\end{eqnarray}
whose range of validity is $\ld \gg d \gg \lgc$.

\begin{figure}%10
\centering
\includegraphics[scale=0.7]{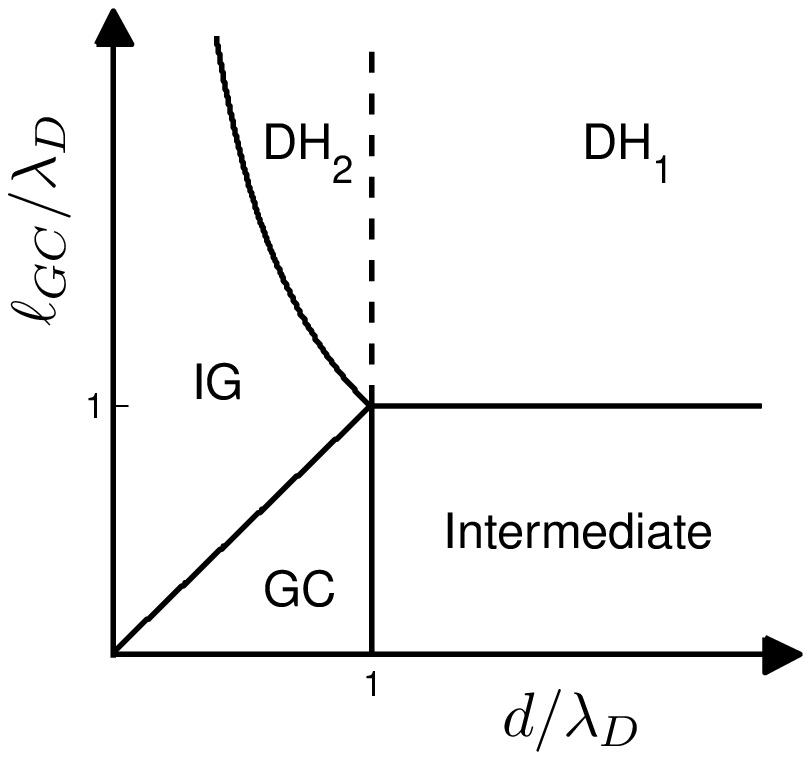}
\caption{ \textsf{Schematic representation of the various regimes of the PB equation
for two flat and equally charged membranes at separation $d$.
We plot the four different pressure regimes: Ideal-Gas (IG), Gouy-Chapman (GC),
Intermediate and Debye-H\"{u}ckel (DH).
The two independent variables are the dimensionless ratios $d/\ld$ and $\lgc/\ld$.
The four regimes are detailed in Table~\ref{table1}.
The DH regime is further divided into two sub-regimes:
${\rm DH}_1$ for large $d/\ld$ and ${\rm DH}_2$ for small $d/\ld$.}
}
\label{figure10}
\end{figure}

The five pressure regimes complete the discussion of the various limits
as function of the two ratios: $\lgc/\ld$ and $d/\ld $.
They are summarized in Table~\ref{table1} and plotted in Fig.~\ref{figure10}.

%%%%%%%%%%%%%%%%%%%%%%%%%%%%%%%%%%%%%%%%%%%%%%%%%%%%%%%%%%%%%%%%%%%%%%%%%%%%%%%%%%%%%
\section{Two Asymmetric Membranes, $\sigma_1 \neq \sigma_2$} \label{sec9.6}
%%%%%%%%%%%%%%%%%%%%%%%%%%%%%%%%%%%%%%%%%%%%%%%%%%%%%%%%%%%%%%%%%%%%%%%%%%%%%%%%%%%%%

For asymmetrically charged membranes, $\sigma_1 \neq \sigma_2$,
the interacting membranes imposes a different boundary condition.
%
%do not have the same charge, $\sigma_1 \neq \sigma_2$.
%Without loss of generality, we choose the total surface charge to be negative, $\sigma_1 + \sigma_2 \leq 0$.
%The PB equation is the same as for the symmetric case, Eq.~(\ref{e1}),
%but with the boundary conditions as in Eq.~(\ref{j1b}).
%
Such a system can model, for example, two surfaces that are coated with two different polyelectrolytes
or two lipid membranes with different charge/neutral lipid compositions.

It is possible to have an overall attractive interaction between two asymmetric membranes,
unlike the symmetric $\sigma_1=\sigma_2$ case.
When $\sigma_1$ and $\sigma_2$ have the same sign, the boundary condition of Eq.~(\ref{j1b})
implies that $\Psi^{\, \prime}(d/2)$ has the opposite sign of $\Psi^{\, \prime}(-d/2)$.
Since $\Psi^{\, \prime}$ is monotonous, it means that there is a point in between the plates for which $\Psi^{\, \prime}=0$.
The osmotic pressure, $\Pi$ of Eq.~(\ref{j8}), calculated at this special point,
has only an entropic contribution and is positive for any inter-membrane separation, $d$,
just as in the $\sigma_1=\sigma_2$ case.
%Because the pressure is fixed between the surfaces, we see that the pressure is purely repulsive, $\Pi > 0$.

However, when $\sigma_1$ and $\sigma_2$ have opposite signs,
$\Psi^{\, \prime}$ is always negative in between the two membranes,
and the sign of the pressure can be either positive (repulsive) or negative (attractive).
A crossover between repulsive and attractive pressure occurs when $\Pi = 0$,
and depends on four system parameters: $\sigma_{1,2}$, $\ld$ and $d$ (see Fig.~\ref{figure12}).

Although the general expression for $\Pi(d)$ cannot be cast in an analytical
form, a closed-form criterion exists for the crossover pressure, $\Pi = 0$,
for any amount of salt (Ben-Yaakov et al. 2007).
The crossover criterion has two rather simple limits: for the linearized
DH (high salt) limit, the general criterion reduces to the well-known result
of Parsegian and Gingell (1972), while in the counter-ion only limit,
another analytical expression has been derived more recently by Lau and Pincus (1999).

%%%%%%%%%%%%%%%%%%%%%%%%%%%%%%%%%%%%%%%%%%%%%%%%%%%%%%%%%%%%%%%%%%%%%%%%%%%%%%%%%%%%%
\subsection{The Debye-H\"uckel Regime} \label{sec9.6.1}
%%%%%%%%%%%%%%%%%%%%%%%%%%%%%%%%%%%%%%%%%%%%%%%%%%%%%%%%%%%%%%%%%%%%%%%%%%%%%%%%%%%%%

The crossover criterion between attraction and repulsion
has an analytical limit for high salinity, (Parsegian and Gingell 1972).
We repeat here the well-known argument (Ben-Yaakov and Andelman 2010) where the starting
point is the linear DH limit, Eq.~(\ref{c2}), of the full PB equation.

The DH equation for planar geometries has a solution, Eq.~(\ref{n1a}),
for which the boundary conditions of Eq.~(\ref{j1b}) yields,
\begin{eqnarray}
\label{p2}
\Psi(z) = \frac{2\pi\lb}{\kd e}\left[ \frac{\sigma_1+\sigma_2}{\sinh(\kd d/2)}\cosh(\kd z)
+ \frac{\sigma_2-\sigma_1}{\cosh(\kd d/2)} \sinh(\kd z)   \right] \, .
\end{eqnarray}
In the DH regime, the pressure expression, Eq.~(\ref{j8}), can be expanded in powers of the
electrostatic potential, $\Psi$.
Keeping only terms of order $\Psi^2$, the pressure can be written as
%The pressure is calculated from Eq.~(\ref{j8}) by expanding
%it to second order in $\psi$ (valid in the DH regime)
%
\begin{eqnarray}
\label{p3}
\Pi \simeq \frac{\kbt}{2\pi\lb\sinh^2(\kd d)}\Bigg[ \frac{1}{\ell_1^2} + \frac{1}{\ell_2^2} \pm \frac{2}{\ell_1\ell_2}\cosh(\kd d) \Bigg] \, ,
\end{eqnarray}
where $l_{1,2} = e/(2\pi\lb|\sigma_{1,2}|)$ are the two Gouy-Chapman lengths corresponding
to the two membranes with $\sigma_1$ and $\sigma_2$, respectively.
The $\pm$ sign of the last term corresponds to
the two situations: $\sigma_1 \cdot \sigma_2 > 0$ and $\sigma_1 \cdot \sigma_2 < 0$, respectively.

This $\Pi$ expression can be simplified in two limits.
For small separation, $d \ll \ld$, the expansion of the hyperbolic functions
yields a power-law divergence $\sim d^{-2}$ for $d \to 0$,
\begin{eqnarray}
\label{p4}
\Pi \simeq \frac{\kbt}{2\pi\lb}\left[ \left( \frac{1}{\kd \ell_1} \pm \frac{1}{\kd \ell_2} \right)^2\frac{1}{d^2}
\pm \frac{1}{\ell_1\ell_2}   \right] > 0 \, .
\end{eqnarray}
Clearly, it is positive definite (hence repulsive) for both $\pm$ signs.
However, when $\sigma_1=-\sigma_2$ (the antisymmetric case with $\sigma_1\cdot\sigma_2<0$),
the pressure goes to a negative constant
(independent of $d$), $\Pi = - \kbt/(2\pi\lb \ell_1 \ell_2)$.

For the opposite limit of large separation $d \gg \ld$,
the pressure decays exponentially, while its sign depends on the sign of $\sigma_1 \cdot \sigma_2$,

\begin{eqnarray}
\label{p5}
\Pi \simeq \pm \frac{\kbt}{\pi\lb \ell_1\ell_2}\e^{-\kd d} \, ,
\end{eqnarray}
thus, it is attractive for $\sigma_1 \cdot \sigma_2 < 0$.

\begin{figure}%12
\centering
\includegraphics[scale=0.7]{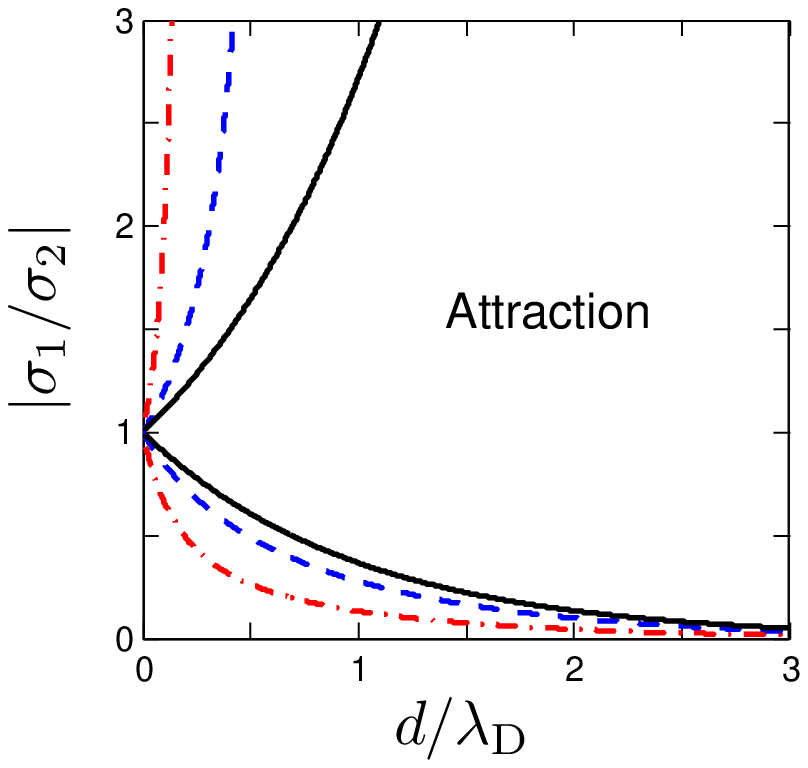}
\caption{ \textsf{Crossover from attraction to repulsion, $\Pi(d) = 0$, for two oppositely
charged membrane, $\sigma_1\cdot\sigma_2 < 0$,
in the $(\left|\sigma_1/\sigma_2\right|,d/\ld)$ plane.
The solid lines shows the crossover in the high-salt DH limit, Eq.~(\ref{p7}).
For lower salinity derived from the full PB theory, Eq.~(\ref{q4}), the attractive region is increased as is seen in
the examples we choose: $\ell_2/\ld = 0.2$ (red dash-dotted line) and $\ell_2/\ld = 0.7$ (blue dashed line).}
}
\label{figure12}
\end{figure}

The attraction/repulsion crossover is calculated from the
zero pressure condition of Eq.~(\ref{p3}),
while keeping in mind that attraction is possible only for oppositely charged membranes,
$\sigma_1 \cdot \sigma_2 < 0$ (see the beginning of this section)
\begin{eqnarray}
\label{p7}
\e^{-\kd d} < \Big|\frac{\sigma_1}{\sigma_2}\Big| < \e^{\kd d} \, .
\end{eqnarray}
This is exactly the result obtained by Parsegian and Gingell (1972).
Interestingly, in the linear DH case, the crossover depends only on the
ratio of the two surface charges $|\sigma_1/\sigma_2|$ and not on their separate values,
as can be seen in Fig.~\ref{figure12}.
For comparison, we plot (with dashed and dash-dotted lines on the same figure)
two examples of low-salt crossovers, as calculated
from the general criterion presented below for the full PB theory (section~\ref{sec9.6.3}).
The low-salt line has a smaller repulsive region.
Increasing the salt concentration increases the repulsion region due to screening of electrostatic interactions.
The repulsive region increases till it reaches the
Parsegian-Gingell result for the DH limit (solid line in Fig.~\ref{figure12}).

%%%%%%%%%%%%%%%%%%%%%%%%%%%%%%%%%%%%%%%%%%%%%%%%%%%%%%%%%%%%%%%%%%%%%%%%%%%%%%%%%%%%%
\subsection{DH Regime with Constant Surface Potential} \label{sec9.6.1.1}
%%%%%%%%%%%%%%%%%%%%%%  Constant potential %%%%%%%%%%%%%%%%%%%%%%%%%%%%%%

So far we have solved the PB equation using the constant charge boundary conditions.
However, constant potential boundary conditions
are appropriate when the surfaces are metal electrodes,
and it is important to understand this case as well.

Let us examine the effect of constant surface potential on the pressure.
For simplicity we will focus on the linearized PB equation (DH)
in the asymmetric membrane case.
We still refer to the setup as in Fig.~\ref{figure6}.
The two membranes at $z=\pm d/2$ are held at different values of
constant surface potential, $\Psi_{1,2}$,
\begin{eqnarray}
\label{r4}
\nonumber& &\Psi\Big|_{z = -d/2}  = \Psi_1 \, ,\\
& &\Psi\Big|_{z = d/2}  \,\,\,\,\,= \Psi_2  \, .
\end{eqnarray}
Applying the DH solution of Eq.~(\ref{n1a})
with the boundary conditions of Eq.~(\ref{r4}) leads to
\begin{eqnarray}
\label{r2}
\Psi(z) = \frac{\Psi_1 + \Psi_2}{2\cosh(\kd d/2)} \cosh(\kd z) +
\frac{\Psi_1 - \Psi_2}{2\sinh(\kd d/2)} \sinh(\kd z) \, ,
\end{eqnarray}
and expanding the pressure $\Pi$ from Eq.~(\ref{j8b}) to second order in powers of $\Psi_{1,2}$ yields,
\begin{eqnarray}
\label{r6}
\Pi \simeq \frac{\kbt n_b }{\sinh^2(\kd d)}\left(2\Psi_{2} \Psi_{1}\cosh(\kd d) - \Psi^2_{2} - \Psi^2_{1} \right) \, .
\end{eqnarray}
This expression is similar to the one obtained for
constant surface charge, Eq.~(\ref{p3}).
Indeed, for large separations, $d\gg\ld$, the relative sign of $\Psi_{2}$
and $\Psi_{1}$ determines the sign of the pressure
\begin{eqnarray}
\label{r7}
\Pi \simeq 2\kbt n_b \Psi_{2} \Psi_{1}\e^{-\kd d} \, ,
\end{eqnarray}
as for the large-separation behavior of the constant-charge case.

However, for small separations, $d \ll \ld$, the pressure is different than for the constant surface-charge case,
\begin{eqnarray}
\label{r8}
\Pi \simeq -\frac{\kbt\left(\Psi_2-\Psi_1\right)^2}{8\pi\lb}\frac{1}{d^2} + \kbt n_b \Psi_{2}\Psi_{1} \, .
\end{eqnarray}
It yields a pure attractive (negative) pressure that diverges as $\sim 1/d^2$,
and does not depend on $n_b$.
For the special symmetric case $\Psi_{2} = \Psi_{1}$, at those small $d$, the pressure does not diverge and
reaches a positive constant,
$\Pi>0$, proportional to $n_b$.
Unlike the constant-charge case, here the counter-ion concentration remains constant near each of the membranes,
because it depends only on the surface potential through the Boltzmann factor (Ben-Yaakov and Andelman 2010).
However, the induced surface charge ($\sigma \propto \Psi_s^{\,\prime}$, Eq.~(\ref{r2}))
diverges when the membranes are brought closer together, resulting in a diverging electrostatic attraction.

Note that the crossover from repulsive to attractive pressure is obtained for zero pressure in Eq.~(\ref{r6})
and is possible only for potentials of the same sign, $\Psi_{2}\cdot\Psi_{1}>0$ including $\Psi_1 = \Psi_2$.
The condition for attraction reads
\begin{eqnarray}
\label{r8a}
\e^{-\kd d} < \frac{\Psi_2}{\Psi_1} < \e^{\kd d} \, .
\end{eqnarray}
For potentials of opposite sign, $\Psi_{2}\cdot\Psi_{1}<0$, the pressure is purely attractive.

%%%%%%%%%%%%%%%%%%%%%%%%%%%%%%%%%%%%%%%%%%%%%%%%%%%%%%%%%%%%%%%%%%%%%%%%%%%%%%%%%%%%%
\subsection{Counter-ions Only} \label{sec9.6.2}
%%%%%%%%%%%%%%%%%%%%%%%%%%%%%%%%%%%%%%%%%%%%%%%%%%%%%%%%%%%%%%%%%%%%%%%%%%%%%%%%%%%%%

In the absence of an external salt reservoir, the only mobile ions in the solution are
counter-ions with density $n(z)$,
such that the system is charge neutral,
\begin{eqnarray}
\label{o3a}
\sigma_1 + \sigma_2  = -e\int_{-d/2}^{d/2}n(z) \, \D z \, .
\end{eqnarray}
For the assumed overall negative charge on the two membranes,
$\sigma_1+\sigma_2<0$, the counter-ions are positive, $z_+=1$.

The PB equation for the two-membrane system is the same as for the single membrane, Eq.~(\ref{e1}),
with the boundary condition as in Eq.~(\ref{j1b}).
The osmotic pressure, Eq.~(\ref{j8}), reduces here to
\begin{eqnarray}
\label{o3}
\Pi = -\frac{\kbt}{8\pi\lb}{\Psi^{\, \prime}}^2(z) + \kbt n_0\e^{-\Psi(z)} \, ,
\end{eqnarray}
where $n_0$ is defined as the reference density for which $\Psi = 0$.
This equation is a first-order ordinary differential equation and can be integrated.
Nevertheless, its solution depends on the sign of the osmotic pressure.
We will not present here the solution of the PB equation, but rather discuss the
crossover between attractive and repulsive pressures.
This crossover is obtained by solving Eq.~(\ref{o3}) with $\Pi = 0$.
As the total surface charge is chosen to be negative, $\sigma_1 + \sigma_2 \leq 0$,
and attraction occurs only for $\sigma_1\cdot\sigma_2 < 0$,
we choose $\sigma_1$ to be negative and $\sigma_2$ to be positive.

Integrating this equation and using the boundary condition at
$z = -d/2$, we obtain the same algebraically decaying profile for the counter-ion
density as in the single membrane case, Eq.~(\ref{f4a}), with a shifted $z$-axis origin: $z \to z+d/2$
and $\lgc \to \ell_1$ (with $\ell_1$ defined as before):
\begin{eqnarray}
\label{o10}
\Psi = \Psi_0 + 2\ln\left( z + \ell_1 + d/2 \right) \, .
\end{eqnarray}
The second boundary condition at $z=d/2$ gives a relation between $d$ and $\sigma_{1,2}$.
The condition for attraction can be expressed in terms of the surface densities (Kandu\v c et al. 2008):
\begin{eqnarray}
\label{o12}
|\sigma_1| - |\sigma_2| < \frac{|\sigma_2\sigma_1|}{\sigma_d}  \, ,
\end{eqnarray}
where $\sigma_d \equiv e/(2\pi\lb d)$.

\begin{figure}%11
\centering
\includegraphics[scale=0.7]{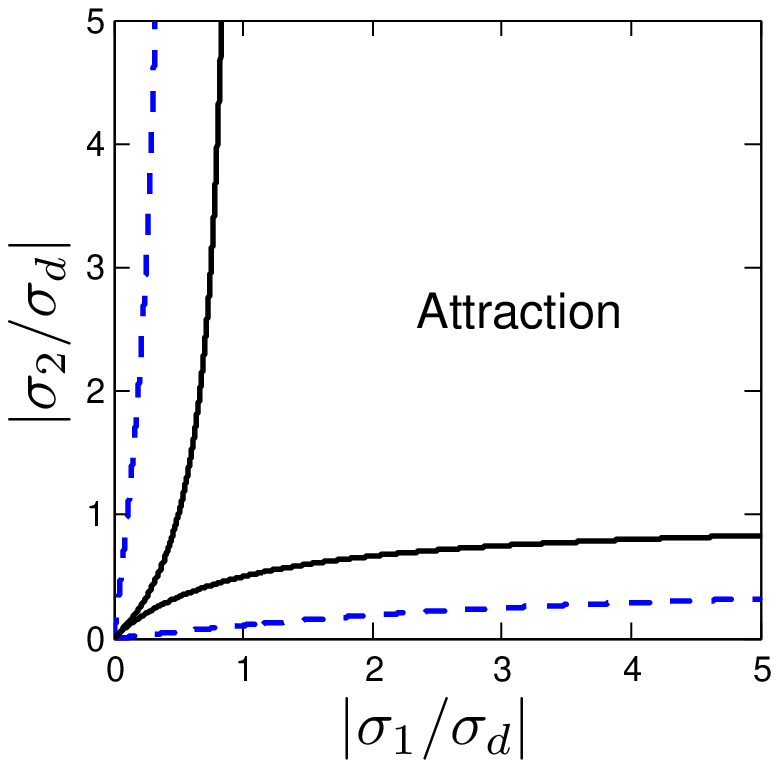}
\caption{ \textsf{Regions of attraction ($\Pi < 0$) and repulsion ($\Pi > 0$) for the counter-ion
only case, plotted in terms of the two rescaled charge densities, $\left|\sigma_1/\sigma_d\right|$
and $\left|\sigma_2/\sigma_d\right|$, where $\sigma_d = e/2\pi\lb d$.
The figure is plotted for the oppositely charged membranes, $\sigma_1\cdot\sigma_2<0$, and
is symmetric about the diagonal $|\sigma_1| = |\sigma_2|$.
The two solid lines delimit the boundary between
repulsion and attraction in the no-salt limit, $n_b\to0$, Eq.~(\ref{o12a}).
For comparison, we also plot the crossover between attraction and
repulsion for finite $n_b$ from Eq.~(\ref{q4}) for $d/\ld = 2.2$ (blue dashed line).
For the case of $\sigma_1\cdot\sigma_2>0$, $\Pi$ is always repulsive and there is no crossover.}
}
\label{figure11}
\end{figure}

The crossover between attraction and repulsion is plotted in Fig.~\ref{figure11}.
Two crossover lines separate the central attraction region
from two repulsion ones.
The upper one lies above the diagonal, $|\sigma_1| > |\sigma_2|$
and corresponds directly to the condition of Eq.~(\ref{o12}).
A second crossover line lies in the lower wedge below the diagonal, $|\sigma_1| < |\sigma_2|$.
It corresponds to the crossover of the complementary problem of an overall
positive surface charge, $\sigma_1+\sigma_2 > 0$, and negative counter-ions.
Note that the figure is symmetric about the principal diagonal,
$|\sigma_1| \leftrightarrow |\sigma_2|$, as expected.
The condition of attraction, irrespectively of the sign of $\sigma_1$ and $\sigma_2$, can be written as,
\begin{eqnarray}
\label{o12a}
|\sigma_1| - |\sigma_2|< \frac{|\sigma_2\sigma_1|}{\sigma_d} < |\sigma_1| - |\sigma_2|  \, ,
\end{eqnarray}
as was obtained by Lau and Pincus (1999).

%%%%%%%%%%%%%%%%%%%%%%%%%%%%%%%%%%%%%%%%%%%%%%%%%%%%%%%%%%%%%%%%%%%%%%%%%%%%%%%%%%%%%
\subsection{Atrraction/Repulsion Crossover} \label{sec9.6.3}
%%%%%%%%%%%%%%%%%%%%%%%%%%%%%%%%%%%%%%%%%%%%%%%%%%%%%%%%%%%%%%%%%%%%%%%%%%%%%%%%%%%%%

We now calculate the general criterion of the attractive-to-repulsive crossover.
The $\Pi = 0$ pressure between two membranes located at $z = \pm d/2$
can be mapped exactly into the problem of a single membrane at $z=0$
in contact with the same electrolyte reservoir.
The only difference is that beside the boundary at $z=0$,
there is another boundary at $z=d$.
The mapping to the single-membrane case is possible as the osmotic pressure
of a single membrane is zero.
This equivalence can be checked by substituting $\Pi=0$ in Eq.~(\ref{j8})
to recover the first integration of the PB equation
for the single membrane system with added 1:1 electrolyte, Eq.~(\ref{h1}).

The potential can be written as in Eq.~(\ref{g1a}) with
$\gamma \to \gamma_1 = \sqrt{1+(\kd l_{1})^2} - \kd \ell_1$.
This solution already satisfies the boundary condition, $\Psi^{\, \prime}(0) = -4\pi\lb\sigma_1/e$,
but another boundary condition at $z=d$ needs to be satisfied as well, $\Psi^{\, \prime}(d) = 4\pi\lb\sigma_2/e$.

Attraction will occur only for charged membranes of different sign, $\sigma_1\cdot\sigma_2 < 0$.
Using the boundary condition at $z=d$ for the two cases, $\sigma_1 < 0$ and $\sigma_1 > 0$,
determines the region of attraction, $\Pi < 0$, by the inequalities (Ben-Yaakov et al. 2007)
\begin{eqnarray}
\label{q4}
\e^{-\kd d} < \frac{\gamma_2}{\gamma_1} < \e^{\kd d} \, ,
\end{eqnarray}
with $\gamma_2 = \sqrt{1+(\kd l_{2})^2} - \kd \ell_2$.
It can be shown that the above general expression, Eq.~(\ref{q4}),
reduces to the expression of Eq.~(\ref{o12a}), in the limit of counter-ion only
and to that of Eq.~(\ref{p7}) in the high-salt (DH) limit.

\begin{figure}%11a
\centering
\includegraphics[scale=0.7]{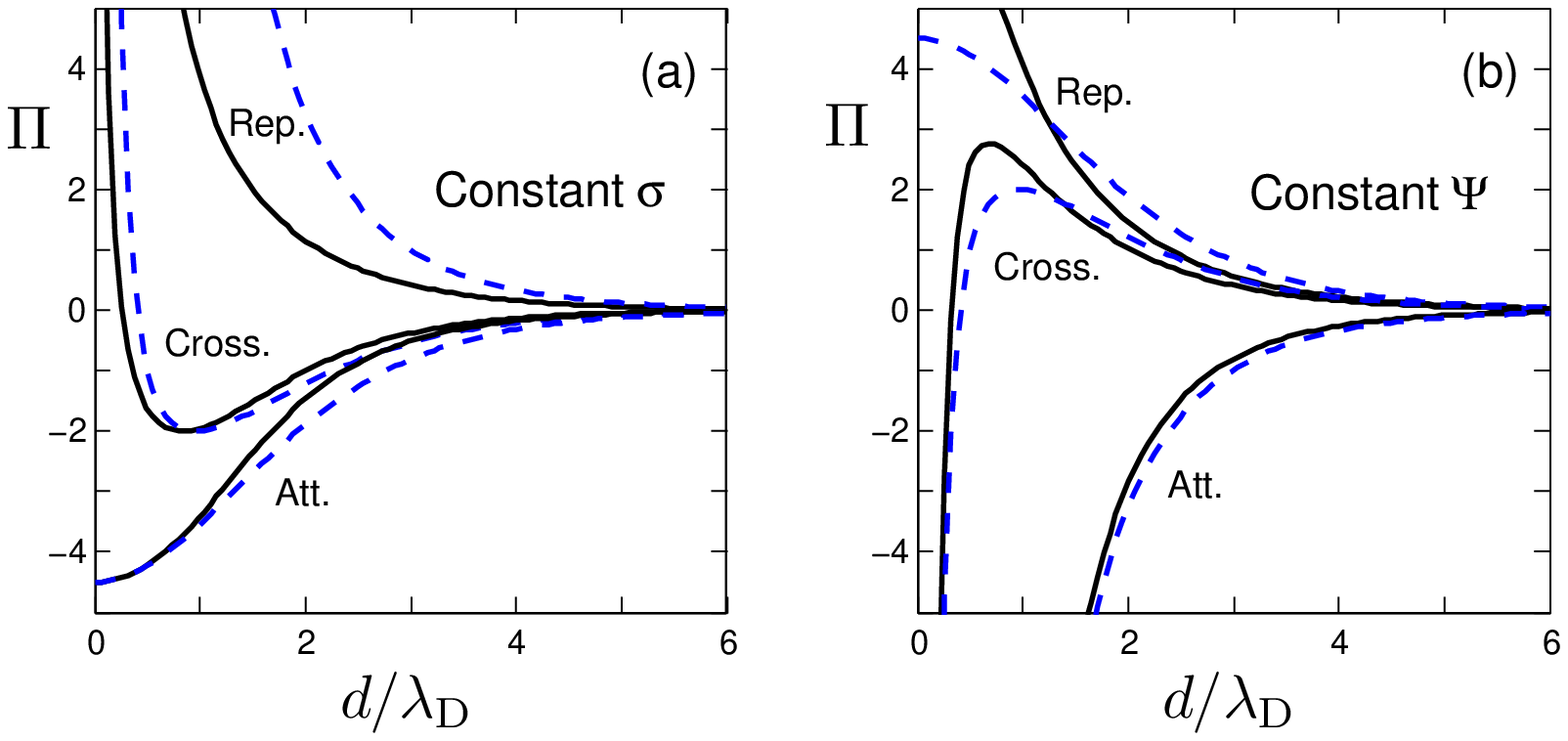}
\caption{\textsf{ Osmotic pressure, $\Pi$, in units of $\kbt/(4\pi\lb\ld^2)$,
as function of the dimensionless inter-membrane separation, $d/\ld$.
We present the solution of the non-linear (solid black lines) PB equation
and the linear (blue dashed lines) DH equation for two boundary conditions:
(a) constant surface charge, and (b) constant surface potential.
In each of the figure parts we show three profiles: repulsive, crossover and attractive.
In (a), the boundary conditions for the repulsive, crossover and attractive profiles are
$\sigma_1 = \sigma_2 = 3$, $\sigma_1 = 3$ and $\sigma_2 = -2$,
$\sigma_1 = 3$ and $\sigma_2 = -3$, respectively,
where $\sigma$ is given in units of $e/(4\pi \lb\ld)$.
In (b), the boundary conditions for the repulsive, crossover and attractive profiles are
$\Psi_1 = \Psi_2 = 3$, $\Psi_1 = 3$ and $\Psi_2 = 2$, $\Psi_1 = 3$ and $\Psi_2 = -3$, respectively.
Note that the non-linear PB solution of $\Pi$ for the symmetric (repulsive) osmotic pressure for
constant potential reaches a constant value as $d\to 0$, like in the DH case, but with different value,
$\Pi(d\to 0)\simeq 9.07$ in units of $\kbt/(4\pi\lb\ld^2)$ (not shown in the figure).
}
}
\label{figure11a}
\end{figure}

A similar general crossover criterion can also be obtained for constant potential
boundary conditions.
As we explained above, the crossover condition maps to the single membrane problem,
yielding the generalized relation of Eq.~(\ref{g3}), $\gamma_{1,2} = \pm\tanh(\Psi_{1,2}/4)$.
The $\pm$ sign is chosen such that $\gamma_{1,2}$ is positive.
For opposite surface potentials, $\Psi_1\cdot\Psi_2 < 0$,
there is a point between the membranes in which the potential vanishes.
The osmotic pressure of Eq.~(\ref{j8}), calculated at this point,
has only the negative Maxwell stress contribution and therefore, it is always {\it attractive}.
%
%The region of attraction
On the other hand, for $\Psi_1\cdot\Psi_2 > 0$,
the following condition on $\Psi_1$ and $\Psi_2$ results in an attraction
%written in terms of the surface potentials, yields
%
\begin{eqnarray}
\label{q4a}
\e^{-\kd d} < \frac{\tanh(\Psi_2/4)}{\tanh(\Psi_1/4)} < \e^{\kd d} \, .
\end{eqnarray}

In Fig.~\ref{figure11a} we show the osmotic pressure, $\Pi$, in units of $\kbt/(4\pi\lb\ld^2)$,
as a function of the (dimensionless) inter-membrane separation, $d/\ld$.
The pressure is calculated for several values of constant charge and constant potential boundary conditions.
Three types of pressure profiles are seen in the figure: attractive, repulsive and the crossover between attraction and repulsion.
%The crossover from purely repulsive to purely attractive pressure to a crossover profile
%is seen clearly in the figure.

%%%%%%%%%%%%%%%%%%%%%%%%%%%%%%%%%%%%%%%%%%%%%%%%%%%%%%%%%%%%%%%%%%%%%%%%%%%%%%%%%%%%%
\section{Charge Regulation} \label{sec9.7}
%%%%%%%%%%%%%%%%%%%%%%%%%%%%%%%%%%%%%%%%%%%%%%%%%%%%%%%%%%%%%%%%%%%%%%%%%%%%%%%%%%%%%

\begin{figure}%11b
\centering
\includegraphics[scale=0.5]{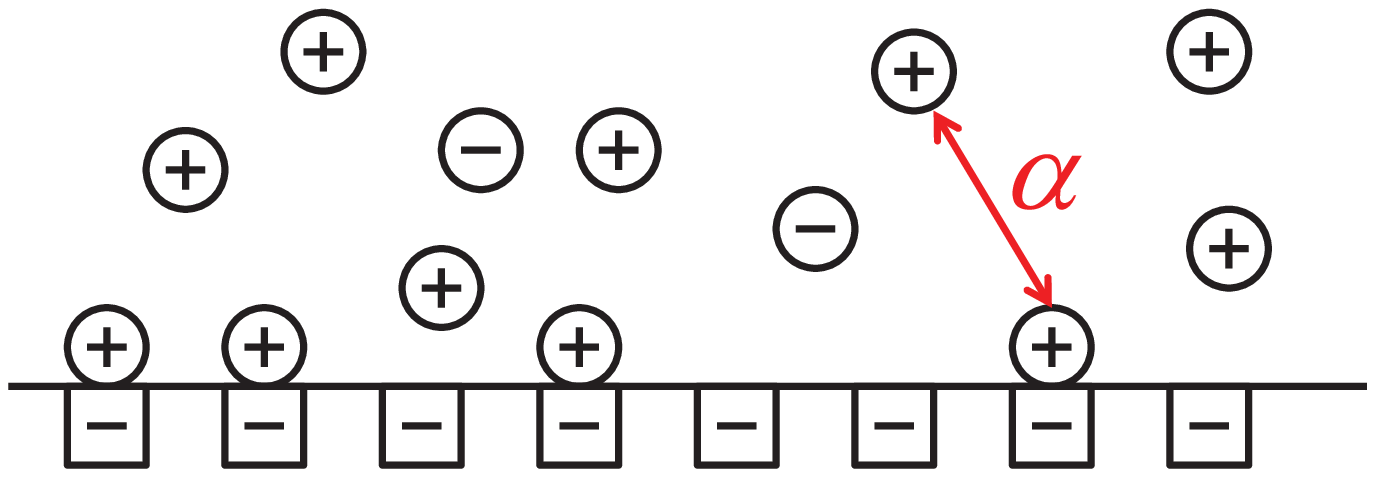}
\caption{  \textsf{Illustration of the charge regulation boundary condition for cation association/dissociation energy $\alpha$.}
}
\label{figure11b}
\end{figure}

As discussed in section~\ref{sec9.6.1.1}, the difference between constant surface potential,
$\Psi_s$, and constant surface charge density, $\sigma$, is large when
the distance between the two membranes is of order of the Debye screening
length, $\ld$, or smaller.
Ninham and Parsegian (1971) considered an interesting intermediate case
of great practical importance.
Membranes with ionizable groups that can release ions into the aqueous
solution or trap them $\large{-}$ a situation intermediate between a constant $\sigma$,
describing inert ionic groups on the membrane,
and constant $\Psi_s$, relevant for a surface (an electrode or a membrane) held at a constant
potential by an external potential source.

Let us consider a system where the membrane is composed of ionizable groups (lipids)
that each can release a counter-ion into the solution.
This surface dissociation/association (see Fig.~\ref{figure11b}) is described by the reaction:
\begin{eqnarray}
\label{s1}
{\rm A}^+ + {\rm B}^- \rightleftharpoons {\rm AB} \, ,
\end{eqnarray}
where A denotes a surface site that can be either ionized $({\rm A}^+)$
or neutral (AB).
The process of membrane association/dissociation is characterized by a kinetic constant $K_d$
through the law of mass action (le Chatelier's principle)
\begin{eqnarray}
\label{s2}
K_d = \frac{[{\rm A}^+][{\rm B}^-]_s}{[{\rm AB}]} \, ,
\end{eqnarray}
where $[{\rm A}^+]$, $[{\rm B}^-]_s$ and [AB] denote the three corresponding surface concentrations
(per unit volume).
We define $\phi_s$ to be the area fraction of the ${\rm A}^+$ ions,
related to $\sigma > 0$, the membrane charge density by $\phi_s = \sigma a^2/e \sim [{\rm A}^+]$
and $1 - \phi_s \sim [{\rm AB}]$, where $a^2$ is the surface area per charge.
The equilibrium condition of Eq.~(\ref{s2}) is then written as
\begin{eqnarray}
\label{s3}
K_d = \frac{\phi_s}{1 - \phi_s}[{\rm B}_s^-] \, .
\end{eqnarray}

As the counter-ions are released into the ionic solution,
the relation between surface and bulk ${\rm B}^-$ concentrations, $[{\rm B}^-]_s$ and $[{\rm B}^-]_{\infty}$,
is obtained via the Boltzmann distribution, $[{\rm B}^-]_s = [{\rm B}^-]_{\infty} \exp(\Psi_s)$
where $[{\rm B}^-]_{\infty} = n_b$ is the bulk salt concentration.
We note that in this section we choose $\sigma>0$ and it implies $\Psi_s > 0$,
and the reference potential in the bulk is set to be $\Psi = 0$.
The concentration of dissociated ${\rm B}^-$ ions at the surface is $[{\rm B}^-]_s$
and it is equal to $n_s^- = n_-(z\to0)$.
Note that this is {\it not} the charge concentration of the membrane itself
(which comprises the bound ionic groups and is proportional to $\phi_s$),
but is the concentration of mobile ions evaluated at the sub-membrane position (just as in section~\ref{sec9.3.3}),
where the standard PB equation holds.
One can then write the area fraction of the membrane ionized sites as
\begin{eqnarray}
\label{s4}
\phi_s = \frac{K_d}{K_d + n_b\e^{\Psi_s}} \, .
\end{eqnarray}

\begin{figure}%13a
\centering
\includegraphics[scale=0.7]{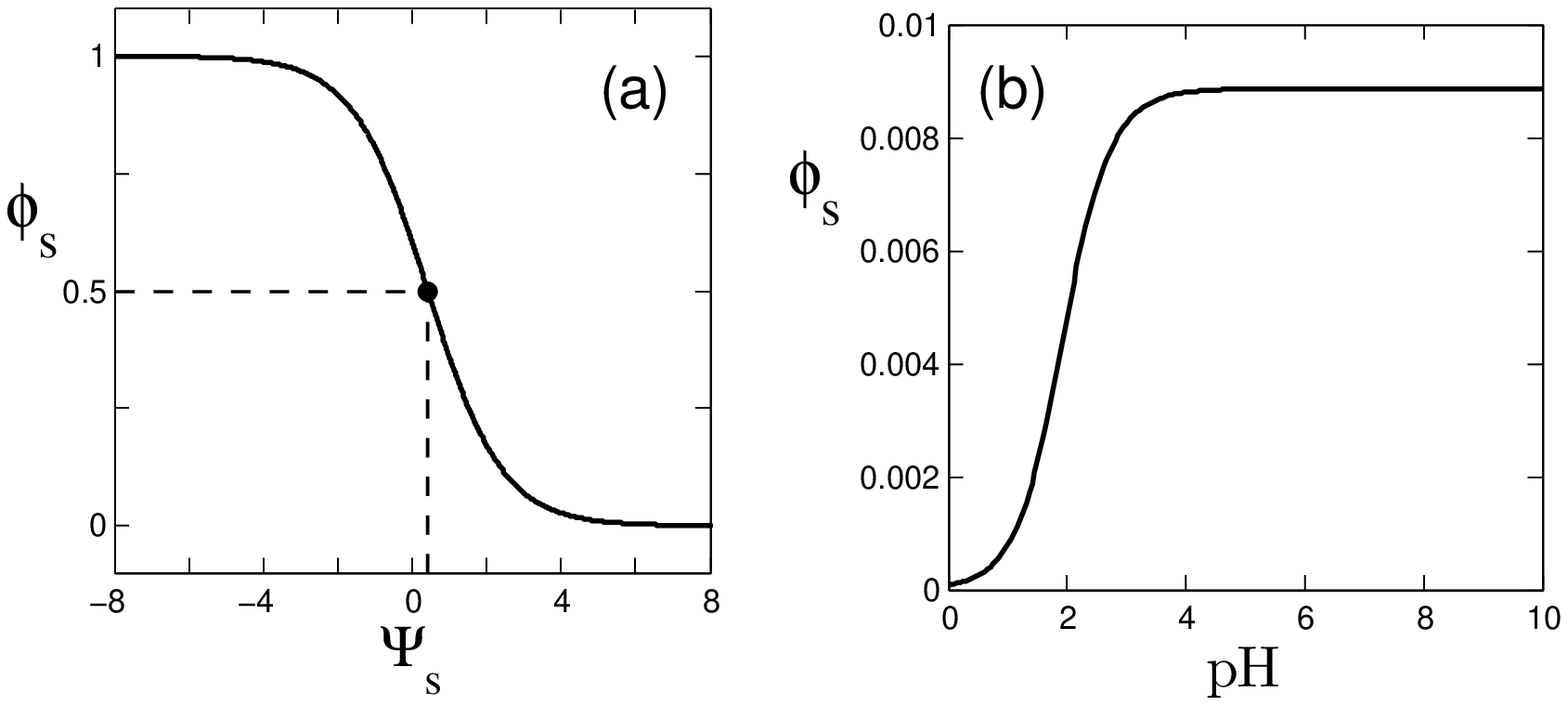}
\caption{\textsf{
In (a) the area fraction of charge groups on the membrane, $\phi_s$,
is plotted as a function of the dimensionless electrostatic potential, $\Psi_s$, using Eq.~(\ref{s5}).
The parameters used are: $a = 0.3 \, {\rm nm}$, $n_b = 0.1\, {\rm M}$ and $\alpha = -6$ (${\rm pK} \simeq 0.82$).
The symmetric point of $\phi_s = 1/2$ occurs at $\Psi_s^* \simeq 0.425$,
and the slope there is exactly $-0.25$.
In (b) we present the area fraction of charge groups on the membrane, $\phi_s$,
as a function of pH for surface binding ${\rm A}^- + {\rm H}^+ \rightleftharpoons {\rm AH}$ (see box below).
The parameters used are: $a = 0.5 \, {\rm nm}$ and $\alpha = -12$ (${\rm pK} \simeq 4.1$).
}
}
\label{figure13a}
\end{figure}

It is useful to introduce a surface interaction parameter $\alpha = \ln(a^3 K_d)$
instead of using the kinetic constant (see Fig.~\ref{figure11b}).
This gives the adsorption isotherm for $\phi_s$ (the fraction of ${\rm A}^+$ groups on the membrane)
\begin{eqnarray}
\label{s5}
\phi_s = \frac{1}{1 + \phi_b\e^{ - \alpha + \Psi_s  }} \, ,
\end{eqnarray}
where $\phi_b = a^3n_b$ as before.
The typical shape of $\phi_s$ as function of $\Psi_s$ is a sigmoid and is shown in Fig.~\ref{figure13a} (a)
for $a = 0.3 \, {\rm nm}$, $n_b = 0.1 \, {\rm M}$ and $\alpha = -6$ (${\rm pK} \simeq 0.82$).
The fraction of charge groups, $\phi_s$, varies between $\phi_s=1$ (fully charged)
to $\phi_s=0$ (neutral) as $\Psi_s$ varies from negative values to positive ones.
For the half-filled surface charge, $\phi_s=0.5$, the surface potential is $\Psi_s^* = \alpha - \ln\phi_b$.
At this special point, the slope of $\phi_s(\Psi_s)$ is exactly $-0.25$.
The differential capacitance as discussed in section~\ref{sec9.4}
can be cast into a simple form for the charge regulation case
\begin{eqnarray}
\label{s5b}
C_{\rm CR} = \frac{\D\sigma}{\D\psi_s} = \frac{e^2}{\kbt a^2}\frac{\D\phi_s}{\D\Psi_s}
= \frac{e^2}{\kbt a^2}\phi_s\left(1-\phi_s\right) \, .
\end{eqnarray}
It follows that for the sigmoid shape of $\phi_s(\Psi_s)$, the differential capacitance
has a unimodal shape with a universal maximum, $C_{\rm CR}=e^2/(4\kbt a^2)$.

Similarly, for the fraction of neutral AB groups on the membrane, $1 - \phi_s$, we write
\begin{eqnarray}
\label{s6}
1-\phi_s = \frac{\phi_b}{\phi_b + \e^{ \alpha - \Psi_s}} \, .
\end{eqnarray}
The above equation for $1-\phi_s$, Eq.~(\ref{s6}), is the {\it Langmuir-Davies isotherm} (Davies 1958),
and is an extension of the Langmuir adsorption isotherm (see, {\it e.g.}, Adamson and Gast 1997)
for charged adsorbing particles. This can be understood because $1-\phi_s$ is the fraction of the membrane AB neutral groups.
Hence, it effectively describes the adsorption of ${\rm B}^-$ ions onto a charged membrane.

%%
%\begin{figure}%13a
%\centering
%%\includegraphics[width=200pt, height=200pt]{Chapters/chapter9/figures/General_Crossover_DH.eps}
%\includegraphics[scale=0.7]{Chapters/chapter9/figures/Area_Fraction.eps}
%\caption{ Typical area fraction of charge groups on the membrane, $\phi_s$,
%as a function of the dimensionless electrostatic potential, $\Psi_s$, using Eq.~(\ref{s5}).
%The parameters used are: $a = 0.3 \, {\rm nm}$, $n_b = 0.1\, {\rm M}$ and $\alpha = -6$ (${\rm pK} \simeq 0.82$).
%The symmetric point of $\phi_s = 1/2$ occurs at $\Psi_s^* \simeq 0.425$,
%and the slope there is exactly $-1/4$.
%}
%\label{figure13a}
%\end{figure}
%%
%
%%
%\begin{figure}%13b
%\centering
%\includegraphics[scale=0.7]{Chapters/chapter9/figures/Phi_of_PH.eps}
%\caption{ Typical area fraction of charge groups on the membrane, $\phi_s$,
%as a function of pH for the process ${\rm A}^- + {\rm H}^+ \rightleftharpoons {\rm AH}$.
%The parameters used are: $a = 0.5 \, {\rm nm}$ and $\alpha = -6$ (${\rm pK} \simeq 1.48$).
%}
%\label{figure13b}
%\end{figure}
%%

%%%%%%%%%%%%%%%%%%%%%%%%%%%%%%%%%%%%%%%%%%%%%%%%%%%%%%%%%%%%%%%%%%%%%%%%%%%%%%%%%%%%%
\begin{shadebox}
\Boxhead{Surface pH and pK}
When an acidic reservoir exchanges
${\rm H}^+$ ions with the membrane, the membrane chemical reaction is
${\rm A}^- + {\rm H}^+ \rightleftharpoons {\rm AH}$,
and the same local equilibrium of Eq.~(\ref{s2}) can be expressed in terms of three
logarithms: ${\rm pK}\equiv -\log_{10} K_d = -\log_{10}\left(a^{-3}\e^{\alpha}\right)$ where the close-packing density $a^{-3}$ is measured in Molar,
${\rm pH}_s \equiv -\log_{10} [{\rm H}^+]_s$,
and ${\rm pH} \equiv -\log_{10}[{\rm H}^+]_\infty$.
The regular pH measures the acidic strength in the reservoir, while the membrane pK
(or $K_d$) is a fixed (and usually unknown) parameter that depends
on the membrane as well as on the binding ${\rm H}^+$ ions.
Equation~(\ref{s3}) can then be written as
\begin{eqnarray}
\label{s14}
\frac{\phi_s}{1-\phi_s} = 10^{-{\rm pK}+{\rm pH}}\e^{\Psi_s} \, .
\end{eqnarray}
When half of the membrane is charged, $\phi_s = 0.5$,
the given pK relates the solution pH with the surface potential:
\begin{eqnarray}
\label{s15}
\Psi_s = \left( {\rm pH} - {\rm pK} \right) \ln 10 \, .
\end{eqnarray}

In Fig.~\ref{figure13a} (b) we show the dependence of $\phi_s$ on ${\rm pH}$ for process of protonation/deprotonation
for the case of one membrane with:
$a = 0.5 \, {\rm nm}$ and $\alpha = -12$ (${\rm pK} \simeq 4.1$).
The typical shape of $\phi_s({\rm pH})$ is a sigmoid and is obtained by solving numerically
Eq.~(\ref{s8}) with the charge regulation boundary condition of Eq.~(\ref{s5}).
\end{shadebox}
%%%%%%%%%%%%%%%%%%%%%%%%%%%%%%%%%%%%%%%%%%%%%%%%%%%%%%%%%%%%%%%%%%%%%%%%%%%%%%%%%%%%%

The Langmuir-Davies isotherm, Eq.~(\ref{s6}), relates the self-adjusting surface charge
fraction $\phi_s$ and surface potential $\Psi_s$ with $K_d$ (or $\alpha$) and the bulk density, $n_b$.
In order to find $\Psi_s$ and $\phi_s$ separately as function of $K_d$ and $n_b$,
one needs to find the electrostatic relation between them.
This relation can be obtained from the Grahame equation~(\ref{h3}),
introduced in section~\ref{sec9.3.3}.
\begin{eqnarray}
\label{s8}
w\phi_s = \sinh(\Psi_s/2) \quad ; \quad w \equiv \frac{2\pi\lb\ld}{a^2} \, ,
\end{eqnarray}
where we expressed the Grahame equation as $\phi_s = \phi_s(\Psi_s)$
in terms of the Bjerrum and Debye lengths via a dimensionless parameter, $w$.
By inverting the relation, $\Psi_s = \Psi_s(\phi_s)$, we get,
\begin{eqnarray}
\label{s9}
\e^{\Psi_s/2} = w\phi_s + \sqrt{1 + (w\phi_s)^2} \, .
\end{eqnarray}
Recall that this Grahame equation~(\ref{h3}) is obtained from
the PB equation for one planar membrane and gives a relationship between the density $\sigma$
at the membrane and $n_s^{\pm}$ at the sub-surface layer.
The Grahame equation depends only on the electrostatics properties
and applies for any boundary condition: constant $\sigma$, constant $\Psi_s$
or the present case of charge regulation
due to association/dissociation of the ionizable groups on the membrane.
These two equations determine complectly the surface potential, $\Psi_s = \Psi_s(K_d,n_b)$,
and the surface charge fraction, $\phi_s = \phi_s(K_d,n_b)$.

It is instructive to take the two opposite limits of large and small $\phi_s$.
When the ionizable surface sites are almost fully dissociated, $\phi_s \simeq 1$,
$K_d$ is large enough so that $K_d \gg n_b\exp(\Psi_s)$.
In this limit, Eqs.~(\ref{s4}) and (\ref{s9}) reduce to
\begin{eqnarray}
\label{s10}
\nonumber& &\phi_s \simeq 1 - \frac{n_b}{K_d}\,\e^{\Psi_s} \, , \\
& &\Psi_s \simeq 2\ln\left( w + \sqrt{1+w^2}  \right) \, ,
\end{eqnarray}
and for the additional requirement of large $w$ (or $\ld \gg a^2/\lb$)
\begin{eqnarray}
\label{s11}
\Psi_s \simeq 2\ln 2w = 2\ln \left( \frac{8\pi\lb\ld}{a^2} \right) \, .
\end{eqnarray}

In the opposite limit, only a small fraction of the ionizable surface groups are
dissociated, $\phi_s \ll 1$, which results in $K_d \ll n_b \exp(\Psi_s)$ and
\begin{eqnarray}
\label{s12}
\phi_s \simeq \frac{K_d}{n_b}\e^{-\Psi_s} \ll 1 \, .
\end{eqnarray}
If, in addition to small $\phi_s$ also $w\phi_s \ll 1$, expanding
the right-hand side of Eq.~(\ref{s9}) gives
\begin{eqnarray}
\label{s13}
\Psi_s \simeq 2w\phi_s \ll 1 \, .
\end{eqnarray}

\begin{figure}%13
\centering
\includegraphics[scale=0.7]{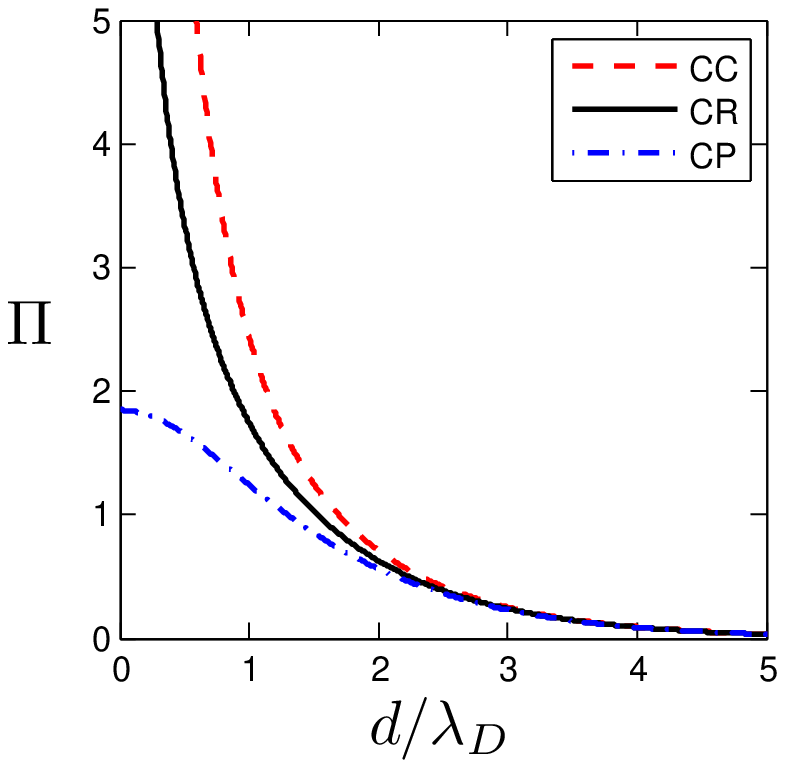}
\caption{\textsf{ The pressure for two symmetric membranes in units of $\kbt/(4\pi\lb\ld^2)$,
in the presence of three different boundary
conditions: constant charge (CC), constant potential (CP) and charge regulation (CR).
The pressure inequality, seen in the figure, $\Pi_{\rm CC} > \Pi_{\rm CR} > \Pi_{\rm CP}$,
is an inequality that holds in general.
The parameters used are: $a = 0.5 \, {\rm nm}$, $n_b = 0.1\, {\rm M}$ and $\alpha = -6$ (${\rm pK} \simeq 1.48$).}
}
\label{figure13}
\end{figure}

As stated in the beginning of this section, the charge regulation boundary condition
lies in between the constant charge and constant potential boundary conditions.
In other words, the osmotic pressure obtained for constant charge (CC) boundary condition, $\Pi_{\rm CC}$,
will always be more repulsive than the pressure, $\Pi_{\rm CR}$, of charge regulation (CR).
Furthermore, the latter is even more repulsive than $\Pi_{\rm CP}$, the pressure for constant potential (CP),
yielding $\Pi_{\rm CC} > \Pi_{\rm CR} > \Pi_{\rm CP}$.
A calculated example manifests this fact and can be seen in Fig.~\ref{figure13}, where we show the osmotic pressure, $\Pi$,
in units of $\kbt/(4\pi\lb\ld^2)$ as a function of the dimensionless inter-membrane separation, $d/\ld$.
We use Eqs.~(\ref{m1}) and (\ref{s5}) to obtain numerically the surface potential, $\Psi_s$,
for the symmetric charge regulation boundary condition for any inter-membrane separation, $d/\ld$.
The midplane potential is obtained from Eq.~(\ref{m3a}), and the pressure is calculated via Eq.~(\ref{k2}).
The difference in the osmotic pressure for the three boundary conditions arises only for small separations, $d \simeq \ld$,
while in the large separation limit, $d\ \gg \ld$, the three pressures coincide.
From the extrapolation of the CR surface potential at large separations, $\Psi_s(d \to \infty)$,
we find the constant surface potential to be, $\Psi_s^{(\rm const)} \simeq -1.7$,
while the constant surface charge is $\sigma_{\rm const} \simeq -e/4.35 \, {\rm nm}^2$.

The  behavior of $\Pi_{\rm CP}$ for small distances, $d \ll \ld$, is very different from the behavior of
$\Pi_{\rm CC}$ and of $\Pi_{\rm CR}$, as $\Pi_{\rm CP}$ saturates at a value of $\Pi_{\rm CP} \simeq 1.85$
(in units of $\kbt/(4\pi\lb\ld^2)$).
For $d/\ld \to 0$ and finite $\lgc$, $\Pi_{\rm CC}$ is always in the ideal gas regime (see Table~\ref{table1}), thus diverging as $\Pi_{\rm CC} \sim 1/d$.
Note that the osmotic pressure of CR also diverges when $d/\ld \to 0$,
but only as $\Pi_{\rm CR} \sim 1/d^{1/2}$, which is weaker than for $\Pi_{\rm CC}$ (Markovich, Andelman and Podgornik, to be published).

%%%%%%%%%%%%%%%%%%%%%%%%%%%%%%%%%%%%%%%%%%%%%%%%%%%%%%%%%%%%%%%%%%%%%%%%%%%%%%%%%%%%%
\subsection{Charge Regulation via Free Energy} \label{sec9.7.1}
%%%%%%%%%%%%%%% Free energy derivation %%%%%%%%%%%%%%%%%%%%

The Langmuir-Davies isotherm, Eq.~(\ref{s6}),
can also be derived from a free-energy minimization (Diamant and Andelman 1996).
It is done by including a surface free energy, $F_s$, to account
for the association/dissociation of ions onto/from the membrane.
The total free energy is $F_t = F_v + F_s$,
where the volume contribution, $F_v$, is the same as in Eq.~(\ref{b13})
and the surface free-energy is
\begin{eqnarray}
\label{u1}
 a^2 F_s / A\kbt = \Psi_s\phi_s
+  \phi_s\ln\phi_s + \left(1-\phi_s\right)\ln\left(1-\phi_s\right)
 + \alpha \left(1-\phi_s\right) \, ,
\end{eqnarray}
where $A$ is the lateral membrane area and $a$ is its thickness.
The first term describes the coupling between the surface charge density $\sigma = e\phi_s/a^2$
(taken as positive), and the surface potential $\Psi_s = e\psi_s/\kbt$.
The second and third terms describe the mixing entropy of dissociated (charged)
surface sites of fraction $\phi_s$, and associated (thus neutral) ones of fraction
$1-\phi_s$.
Finally, the fourth term is proportional to $1-\phi_s$, the amount of ${\rm B}^-$
ions adsorbing onto the membrane.
It accounts for the excess surface interaction as an ion binds onto the membrane
creating a neutral AB group.
The surface interaction parameter, $\alpha$, is the same as the one defined after Eq.~(\ref{s4}).
%
%Finally, the fourth term includes a cooperativity between
%counter-ions at the surface.
%This is modeled by adding a square term in the coverage.
%Usually this term is negative, $b<0$, which reflects an attraction between counter-ions on the surface.

In order to obtain the surface isotherm, one needs to take
the variation of $F_t$ with respect to $\phi_s = \sigma a^2/e$,
$\delta F_t/\delta\phi_s = \delta F_s/\delta\phi_s = \mu_s$, giving
\begin{eqnarray}
\label{u2}
\Psi_s  + \ln\left(\frac{\phi_s}{1-\phi_s}\right) + \left(\beta\mu_s-\alpha\right) = 0 \, .
\end{eqnarray}
For a dilute ionic solution, the chemical potential of the ions is related
to their bulk density by $\mu_b = \kbt\ln \phi_b$, as explained after Eq.~(\ref{b3}).
In thermodynamical equilibrium the chemical potential is equal throughout the solution, hence, $\mu_s = \mu_b$,
and by rearranging Eq.~(\ref{u2}) the Langmuir-Davies isotherm emerges,
\begin{eqnarray}
\label{u3}
1-\phi_s = \frac{\phi_b}{\phi_b + \e^{ \alpha - \Psi_s}} \, .
\end{eqnarray}
The above equation is exactly the Langmuir-Davies isotherm of Eq.~(\ref{s6}).

The advantage of the free-energy formulation presented in this section
over the chemical equilibrium one presented earlier is that the former can be generalized to other cases of surface interaction,
such as cooperativity between the surface sites modeled by adding a $b\left(1-\phi_s\right)^2$
term to $F_s$, adsorbing of several ion types with different ion-surface interactions, $\alpha$,
and other extensions of the simple charge regulation mechanism (Diamant and Andelman 1996, Ariel, Diamant and Andelman 1999).
%

%%%%%%%%%%%%%%%%%%%%%%%%%%%%%%%%%%%%%%%%%%%%%%%
\section{Van der Waals Interactions}  \label{sec9.8}
%%%%%%%%%%%%%%%%%%%%%%%%%%%%%%%%%%%%%%%%%%%%%%%

Long-range van der Waals (vdW) interactions between molecules are universal and result from the
molecular dipolar fluctuations (for details see Parsegian 2005, Bordag et al. 2009).  These fluctuations can
have different origins. For polar molecules with permanent dipoles their orientational fluctuations lead to Keesom interaction.
When orientational fluctuations of a permanent dipole induce a dipole in another non-polar but polarizable molecule, the induced dipole
leads to Debye interaction.  In all remaining cases, transient dipoles of nonpolar polarizable molecules
induce other transient dipoles and lead to London dispersion interactions.
In general, one can write the total vdW interaction potential $V({\bf r})$ between two molecules at positions ${\bf r}_1$ and ${\bf r}_2$ separated by
${\bf r} = {\bf r}_1 - {\bf r}_2$, in the form
\begin{equation}
V(\vecr) = - \frac{\cal C}{{r}^6} = - \frac{3 \kbt\alpha_1 \alpha_2}{(4\pi\varepsilon_0)^2}\frac{1}{r^6} \, .
\end{equation}
but we get a factor
The ${1}/{{r}^6}$ interaction reflects the dipolar nature of vdW interaction: dipolar
field of the first molecule decays as ${1}/{{r}^3}$, interacts with the second molecule, and
then propagates back to the first molecule, yielding a squared dipolar interaction, ${1}/{{r}^6}$.
This argument does not take into account the relativistic corrections and, thus, corresponds
to the non-retarded case.

The prefactor $\cal C$ gives the strength of the interaction and is proportional
to the product of polarizabilities, $\alpha_1 \alpha_2$, of the two molecules and, thus, also to the product of molecular
volumes. If the molecules interact in a medium of dielectric constant $\varepsilon_w$,
then the strength of the interaction is proportional to $(\alpha_1 \alpha_2)/\varepsilon_w^2$.

%%%%%%%%%%%%%%%%%%%%%%%%%%%%%%%%%%%%%%%%%%%%%%
\subsection{The Hamaker Pairwise Summation}  \label{sec9.8.1}
%%%%%%%%%%%%%%%%%%%%%%%%%%%%%%%%%%%%%%%%%%%%%%%

The simplest way to account for the vdW interactions between two large macroscopic bodies is called the {\it Hamaker summation}.
It results from the pairwise summation of the molecular vdW interactions, yielding for the interaction free energy
\begin{equation}
\mathcal{F} = - \frac{3 \kbt}{(4\pi\varepsilon_0)^2} \int_{V_1}\D^3 r_1\int_{V_2}\D^3 r_2
\frac{\alpha_1 n_1(\vecr_1)\alpha_2 n_2(\vecr_2)}{\vert \vecr_1 - \vecr_2\vert^6} \, ,
\end{equation}
where $n_{1,2}$ are the densities of the molecules in the two bodies,
and the volume integrals go over the volumes of the two bodies, $V_1$ and $V_2$.

\begin{figure}%14a
\centering
\includegraphics[scale=0.3]{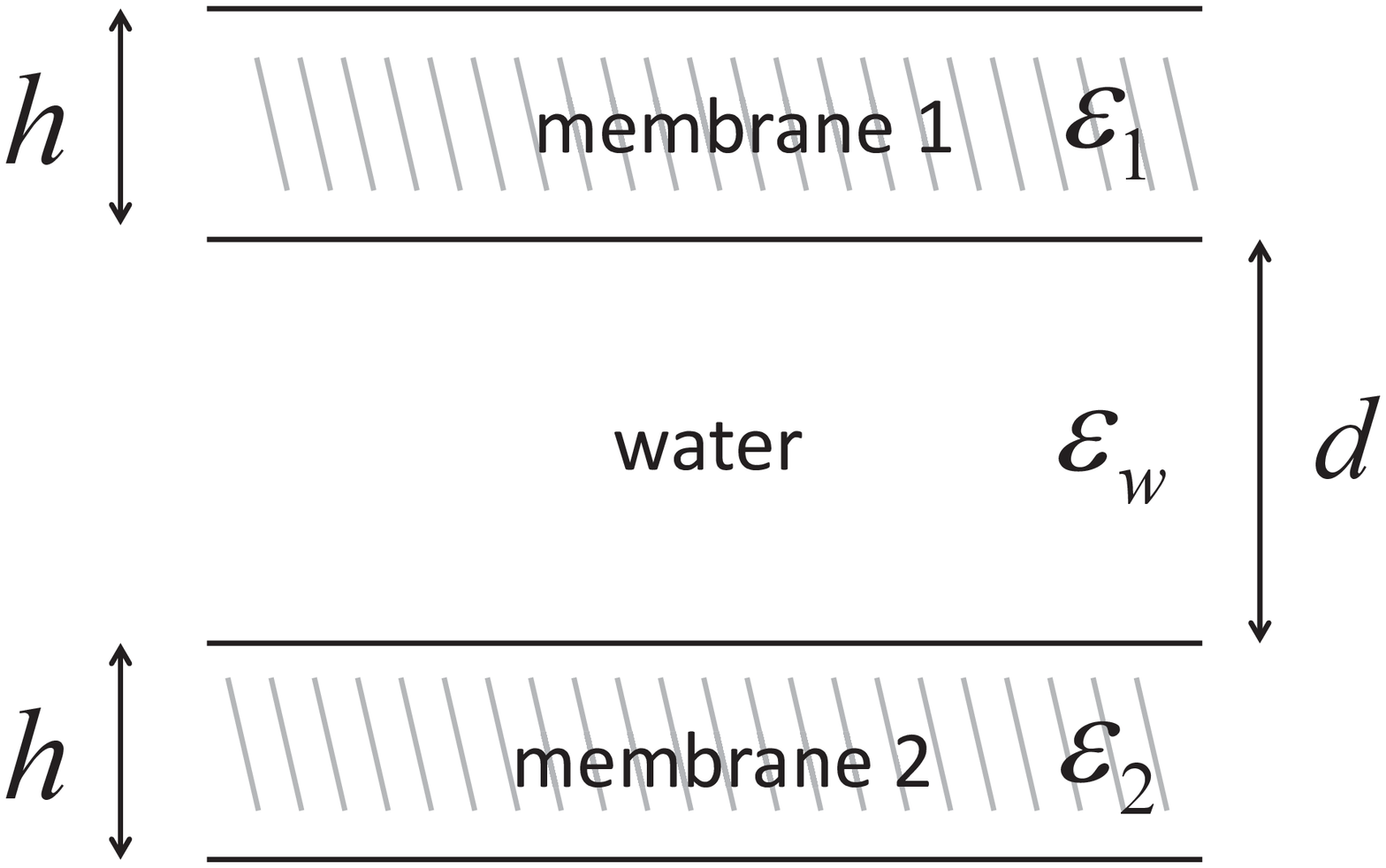}
\caption{\textsf{Illustration of two planar membranes of thickness $h$ each and with dielectric constants $\varepsilon_1$
and $\varepsilon_2$ interacting across a water slab of thickness $d$ of dielectric constant $\varepsilon_w$.}
}
\label{figure14a}
\end{figure}

For two planar membranes of constant molecular density $n_1$ and $n_2$,  each of finite thickness $h$
at a separation $d$ (see Fig.~\ref{figure14a}), the above Hamaker integral yields an interaction energy per unit surface area $A$ of the form
\begin{equation}
\frac{\mathcal{F}(d, h)}{A} = - \frac{3 \kbt \alpha_{1}n_1\alpha_{2}n_2}{(4\pi\varepsilon_0\varepsilon_w)^2}
 \int_{d/2}^{d/2 + h} \D z_1 \int^{-d/2}_{-d/2 - h} \D z_2
\int_0^\infty \frac{2\pi\rho \D\rho}{\left[ (z_1 - z_2)^2 + \rho^2\right]^3} \, ,
\end{equation}
where $\vecr = (z,{\bf \rho})$ on cylindrical coordinates.
In this case of two planar dielectric media (membranes) interacting across a gap of dielectric constant $\varepsilon_w$ (water),
the excess polarizabilities are given by $n_{1,2} \alpha_{1,2} = 2 \varepsilon_0 \varepsilon_w
(\varepsilon_{1,2} - \varepsilon_w)/(\varepsilon_{1,2} + \varepsilon_w)$
(Israelachvili 2011).
The integrals over $z_1$ and $z_2$ can be evaluated analytically, yielding
\begin{equation}
\frac{\mathcal{F}(d, h)}{A} = - \frac{{\cal H}}{12\pi }\left( \frac{1}{d^2} - \frac{2}{(d+h)^{2}}  +\frac{1}{(d + 2h)^{2}}\right) \, ,
\label{bgjkrw}
\end{equation}
with a prefactor defined as the {\it Hamaker constant}
\begin{equation}
{\cal H} = \frac{3\kbt}{4}\left( \frac{\varepsilon_1 - \varepsilon_w}{\varepsilon_1 + \varepsilon_w} \right)
\left( \frac{\varepsilon_2 - \varepsilon_w}{\varepsilon_2 + \varepsilon_w} \right) \, .
\end{equation}
From this general expression one can derive two interesting scaling limits for small and large inter-membrane separation, $d$.

For small separations, $d \ll h$, corresponding to the
vdW interaction between two semi-infinite media separated by distance $d$,
\begin{equation}
\frac{\mathcal{F}(d)}{A} \simeq - \frac{{\cal H}}{12\pi d^2} \sim \frac{1}{d^2} \, ,
\label{appro-1}
\end{equation}
while for large separation, $d \gg h$, corresponding to the interaction of two thin sheets,
\begin{equation}
\frac{\mathcal{F}(d,h)}{A} \simeq - \frac{{\cal H} h^2}{2\pi d^4} \sim \frac{1}{d^4} \, .
\label{gefh}
\end{equation}
The $d$ dependence obtained from the Hamaker summation is, to the lowest order, the same as obtained in more
sophisticated approaches.
However, the Hamaker constant, ${\cal H}$, can only be taken heuristically and, in fact,
cannot be obtained from the simple pair-wise summation procedure.

%

%%%%%%%%%%%%%%%%%%%%%%%%%%%%%%%%%%%%%%%%%%%%%%%%%%%%%%%%%%%%%%%%%%%%%%%%%%%%%%%%%%%%%
\subsection{The Lifshitz Theory}  \label{sec9.8.2}
%%%%%%%%%%%%%%%%%%%%%%%%%%%%%%%%%%%%%%%%%%%%%%%%%%%%%%%%%%%%%%%%%%%%%%%%%%%%%%%%%%%%%

A more sophisticated theory for vdW interactions between macroscopic bodies was developed by J. M. Lifshitz in the 1950s (see Parsegian 2005).
In the Lifshitz theory, the vdW interactions are {\it electromagnetic fluctuation interactions}  and
the Hamaker coefficient, ${\cal H}$, is a functional of the frequency-dependent dielectric permeabilities of the interacting media.
It can be evaluated from either experimentally determined dispersion properties or calculated dispersion spectra of the interacting materials.
It consistently includes the relativistic retardation effects due to the finite velocity of light propagation and finite temperature effects.

In the Lifshitz theory, the free-energy of two planar semi-infinite bodies is
typically cast into the Hamaker-type form (Safran 1994, Parsegian 2005)
\begin{equation}
\frac{\mathcal{F}(d,h)}{A} = - \frac{{\cal H}(d,h)}{12 \pi d^2} \sim \frac{1}{d^2} \, ,
\label{gbfhjwd}
\end{equation}
as in Eq.~(\ref{appro-1}),
but with the important difference that the $d$-dependent ${{\cal H}(d)}$  can now be calculated explicitly via the Lifshitz formalism, when
dielectric frequency-dependent properties of the interacting materials are available.

The most important characteristics of vdW interactions in the context of bio-matter and, specifically,
inter-membrane interactions comes from the presence of solvent, {\it i.e.}, water (Ninham and Parsegian 1970), and can be consistently taken
into account within the Lifshitz theory.
The calculated non-retarded Hamaker constant between lipid bilayers in water is found to be in the
range $10^{-20} - 10^{-21} \, {\rm J}$.
The high static dielectric constant of water
and the low static dielectric constant of hydrocarbons (consisting of the membrane core region)
leads to an anomalously large contribution to the entropic part of the vdW free-energy,
which remains unretarded at all separations. Some characteristic values of the Hamaker constant for interaction of different materials
across a water layer are given in Table~\ref{table2}.

\begin{table} [!h]
\center
\begin{tabular}{ c  c}
 \hline\noalign{\medskip}
 Material \,\,  & Hamaker constant $\times 10^{21}$ [J] \,\,  \\ \noalign{\medskip}\hline \\
 Polystyrene &  13 \\
 Polycarbonate & 3.5 \\
 Hydrocarbons & 3.8 \\
 Polymethyl mathecrylate & 1.47 \\
 Proteins & 5-9   \\ \\ \hline
\end{tabular}
\caption{ \label{table2}
\textsf{Values of the Hamaker constant for different materials interacting across water (from Parsegian 2005).}}
\end{table}

%%%%%%%%%%%%%%%%%%%%%%%%%%%%%%%%%%%%%%%%%%%%%%%%%%%%%%%%%%%%%
\begin{shadebox}
\Boxhead{Lifshitz Theory}
The vdW interaction free-energy is obtained in the Lifshitz theory as a sum, Eq.(\ref{cgsrlhjk}), over discrete imaginary Matsubara frequencies,
\begin{equation}
\zeta_n=\frac{2\pi\kbt n}{\hbar} \, , \qquad n = 0, 1, 2 \dots \, ,
\end{equation}
where $\hbar=h/2\pi$ and $h$ is the Planck constant.
The terms of this sum involve the dielectric response functions of the different media:
the two lipid bilayers separated by a slab of aqueous medium.
The Hamaker constant in Eq.~(\ref{gbfhjwd}) depends on the dielectric response functions, and can be obtained quantitatively.

The dielectric response function at imaginary frequencies, $\varepsilon(i \zeta)$,
is given formally by the Kramers-Kronig relation (Smith 1985)
\begin{equation}
\varepsilon(i \zeta) = 1 + \frac{2}{\pi} \int_0^{\infty} \frac{\omega ~{\varepsilon'' (\omega)}}{\omega^2 + \zeta^2}\D\omega \, ,
\label{cgsrlhjk}
\end{equation}
with $\varepsilon''(\omega)$ being the imaginary part of the complex frequency dependent-dielectric function,
$ \varepsilon(\omega) = \varepsilon'(\omega) + i\varepsilon''(\omega)$.
Quite generally, $\varepsilon(i \zeta)$ is a real, monotonically decreasing function of its argument, $\zeta$.
The  Kramers-Kronig relation also establishes the connection
between the vdW interactions and the measurable dispersion
part of the dielectric response functions, $\varepsilon''(\omega)$.
For this reason the vdW interactions are also referred to as the {\it dispersion interactions}.
The imaginary frequencies can be  rationalized intuitively as follows:
just as $\varepsilon(\omega)$ characterizes the temporal response of a material
to an external oscillating electric-field $\sim \exp(i \omega t)$, $\varepsilon(i \zeta)$
characterizes the spontaneous time decaying  fluctuation $\sim \exp(-\zeta t)$.
\end{shadebox}
%%%%%%%%%%%%%%%%%%%%%%%%%%%%%%%%%%%%%%%%%%%%%%%%%%%%%%%%%%%%%

From the full Lifshitz formula for interacting lipid membranes, the limit of thick membranes,
$h \gg d$, Eq. (\ref{appro-1}),
without any retardation effects, yields the Hamaker constant in the form
\begin{equation}
%{\cal F}(d)/A \simeq - \frac{{\cal H}}{12 \pi d^2} \quad {\rm where} \quad
{\cal H} = \frac{3}{2} k_BT ~{\sum_{n=0}^{~~\infty_{~~\prime}}} \, \left[\,\Delta(i \zeta_{n})\right]^2 \, ,
\label{kigjoyp}
\end{equation}
%\footnote{DA: Since the sum with the prime give an extra factor of 1/2 for the $n=0$ mode, this eq. doesn't seem
%consistent with eq. (1.131). This is not related to the previous question about 1/2. But is another question. }
%
where the dielectric contrast is defined as
\begin{equation}
\Delta (i \zeta_{n}) =
\frac{\epsilon_{L}(i \zeta_{n}) -  \epsilon_{w}(i \zeta_{n}) }
     {\epsilon_{L}(i \zeta_{n}) +   \varepsilon_{w}(i \zeta_{n}) }  \, ,
\end{equation}
in terms of the dielectric response function between the interacting media, {\it i.e.} the lipid $\epsilon_{L}$ and water $\epsilon_{w}$, at imaginary frequencies.
Note that the prime in the summation of Eq.~(\ref{kigjoyp}), $\sum^{\prime}$,
means that we have taken the lowest $n = 0$ term with weight $1/2$.

Standard forms for these dielectric responses can be used (Mahanty and Ninham 1976, Dagastine, Prieve and White 2000),
where the dielectric response of water is described by twelve different $\zeta$ frequencies:
one microwave relaxation frequency, five infrared relaxation frequencies and six ultraviolet relaxation frequencies.
Similarly, the hydrocarbons materials (lipid membrane) is modeled by four $\zeta$
ultraviolet relaxation frequencies (for details see Parsegian 2005).
In this limit of thick membranes, $h \gg d$, the Hamaker constant is indeed a constant, independent of the separation $d$,
but becomes $d$-dependent, ${\cal H}  = {\cal H} (d)$, if the finite velocity of light is taken into account.
These retardation effects change the scaling
of the interaction free-energy from the $1/d^{2}$ into a $1/d^{3}$,
at separations of $d \simeq 10 - 100 ~\rm nm$,
usually too large to be of practical importance for interacting lipid membranes.
We further stress that the summation over the discrete frequencies set
is something that cannot be derived from a simple Hamaker summation procedure of section~\ref{sec9.8.1}.

Using model expressions for the dielectric response of the hydrocarbon
core of lipid bilayers and the aqueous medium, one ends up with the value of $4.3 \times 10^{-21} {\rm J}$
for the relevant Hamaker coefficient (Podgornik, French and Parsegian 2006).
A comparable value has been obtained in experiments for small membrane spacings.
For example, it was found that the values of the Hamaker constant
for dimyristoyl phosphatidylcholine (DMPC) and for dipalmitoyl phosphatidylcholine (DPPC)
membranes forming a multi-stack is in the range of  $2.87-9.19 \times 10^{-21} \, {\rm J}$ (Petrache et al. 1998).

For thin membranes, $h \ll d$, the Lifshitz formula is
valid in the non-retarded limit.
The Hamaker constant of Eq.~(\ref{kigjoyp}) has the same scaling of $1/d^4$ as in the Hamaker summation of Eq.~(\ref{gefh}).
The same Hamaker constant is obtained also for membranes of finite thickness, see Eq.~(\ref{bgjkrw}).

VdW interactions for symmetric bodies, {\it e.g.}, two identical membranes interacting across a finite gap, are always attractive,
just as electrostatic interactions between two symmetric bodies in the PB theory are always repulsive (Neu 1999).
For asymmetric bodies, {\it e.g.}, material `$1$' interacting with material `$2$' across water `${\rm w}$',
the Hamaker constant is given by a generalization of Eq.~(\ref{kigjoyp})
\begin{equation}
{\cal H} = \frac{3}{2} k_BT ~{\sum_{n=0}^{~~\infty_{~~\prime}}} \,
\Delta_{1w}(i \zeta_{n}) \Delta_{w2}(i \zeta_{n}) \, .
\end{equation}
This form of the Hamaker constant allows us to propose approximate {\it combining relations}
that allow to extract unknown Hamaker constants from known ones (Israelachvili 2011).
For the case of two media of material `$1$' interacting over material `$3$',
the combining relation assumes the simple form ${\cal H}_{131} \simeq {\cal H}_{11 } + {\cal H}_{33 } - 2 {\cal H}_{13 }$,
where ${\cal H}_{13 }$ is for media `$1$' and `$3$' interacting across vacuum\footnote{Note that the combining relations can be also obtained
from the simpler Hamakar pair-wise summation of section~\ref{sec9.8.1}  }.
The Lifshitz form of the Hamaker constant for asymmetric bodies also suggests that the vdW interaction  can change sign, becoming repulsive.
Sometimes this repulsive vdW interaction is referred to as quantum levitation (Munday, Capasso and Parsegian 2009).

%%%%%%%%%%%%%%%%%%%%%%%%%%%%%%%%%%%%%%%%%%%%%%%%%%%%%%%%%%%%%%%%%%%%%%%%%%%%%%%%%%%%%
\subsection{The Derjaguin-Landau-Verwey-Overbeek (DLVO) Theory}  \label{sec9.8.3}
%%%%%%%%%%%%%%%%%%%%%%%%%%%%%%%%%%%%%%%%%%%%%%%%%%%%%%%%%%%%%%%%%%%%%%%%%%%%%%%%%%%%%

In the Derjaguin-Landau-Verwey-Overbeek (DLVO) theory (Verwey and Overbeek 1948), the total interaction energy
between charged bodies (colloidal particles or membranes) is assumed to be a simple sum of the electrostatic and vdW interactions.
For interacting planar bilayer membranes the total interaction free-energy is:
\begin{equation}
 \frac{1}{A}\mathcal{F}(d,h) = \frac{1}{A}\left[ \mathcal{F}_{\rm el}(d,h) +  \mathcal{F}_{\rm vdW}(d,h)\right] \, ,
 \label{bhfdjwk}
\end{equation}
where the electrostatic part, $\mathcal{F}_{\rm el}(d,h)$, is calculated in the PB framework as in section~\ref{sec9.5a}.
Since the PB osmotic pressure is easier to evaluate, the corresponding free-energy can be
obtained via the integral
\begin{equation}
\frac{1}{A}\mathcal{F}_{\rm el}(d,h) = \int_d^{\infty} \D\ell ~\Pi(\ell) \, ,
\end{equation}
with $\Pi$ calculated in section~\ref{sec9.7}.
%after Fig.~\ref{figure13}.
The vdW interaction free-energy, $\mathcal{F}_{\rm vdW}$,
is calculated either from the Hamaker summation procedure or, more appropriately, from the Lifshitz theory.
%with ${\cal H}_{\rm eff} \simeq 4.3 \, {\rm zJ}$.
%
\begin{figure}%14
\centering
\includegraphics[scale=0.7]{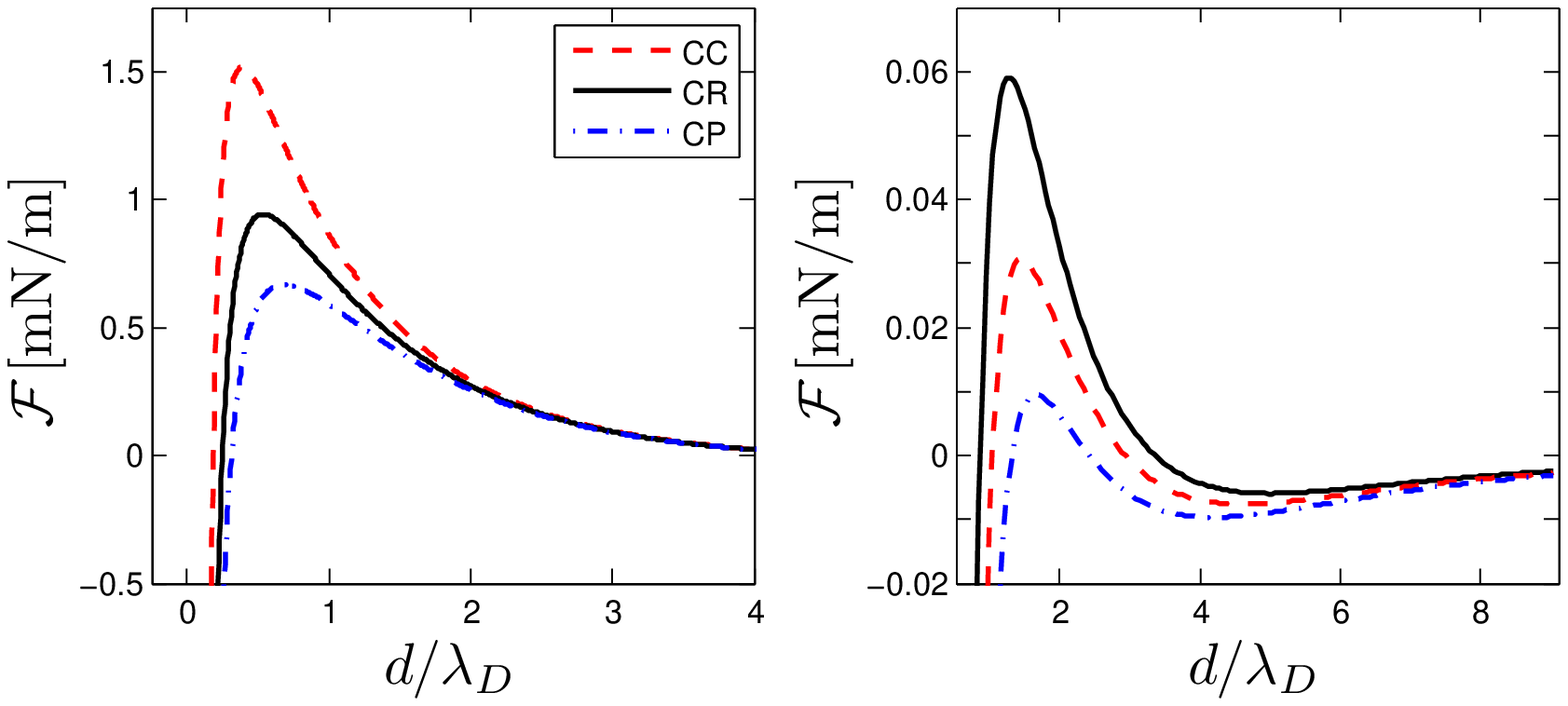}
\caption{ \textsf{The interaction free-energy density for two symmetric membranes,  $\mathcal{F}(d,h)/A$,  Eq.~(\ref{bhfdjwk}).
In (a) the electrostatic part is calculated with three different boundary
conditions: constant charge (CC), constant potential (CP) and charge regulation (CR).
The inequality, $\mathcal{F}_{\rm CC} > \mathcal{F}_{\rm CR} > \mathcal{F}_{\rm CP}$, seen here is a general relation.
The parameters used are: $a = 5 \, {\rm \AA}$, $n_b = 0.1\, {\rm M}$ and $\alpha = -6$ (${\rm pK} \simeq 1.48$).
In (b) we calculate the electrostatic part of the free energy for constant charge boundary conditions
with $\sigma = -e/7 \, {\rm nm}^2$ for three different bulk ionic concentrations, $n_b = 0.32\, {\rm M}$ (solid black line),
$n_b = 0.35\, {\rm M}$ (dotted red line) and $n_b = 0.38\, {\rm M}$ (dash-dotted blue line).
The membrane width in (a) and (b) is $h = 4 \, {\rm nm}$.
} }
\label{figure14}
\end{figure}
%

%The details of the total interaction free energy in the DLVO approximation of course depend on the charge model used to calculate the electrostatic contribution and the dielectric response model invoked in the calculation of the Hamaker coefficient.

Because of the $1/d^2$ dependence of the vdW free-energy, the total
interaction free-energy exhibits a universal {\em primary minimum} at vanishing spacings. However, since the assumption of the
continuum solvent is bound to break down in this small $d$ limit, this minimum is more related to the continuum theory assumption inherent in the
Lifshitz theory than to the physical interactions between membranes. In fact, a strong {\it hydration interaction}
(related to the breakdown of the continuum model of the solvent) usually obliterates the primary minimum,
and results in a monotonic repulsive interaction even for very small inter-membrane separations (Parsegian, Fuller and Rand 1979).

Apart from the  primary minimum at small separations, a secondary minimum can emerge at larger separations,
depending on the system parameters.
In Fig.~\ref{figure14} we plot some of these scenarios for various assumptions
on the electrostatic interactions.
The curves shown in  Fig.~\ref{figure14} embody the essence of the DLVO theory.
They can either show a monotonic repulsion extending over the whole ranges of separation,
or attraction that is turned into repulsion or vice versa.
In Fig.~\ref{figure14} (a), the DLVO free-energy is shown for the three different boundary conditions:
constant charge (CC), constant potential (CP) and charge regulation (CR).
Two minima are clearly seen: a minimum at $d/\ld \to 0$, which is the non-physical primary minimum mentioned earlier,
and a minimum at $d/\ld \to \infty$.

In Fig.~\ref{figure14} (b) we show the DLVO free-energy for CC boundary conditions with three different ionic concentrations:
$n_b = 0.32\, {\rm M}$ $(\ld \simeq 0.54 \, {\rm nm})$, $n_b = 0.35\, {\rm M}$ $(\ld \simeq 0.52 \, {\rm nm})$
and $n_b = 0.38\, {\rm M}$ $(\ld \simeq 0.5 \, {\rm nm})$.
As shown in the figure, for these bulk salt concentration a secondary, very shallow, minimum appears.
Increasing $n_b$ lowers the energy barrier and strengthen the shallow secondary minimum.
The appearance of this secondary minimum is an essential ingredient in the explanation of the
stability of colloidal particles and interacting membranes.
They come into stable equilibrium at this secondary minimum.
The high energy barrier between the secondary and the primary minimum makes it stable.
%

%%%%%%%%%%%%%%%%%%%%%%%%%%%%%%%%%%%%%%%%%%%%%%%%%%%%%%%%%%%%%%%%%%%%%%%%%%%%%%%%%%%%%
\section{Limitations and Generalizations}  \label{sec9.9}
%%%%%%%%%%%%%%%%%%%%%%%%%%%%%%%%%%%%%%%%%%%%%%%%%%%%%%%%%%%%%%%%%%%%%%%%%%%%%%%%%%%%%

The DLVO theory relies on approximations that have a finite range of validity (Naji et al. 2013).
First, the vdW interaction is not really decoupled from the PB mean-field formulation, and should be correspondingly modified.
And second, one of the central results of PB theory that symmetric bodies always repel each other, is {\it incorrect}.
For physically interesting situations involving highly charged interfaces, or multivalent mobile ions,
the electrostatic interaction can, in fact, be attractive.

These drawbacks of the classical DLVO theory, describing interactions between charged colloidal bodies
or interacting membranes, can be amended.
Recently, a new paradigm introduced a transparent systematization of the electrostatic interactions between charged bodies
in terms of two useful regimes: {\it weak coupling} (WC) and {\it strong coupling} (SC) (Boroudjerdi et al. 2005).
This allows a more accurate evaluation of the electrostatic interactions and the coupling between electrostatic and vdW interactions.
%The two limiting regimes also add  to different approaches tending to upgrade
%or supplement the traditional PB way (Grosberg, Nguyen and Shklovskii 2002).
%In fact, it can be shown that the zero-frequency vdW can be obtained from the fluctuations around the mean-field (PB) in the WC regime.

In order to introduce the WC and SC regimes one needs to consider
the relative strength of electrostatic interactions as compared
to the background thermal energy.
The thermal energy can be compared either with Coulomb interaction between two $q=ze$ charges of valency $z$
giving rise to a modified Bjerrum length, $z^2\lb$, as well as a modified Gouy-Chapman length, $\lgc/z$,
quantifying the strength of the electrostatic interaction between
a point charge ($q=ze$) and a surface charge density, $\sigma$.
Dividing the two lengths leads to a fundamental dimensionless
{\it electrostatic coupling parameter} introduced by Netz (2001)
\begin{equation}
\Xi = {z^2\lb}/{(\lgc/z)} = 2\pi z^3\ell_{\mathrm{B}}^2 |\sigma| \, .
\label{aclcadfls}
\end{equation}

For a system composed purely of counter-ions, the regimes of WC and SC can be understood in the following way.
When the coupling parameter is small, {\it i.e.}, $\Xi\ll 1$,
one goes back to the PB theory, with an addition of thermal fluctuations contributing
a vdW-like interaction that scales linearly with $\Xi$, and partially replaces the Lifshitz theory results.
%All the higher order, dispersion terms in the vdW interaction remain intact. Only the zero frequency is affected in this respect.
This clearly establishes a connection between electrostatic and vdW interactions.
%Furthermore, in this regime the vdW-like term scales linearly with $\Xi$ and is thus almost always subdominant,
%except in cases, such as antisymmetric interaction geometry, where the PB part is vanishing and the only remaining part os vdW.

\begin{figure}%15
\centering
\includegraphics[scale=0.7]{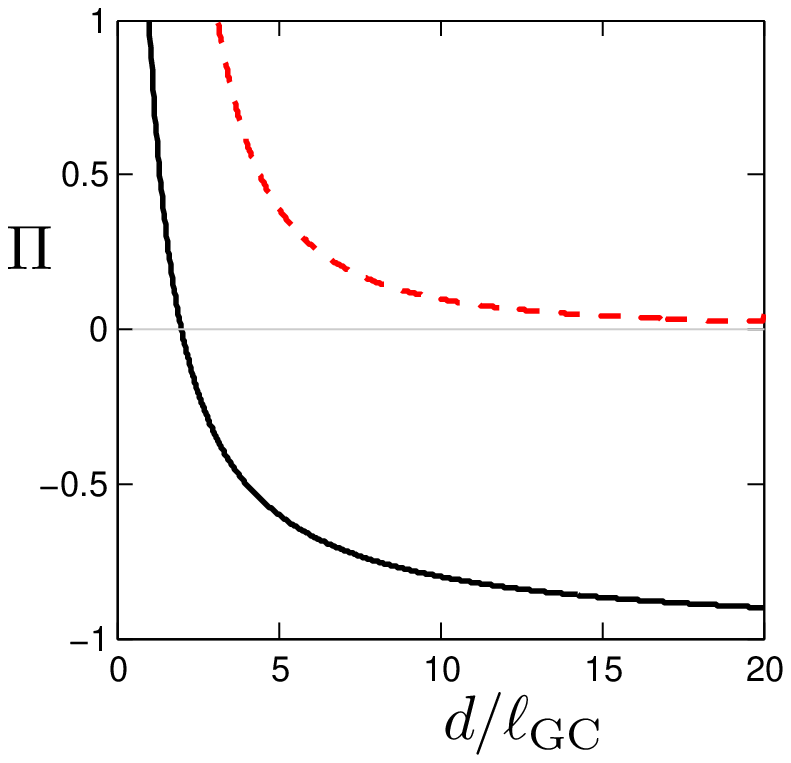}
\caption{ \textsf{The pressure for two symmetric membrane in units of $\kbt/(2\pi\lb\lgc^2)$,
as function of the dimensionless distance $d/\lgc$.
The calculation is done for the counter-ions only case as obtained from the PB theory (dashed red line), Eqs.~(\ref{l3}) and (\ref{l4}),
in comparison with the strong coupling (SC) result (solid black line), $\Pi = -1 + 2\lgc/d$ in units of $\kbt/(2\pi\lb\lgc^2)$
(Ghodrat et al. 2015). }
}
\label{figure15}
\end{figure}

In the opposite limit of large $\Xi\gg 1$, one observes a very different behavior
and important deviations from the standard DLVO paradigm.
Here, the mobile ions become strongly-correlated.
It leads to a fundamental consequence that the interactions between nominally equally-charged surfaces can become {\it attractive}.
This is shown clearly in Fig.~\ref{figure15} that compares the WC and SC results for the osmotic pressure
between two equally charged membranes.
Extensive Monte-Carlo simulations show that the WC to SC crossover is
associated with a hump in the heat capacity and the appearance
of short-range correlations between counter-ions for the coupling range $10<\Xi<100$ (Naji et al. 2013).
For very high values of the coupling parameter, $\Xi \simeq 3 \cdot 10^4$,
a transition to a Wigner crystalline phase (characterized by a diverging heat capacity) occurs.
%
%The latter exhibits a range of inter-membrane separations with an attractive pressure.
%The origin of this attraction are actually short-range correlations
%between counter-ions for the coupling range $10<\Xi<100$.\footnote{ TM: why not $10<\Xi$?}
Therefore, the whole DLVO idea that the interaction is composed of repulsive electrostatic interaction
and an attractive vdW one needs a serious revision for $\Xi \gg 1$ (SC regime).

%

%%%%%%%%%%%%%%%%%   HERE   %%%%%%%%%%%%%%%%%%%%%%%%%%%
%%%%%  The lat paragraph needs a revision %%%%%%%%%%%%%
% SC/WC   -----  DONE
% fluctuations + correlations   ------   DONE
% non-electrostatic interactions ----  DONE
% solvent effects: mixtures, $\varepsilon (n)$, $\varepsilon (\phi)$, dielectrophoretic   --- DONE
% polarizable ions, higher than dipolar ions,   -----  DONE
% solvent: dipolar    ---- DONE
% membrane structure: heterogeneities, undulations etc.   ---- DONE
%%%%%%%%%%%%%%%%%%%%%%%%

The WC/SC paradigm is clear for the simple counter-ions only case,
but becomes more complex in the presence of additional mobile charge components, {\it e.g.} polyvalent plus monovalent salt,
surface charge heterogeneity, mobile ion with multipolar structure and polarizable mobile ions.
These different components introduce new coupling parameters (similar to $\Xi$) that lead to additional features
making the general form of the electrostatic interactions more complicated than in the simple PB and DLVO framework.
%Nevertheless, one can always identify a principal coupling parameter, in analogy to the counter-ion only case, that implies a SC-like fixed point.
%The thusly defined SC fixed point then displays a fine structure and can bifurcate into other strongly coupled states governed by these additional
%length scales and the ensuing coupling parameters.

In this review we have presented in detail the PB treatment for mobile ions in solutions for planar geometries
as applicable to charged membranes.
The PB equation is a mean-field equation that is obtained from the zeroth-approximation in the WC regime
(Podgornik and \v{Z}ek\v{s} 1988, Borukhov, Andelman and Orland 1998, 2000, Netz and Orland 2000, Markovich, Andelman and Podgornik 2014, 2015)
Our chapter does not treat the fluctuations around the mean-field solution explicitly, but only indirectly via the vdW interactions.
Thermal fluctuations, which represent ion-ion correlations, have a key role in many interesting phenomena.
Among them, we mention surface tension of electrolyte solutions (Onsager and Samaras 1934, Markovich, Andelman and Podgornik 2014, 2015),
and their dielectric decrement (Ben-Yaakov, Andelman and Podgornik 2011, Levy, Andelman and Orland 2012).
Furthermore, we do not cover the SC regime, but only gives some of its interesting results in this last section.

Apart from a more accurate treatment of electrostatic interactions, other simplifying assumptions were made at the base
of the PB and DLVO theories presented in this review.
As noted in the beginning of the chapter, the membrane structure was completely ignored.
Therefore, membrane heterogeneities, curvature and undulations were not discussed.
The solvent (water) was treated as a featureless media with dielectric constant $\varepsilon_w$
and the effect of the mobile ions and the solvent structure ({\it e.g.} water permanent dipole)
on the decrement of the solvent dielectric constant was ignored (Ben-Yaakov, Andelman and Podgornik 2011, Levy, Andelman and Orland 2012).
This effect also give rise to a dielectrophoretic saturation of the counter-ions close to the membrane,
similar to the steric mPB saturation in section~\ref{sec9.4} (Nakayama and Andelman 2015).
Other extensions of the PB theory can be done by considering mixture of solvents (Ben-Yaakov et al. 2009)
or non-electrostatic interaction between the
mobile ions themselves such as hydration interactions (Burak and Andelman 2000)
or between the membrane and the mobile ions (Markovich, Andelman and Podgornik 2014, 2015).
For simplicity, the ions were treated as point-like particles which neglects their internal structure
and polarizability (see, {\it e.g.}, D\'emery, Dean and Podgornik 2012 and references therein).
Unfortunately, these interesting developments lie beyond the scope of the present
review and will be covered elsewhere (Burak, Orland and Andelman, to be published).
%The interactions between charged membranes therefore exhibit conceptual challenges and intriguing results,
%which have been corroborated mostly by computer simulations and, less often if at all, by experimental evidence.

%%%%%%%%%%%%%%%%%%  Table of Symboles  %%%%%
%%%%%%%%%%%%%%%%%%%%%%%%%%%%%%%%%%%%%%%%%%%%%%%%%%%
\begin{table} [!h]
%\center
\begin{tabular}{| l | l |}
  \hline \\
  \multicolumn{1}{| c |}{Symbol}  &    \multicolumn{1}{| c |}{Interpretation}
  \\  \\ \hline
  $a$   &   microscopic molecular size
  \\
  $\alpha = \ln(a^3 K_d)$   &   surface interaction parameter
  \\
  $\alpha_i$   &   polarizability of the $i^{th}$ molecule
  \\
  $\beta = 1/\kbt $   & inverse thermal energy
  \\
  $C_{\rm PB}$   &   differential capacitance of the PB model
  \\
  $C_{\rm mPB}$   &   differential capacitance of the mPB model
  \\
  $d$   &   inter-membrane separation
  \\
  $\varepsilon_w$   &   water dimensionless dielectric constant
  \\
  $\varepsilon_L$   &   lipid dimensionless dielectric constant
  \\
  $F$   &   Helmholtz free energy
  \\
  $\mathcal{F} = F(d) - F(d\to\infty)$   &   excess Helmholtz free energy
  \\
  $\phi_b = a^3 n_b$   &   bulk volume fraction for monovalent electrolyte
  \\
  $\phi_s = |\sigma|a^2/e$   &   surface area fraction
  \\
  $h$   &   membrane thickness
  \\
  ${\cal H}$   &   Hamaker constant
  \\
  $I = \half\sum_{i=1}^M z_i^2 n_i^{(b)}$   &   Ionic strength
  \\
  $K_d$   &   kinetic constant
  \\
  $\lb=e^2/(4\pi\varepsilon_0\varepsilon_w\kbt)$   &   Bjerrum length
  \\
  $\lgc = e/(2\pi\lb|\sigma|) \, ; \, \ell_{1,2} = e/(2\pi\lb|\sigma_{1,2}|)$   &   Gouy-Chapman length
  \\
  $\ld = \kd^{-1} =  (8\pi\lb n_b)^{-1/2}$   &   Debye length
  \\
  $\mu_i$   &   intrinsic chemical potential
  \\
  $\mu_i^{\rm tot}$   &   total chemical potential
  \\
  $n_i(\vecr)$   &   concentration of the $i^{th}$ ionic species
  \\
  $n_i^{(b)}$   &   bulk concentration of the $i^{th}$ ionic species
  \\
  $n_b$   &   bulk concentration of monovalent electrolyte
  \\
  $n_0$   &   reference density, taken at zero potential
  \\
  $n_i^{(s)} \, ; \, n_s$   &   surface density
  \\
  $n_i^{(m)} \, ; \, n_m$   &   midplane density (two membranes)
  \\
  $P$   &   pressure
  \\
  $\Pi = P_{\rm in} - P_{\rm out}$   &   osmotic pressure
  \\
  $\psi(\vecr)$   &   electrostatic potential
  \\
  $\Psi(\vecr) = \beta e\psi(\vecr)$   &   dimensionless electrostatic potential
  \\
  $\Psi_s \, ; \, \psi_s$   &   surface potential
  \\
  $\Psi_m \, ; \, \psi_m$   &   midplane potential
  \\
  $q_i = z_i e$   &   charge of the $i^{th}$ ionic species
  \\
  $\rho(\vecr)$   &   charge density of mobile ions
  \\
  $\rho_f(\vecr)$   &   charge density of fixed charges
  \\
  $\rho_{\rm tot}(\vecr)$   &   total charge density
  \\
  $\sigma \, ; \, \sigma_{1,2}$   &   surface charge density
  \\
  $V$   &   volume
  \\
  $z_i$   &   valency of the $i^{th}$ ionic species
  \\ \\ \hline
\end{tabular}
\caption{\textsf{ Table of symbols}}
\label{table3}
\end{table}
%%%%%%%%%%%%%%%%%%%%%%%%%%%%%%%%%%%%%%%%%%%%%%%%%%%%%

%%%%%%%%%%%%%%%%%%%%%%%%%%%

%%%%%%%%%%%%%%%%%%%%%

\thispagestyle{empty}

\ \pagebreak

\printindex

\end{document}